\documentclass[letterpaper,titlepage,11pt]{article}
\pdfoutput=1
\usepackage[colorlinks=true,urlcolor=blue,citecolor=magenta]{hyperref}
\usepackage{amssymb,amsmath,amsfonts}
\usepackage{color}
\usepackage{graphicx}
\usepackage{array}
\usepackage{epsfig}
\usepackage{epstopdf}
\usepackage{caption}
\usepackage{subcaption}
\usepackage{epsfig}
\usepackage{epstopdf}
\usepackage{eso-pic}
\usepackage{tensor}
\usepackage{float}

\def\clock{{\count0=\time
           \divide\count0 60
           \ifnum\count0<10 0\fi\the\count0
           \multiply\count0 -60 \advance\count0 \time
           :\ifnum\count0<10 0\fi \the\count0
         }}
\newcommand{\timestamp}{{\small\vbox{\hbox{\tt\jobname.tex}
\hbox{\the\day/\the\month/\the\year, \clock}}}}

\newcommand{\beq}{\begin{equation}}
\newcommand{\eeq}{\end{equation}}

\let\oldsqrt\sqrt
\def\sqrt{\mathpalette\DHLhksqrt}
\def\DHLhksqrt#1#2{%
\setbox0=\hbox{$#1\oldsqrt{#2\,}$}\dimen0=\ht0
\advance\dimen0-0.2\ht0
\setbox2=\hbox{\vrule height\ht0 depth -\dimen0}%
{\box0\lower0.4pt\box2}}

\numberwithin{equation}{section}
\begin{document}

\hypersetup{pageanchor=false}
\begin{titlepage}
 \vskip 1.8 cm

\centerline{\Huge \bf Black Probes of Schr\"{o}dinger Spacetimes}
\vskip 1.2cm

\centerline{\large {\bf Jay Armas and Matthias Blau }}

\vskip 1.0cm

\begin{center}
\sl  Albert Einstein Center for Fundamental Physics \\
\sl Institute for Theoretical Physics, University of Bern \\
\sl  Sidlerstrasse 5, 3012-Bern, Switzerland
\end{center}
\vskip 0.4cm

\centerline{\small\tt jay@itp.unibe.ch, blau@itp.unibe.ch}

\vskip 1.3cm \centerline{\bf Abstract} \vskip 0.2cm \noindent

We consider black probes of Anti-de Sitter and Schr\"{o}dinger spacetimes embedded in string theory and M-theory and construct perturbatively new black hole geometries. We begin by reviewing black string configurations in Anti-de Sitter dual to finite temperature Wilson loops in the deconfined phase of the gauge theory and generalise the construction to the confined phase. We then consider black strings in thermal Schr\"{o}dinger, obtained via a null Melvin twist of the extremal D3-brane, and construct three distinct types of black string configurations with spacelike as well as lightlike separated boundary endpoints. One of these configurations interpolates between the Wilson loop operators, with bulk duals defined in Anti-de Sitter  and another class of Wilson loop operators, with bulk duals defined in Schr\"{o}dinger. The case of black membranes with boundary endpoints on the M5-brane dual to Wilson surfaces in the gauge theory is analysed in detail. Four types of black membranes, ending on the null Melvin twist of the extremal M5-brane exhibiting the Schr\"{o}dinger symmetry group, are then constructed. We highlight the differences between Anti-de Sitter and Schr\"{o}dinger backgrounds and make some comments on the properties of the corresponding dual gauge theories.

\end{titlepage}
\hypersetup{pageanchor=true}

%\pagestyle{empty}
%\small
%\begin{spacing}{1}
%\tableofcontents
%\end{spacing}
%\tableofcontents
%\normalsize
%\newpage
%\pagestyle{plain}
%\setcounter{page}{1}

\tableofcontents

%%%%%%%%%%%%%%%%%%%%%%%%%%%%%%%%%%%%%%%%%%%%%%%%%%%%%%%%%%%%%%

\section{Introduction}
The in-depth study of Schr\"{o}dinger spacetimes and its physical applications began with the recent work of \cite{Balasubramanian:2008dm, Son:2008ye, Maldacena:2008wh, Adams:2008wt, Herzog:2008wg}, motivated by possible holographic applications to the description of condensed matter systems with non-relativistic symmetries.\footnote{See also \cite{Duval:2008jg} for earlier work on Schr\"{o}dinger spacetimes.} The embedding of these spacetimes into type IIB string theory and their relation to the AdS/CFT correspondence via the null Melvin twist of known supergravity branes \cite{Maldacena:2008wh, Adams:2008wt, Herzog:2008wg}, led to the exciting idea that perhaps holography could be understood in a spacetime with a different geometry than Anti-de Sitter (AdS). This idea consequently led to a series of exploratory works on Schr\"{o}dinger holography (see \cite{Hartong:2013cba} for a recent review and references therein).  

Holography, however, can only be properly tested when the bulk description and the boundary theory are simultaneously available. In the latter case - the dual field theory side - the authors of \cite{Maldacena:2008wh, Adams:2008wt, Herzog:2008wg} argued that Schr\"{o}dinger geometries (Sch) embedded in IIB string theory via the null Melvin twist should have a dual quantum field theory description in terms of a variant of Discrete Light Cone Quantization (DLCQ) (in the language of \cite{Adams:2008wt}, this is referred to as DLCQ$_\beta$ where $\beta$ is the parameter describing the deformation from null AdS to Schr\"{o}dinger), yielding a non-relativistic conformal field theory (CFT), e.g. of $\mathcal{N}=4$ Super Yang-Mills (SYM). However, the exact details and the precise prescription for performing this variant of DLCQ of SYM are still lacking, and it is therefore important to keep probing the properties of this dual field theory by means of well defined bulk calculations.

Thus taking a purely gravitational perspective, and noting that certain Schr\"{o}dinger geometries can be obtained from AdS via the null Melvin twist \cite{Maldacena:2008wh, Adams:2008wt, Herzog:2008wg, Imeroni:2009cs, Bobev:2011qx, Detournay:2012dz}, one is naturally led to trying to understand which geometrical and gravitational properties of AdS carry over to (or are inherited by) Schr\"{o}dinger and which ones do not (this was also the point of view adopted e.g. previously in \cite{Blau:2009gd, Blau:2010fh, Hartong:2010ec, Hartong:2013cba}).

It has been argued in \cite{Maldacena:2008wh} that this null Melvin twist (or null dipole, or DLCQ$_\beta$) deformation leaves invariant certain observables and correlation functions in the dual field theory, in particular those of local operators without external momenta. On the other hand, examples of observables widely studied in the context of AdS/CFT which do not fall into this class are expectation values of the Wilson loop operators \cite{Maldacena:1998im, Rey:1998ik}. It is therefore of particular interest to study the potential gravitational duals of such Wilson loop (and Wilson surface) operators in the Schr\"{o}dinger context, and this is the aim and content of this paper.

In the bulk theory on AdS$_5\times S^{5}$, a Wilson loop is described by a configuration where a string is stretched into the bulk and whose endpoints, representing a quark-antiquark pair, are fixed on the AdS boundary. In particular, the free energy $\mathcal{F}_{\text{loop}}$ of these configurations, obtained via a bulk calculation, has the interpretation of the quark-antiquark potential, which in this case, at zero temperature, is Coulomb-like and hence proportional to the inverse length $L$ of the quark-antiquark pair \cite{Maldacena:1998im, Rey:1998ik}. This dependence of the free energy on $L$ is in accordance with the expected form for a conformally invariant dual theory. Therefore, Wilson loops can be used as probes of some of the properties of the dual field theory and hence it is interesting to search for an analogous statement in Sch$_5\times S^{5}$, obtained via the null Melvin twist of the extremal D3-brane \cite{Maldacena:2008wh, Adams:2008wt, Herzog:2008wg}. In fact, we will see that the free energy of these configurations is no longer Coulomb-like, but instead interpolates between a result proportional to the inverse length $L$ and a result proportional to the inverse square of the length for large enough $\ell=\beta R^2$ - the parameter describing the deformation from AdS$_5$ with radius $R$ into Sch$_5$. Similar arguments apply to the case of the expectation value of Wilson surface operators and their bulk duals defined in AdS$_7\times S^{4}$ \cite{Maldacena:1998im}, and correspondingly in Sch$_7\times S^{4}$ - their null Melvin twisted version \cite{Mazzucato:2008tr}.

Wilson loop operators, dual to strings with endpoints on the AdS$_5\times S^{5}$ boundary, have a supergravity description in terms of fully backreacted fundamental strings via the open/closed string duality. A similar example is that of point particles blown up into D3-branes orbiting a great circle of the $S^{5}$, the so called giant gravitons \cite{McGreevy:2000cw}, which when fully backreacted have a supergravity description in terms of LLM geometries \cite{Lin:2004nb}. Therefore, constructing bulk configurations with holographic duals can be thought of as constructing supergravity solutions. An interesting question that arises is how these configurations are modified when the background is at finite temperature. In this context, finite temperature corrections to the quark-antiquark potential have been obtained in \cite{Brandhuber:1998bs, Rey:1998bq} by solving the Nambu-Goto action, i.e. by placing an extremal, zero-temperature, probe on a black hole background. This method however does not take into account the thermal excitations of the string degrees of freedom, which in practice means that the probe is not in thermal equilibrium with the background. A way to take this into account in the strong coupling, supergravity regime,\footnote{See the recent analysis of \cite{Grignani:2013ewa} for the thermal DBI action at weak coupling for the D3-brane.} is to use the blackfold approach \cite{Emparan:2009cs, Emparan:2009at}. This method, which we briefly recall in Sec.~\ref{method} consists in placing probe black fundamental strings or black branes, hence with an intrinsic temperature, on a given background and requiring thermodynamic equilibrium \cite{Grignani:2010xm, Grignani:2011mr, Grignani:2012iw, Armas:2012bk, Niarchos:2012pn, Niarchos:2013ia, Niarchos:2012cy, Armas:2013ota}. In the case of Wilson loops, this has been analysed in \cite{Grignani:2012iw} and led to quantitive and qualitative new features which were not captured by the analysis of \cite{Brandhuber:1998bs, Rey:1998bq}. Here, we describe the first application of this method to Schr\"{o}dinger spacetimes and construct perturbatively supergravity black strings and black membranes. We also consider black membranes, dual to Wilson surface operators, in AdS$_7\times S^{4}$ using this black probe method and find new features which were not observed in \cite{Chen:2008ds}.

This paper is organised as follows. In Sec.~\ref{sec1} we begin by reviewing the black probe method based on the blackfold approach for supergravity/M-theory \cite{Grignani:2010xm, Emparan:2011hg, Armas:2012bk} which will allow us to construct black string and black membrane geometries dual to certain gauge theory operators. Since the Schr\"{o}dinger spacetimes that we consider are obtained by performing the null Melvin twist of known supergravity solutions and consequently taking their near horizon limit, the resulting geometry is still a solution of the complete set of supergravity equations and hence we can apply the black probe method developed for supergravity/M-theory \cite{Grignani:2010xm, Emparan:2011hg, Armas:2012bk}.  In Sec.~\ref{adsstring} we review the work of \cite{Grignani:2012iw}, where Wilson loops at finite temperature were analysed, and extend it to thermal AdS. In Sec.~\ref{stringschrodinger} we consider analogous configurations in Sch$_5\times S^{5}$ and find three distinct types of configurations. One of these interpolates between the AdS configuration and another Schr\"{o}dinger configuration where the string has boundary endpoints along a spatial direction. Another of these configurations consists of a string with boundary endpoints separated along a null direction. We compare these configurations with the AdS case. In Sec.~\ref{Mtheory} we apply the method to bulk duals to Wilson surfaces in AdS$_7\times S^{4}$ and find dominant temperature corrections compared to the work of \cite{Chen:2008ds}. Afterwards we construct several such configurations in Sch$_7\times S^{4}$. In Sec.~\ref{discussion} we discuss open problems and future research directions. App.~\ref{app} contains some details on the solution space of black membranes in Sch$_7\times S^{4}$.

%%%%%%%%%%%%%%%%%%%%%%%%%%%%%%%%%%%%%%%%%%%%%%%%%%%%%%%%%%%%%%

\section{Black strings in Anti-de Sitter} \label{sec1}
In this section we begin by reviewing the black probe method based on the blackfold approach for supergravity and M-theory developed in \cite{Grignani:2010xm, Emparan:2011hg, Armas:2012bk} which allows us to perturbatively construct the black hole configurations presented throughout this work. We then review how this method was applied in order to construct black strings in AdS$_{5}\times S^{5}$ in the deconfined phase dual to finite temperature Wilson loops \cite{Grignani:2012iw} and generalise this construction to the confined phase. The fact that the leading order temperature contribution to the quark-antiquark potential is due to the blackness of the probe, and not due to the characteristics of the background, is highlighted. We further notice that previous results based on the method of extremal, zero-temperature, probes \cite{Brandhuber:1998bs, Rey:1998bq} are recovered by taking a non-trivial double scaling limit where the resulting finite temperature configuration is composed of locally flat, zero-temperature, fundamental strings appropriately redshifted. This will then serve as a means of comparison with the same configuration in Schr\"{o}dinger spacetime which will be analysed in Sec.~\ref{stringschrodinger}.

%%%%%%%%%%%%%%%%%%%%%%%%%%%%%%%%%%%%%%%%%%%%%%%%%%%%%%%%%%%%%%

\subsection{Black probe method} \label{method}
The open/closed string duality states that Dp-branes and fundamental $F_1$ strings are described either by a low-energy effective action at weak coupling, such as the Dirac-Born-Infeld (DBI) action and the Nambu-Goto (NG) action, or by a fully backreacted supergravity solution at strong coupling. When the low-energy action is used in the context of AdS/CFT in order to find Dp-brane or $F_1$ configurations in a closed string background, they can be related to certain operators in the dual zero-temperature gauge theory and simultaneously have a supergravity regime where the Dp-branes or $F_1$ strings are fully backreacted. The clearest example of this is that of giant gravitons which are solutions of the DBI action coupled to the background Ramond-Ramond field \cite{McGreevy:2000cw} dual to large N gauge theory operators \cite{Corley:2001zk, Balasubramanian:2001nh} and when fully backreacted give rise to the LLM geometries \cite{Lin:2004nb}. 

If the closed string background has a finite temperature, the usual DBI or NG action is no longer a good description of the weak coupling dynamics. Instead, for small temperatures one should quantise the world volume theory and include the thermalisation of the brane/string degrees of freedom which heat up as the brane is immersed in a background at finite temperature. In general, the thermal version of the DBI action at weak coupling is not known but it has recently been derived for the case of the D3-brane \cite{Grignani:2013ewa}. On the other hand, finite temperature configurations at weak coupling will have a strong coupling, fully backreacted, regime where the corresponding objects will be non-extremal, finite-temperature, black hole geometries.

In this paper we are interested in the strong coupling regime $g_s N_{(p)}\gg1$ with $g_s\ll1$ being the string coupling and $N_{(p)}\gg1$ the number of probe Dp-branes (in the case $p=1$ we write $N_{(1)}=k$ to designate the number of probe $F_1$ strings), where we have a supergravity/M-theory description. In this case, the black hole solutions can be constructed order-by-order in a derivative expansion using the blackfold approach \cite{Emparan:2009cs, Emparan:2009at,Grignani:2010xm, Emparan:2011hg, Armas:2012bk} and their construction is independent of there being a weakly coupled description of the same object or not. When considering AdS backgrounds we will be constructing the finite temperature dual of the gauge theory Wilson loops and Wilson surfaces operators but when considering Schr\"{o}dinger backgrounds we take a more gravitational point of view and construct different black hole geometries, regardless of whether or not these configurations are dual to some gauge theory operators in a non-relativistic CFT.

The blackfold approach consists of placing a number of $N_{(p)}$ black probe branes in a background spacetime with metric $g_{\mu\nu}$ and wrapping them along a world volume $\mathcal{W}_{p+1}$ with coordinates $\sigma^{a}$, whose position in the ambient space is parametrised by the mapping functions $X^{\mu}(\sigma^{a})$. In terms of a metric, we are constructing solutions for which their near-horizon geometry is that of a supergravity/M-theory black brane and the far away asymptotics is given by the arbitrary metric $g_{\mu\nu}$ via a matched asymptotic expansion procedure.\footnote{For an example of the implementation of this procedure for neutral black branes in pure vacuum see \cite{Emparan:2007wm}.} It has been shown that for this procedure to yield a valid gravity solution, in the case of neutral black branes it must satisfy, for stationary configurations, the equilibrium condition \cite{Emparan:2007wm, Camps:2012hw}
\beq \label{ext1}
T^{ab}{K_{ab}}^{\rho}=0~~,
\eeq
which, when ignoring possible background dilatonic fields and fluxes, can also be shown to hold for supergravity/M-theory charged branes \cite{Armasnew}. In Eq.~\eqref{ext1}, we have introduced the effective stress-energy tensor $T^{ab}$ describing the black probes and the extrinsic curvature of the geometry ${K_{ab}}^{\rho}$ which can be obtained via the relation ${K_{ab}}^{\rho}=\nabla_{a}\partial_b X^{\rho}$ where the covariant derivative $\nabla_{a}$ is compatible with $g_{\mu\nu}$ as well as with the induced metric $\gamma_{ab}=g_{\mu\nu}\partial_a X^{\mu}\partial_b X^{\nu}$. When the world volume has boundaries, Eq.~\eqref{ext1} must be supplemented with boundary conditions, which in this case are \cite{Armas:2012bk}
\beq \label{b1}
T^{ab}\eta_{b}|_{\partial\mathcal{W}_{p+1}}=0~~,~~J^{ba_1...a_{p}}\eta_{b}|_{\partial\mathcal{W}_{p+1}}=0~~,
\eeq
where $\eta_{b}$ is a unit normalised orthogonal covector to the boundary $\partial\mathcal{W}_{p+1}$ and $J^{a_1...a_{p+1}}$ denotes the $(p+1)$-dimensional effective current carrying the conserved total charge $Q_p$ of the black probe satisfying $\partial_a Q_p=0$.

For supergravity/M-theory probes the effective stress tensor takes the form of a perfect fluid \cite{Emparan:2011hg}, 
\beq
T^{ab}=P\gamma^{ab}+(\epsilon+P)u^{a}u^{b}~~,~~\epsilon+P=\mathcal{T}s~~,
\eeq
while the effective current reads
\beq
J^{a_1...a_{p+1}}=Q_p\thinspace\varepsilon^{a_1...a_{p+1}}~~,
\eeq
where $P,\epsilon,\mathcal{T},s$ denote the local pressure, energy density, temperature and entropy density respectively while $u^{a}$ is the normalised ($u^{a}u_{a}=-1$) fluid velocity which, for stationary configurations, must be aligned with a world volume Killing vector field $\textbf{k}^{a}$ such that $u^{a}=\textbf{k}^{a}/\textbf{k}$, with $\textbf{k}=|-\gamma_{ab}\textbf{k}^{a}\textbf{k}^{b}|^{\frac{1}{2}}$. The indices $a,b,c...$ label the $(p+1)$ world volume directions while the greek indices $\mu,\nu,...$ span the entire spacetime dimensions and $\varepsilon^{a_1...a_{p+1}}$ is the Levi-Civita tensor on $\mathcal{W}_{p+1}$. 

The fact that the probe is composed of a fluid with non-zero temperature gives rise to the required degrees of freedom that need to be heated up when placing the configuration in a background spacetime at finite temperature. The thermodynamic fluid variables, as well as the charge $Q_p$ and corresponding chemical potential $\Phi_p$, are given in terms of the horizon radius $r_0$ and the charge parameter $\alpha$ of the black brane by \cite{Emparan:2011hg}
\beq \label{t1}
\epsilon=\frac{\Omega_{(n+1)}}{16\pi G}r_0^{n}(n+1+n\sinh^2\alpha)~~,~~P=-\frac{\Omega_{(n+1)}}{16\pi G}r_0^{n}(1+n\sinh^2\alpha)~~,
\eeq
\beq
\mathcal{T}=\frac{n}{4\pi r_0\cosh\alpha}~~,~~s=\frac{\Omega_{(n+1)}}{4 G}r_0^{n+1}\cosh\alpha~~,
\eeq
and
\beq\label{t3}
Q_{p}=\frac{\Omega_{(n+1)}}{16\pi G}nr_0^n\sinh\alpha\cosh\alpha~~,~~\Phi_p=\tanh\alpha~~.
\eeq
Here, we we have focused on $D=10$ and $D=11$ supergravity/M-theory branes and introduced Newton's constant $G$ and the volume of an $(n+1)$-sphere $\Omega_{(n+1)}$ where the dimension $n$ is defined via $D=n+p+3$.

In the case of stationary configurations the equation of motion \eqref{ext1} can be integrated to an action \cite{Emparan:2011hg}
\beq \label{act1}
I[X^{\mu}(\sigma^{a})]=\int_{\mathcal{W}_{p+1}}d^{p+1}\sigma\sqrt{-\gamma}\thinspace P~~,
\eeq
where $\gamma$ is the determinant of the induced metric. The branes characterised by these thermodynamic quantities admit an extremal limit where the local temperature $\mathcal{T}\to0$. This limit is obtained by sending $\alpha\to\infty$ such that $\mathcal{T}\to0$ and $r_0\to0$ while keeping the total charge $Q_p$ constant. It is easy to see that in this case $P\to-Q_{p}$, i.e.,
\beq
\alpha\to\infty~~,~~r_0\to0~~,~~P\to-Q_p~~.
\eeq
In the context of AdS/CFT where the charge is quantised such that $Q_p=N_{(p)} T_{(p)}$ with $T_{(p)}$ being the tension of the Dp-brane (or in the case $p=1$,  $T_{(1)}=T_{F_1}$ being the tension of the fundamental $F_{1}$ string), the action \eqref{act1} yields the zero-temperature extremal action
\beq \label{act2}
I[X^{\mu}(\sigma^{a})]_{\text{ext}}=-N_{(p)} T_{(p)}\int_{\mathcal{W}_{p+1}}d^{p+1}\sigma\sqrt{-\gamma}~~,
\eeq
which differs from the weak coupling DBI description, valid for $g_s\ll1$ and $N_{(p)}=1$, by a multiplicative factor of $N_{(p)}$. The fact that we have obtained $N_{(p)}$ times the DBI action reflects the supersymmetry of the solutions.\footnote{For supersymmetric deformations of blackfold geometries see \cite{Niarchos:2014maa}.} 

Since solutions of \eqref{act1} are time independent, the problem can be posed in an Euclidean version, where the action \eqref{act1} is Wick rotated such that $\sigma^0=\tau\to i\tau$ and integrated over the time circle of radius $1/T$ where $T$ is the background temperature related to the local brane temperature via 
\beq \label{tT}
T=\textbf{k}\thinspace\mathcal{T}~~.
\eeq
This relation is one of the main improvements of the black probe method compared to the extremal probe method as it expresses that the probe temperature $\mathcal{T}$ must be in equilibrium with the background temperature $T$ via a redshift factor $\textbf{k}$. The resulting quantity, after the Wick rotation, is the Gibbs free energy
\beq \label{freeenergy}
\mathcal{F}[X^{\mu}(\sigma^{a})]=-\int_{\mathcal{B}_{p}}dV_{(p)}R_0\thinspace P~~,
\eeq
where $dV_{(p)}$ is the volume form on the $p$-spatial world volume $\mathcal{B}_{p}$ and $R_0$ is the local redshift factor obtained by computing the norm of a timelike background Killing vector $\xi$ on the world volume, i.e., $R_0^2=-\xi^{\mu}\xi_{\mu}|_{\mathcal{W}_{p+1}}$. The entropy $S$ and the energy/mass $M$ can be obtained from \eqref{freeenergy} via the relations
\beq \label{thermo}
S=-\frac{\partial \mathcal{F}}{\partial T}~~,~~M=\mathcal{F}-TS~~,
\eeq
where we have used the first law of thermodynamics $d\mathcal{F}=dM-TdS$ and assumed the fluid to be static, as it will be such cases that we will deal with in this paper. In order to obtain different black hole configurations we will be varying the free energy \eqref{freeenergy} and solving the equations of motion \eqref{ext1} for different configurations of the scalars $X^{\mu}(\sigma^{a})$ in several ambient spacetimes.

%%%%%%%%%%%%%%%%%%%%%%%%%%%%%%%%%%%%%%%%%%%%%%%%%%%%%%%%%%%%%%

\subsection{Black strings dual to finite temperature Wilson loops} \label{adsstring}
In this section we review the finite temperature Wilson loops in the deconfined phase constructed in \cite{Grignani:2012iw} by wrapping black strings using the method described above and extend it to the confined phase. The construction in the confined phase, even though of less physical interest since it does not exhibit a phase transition and hence no Debye screening effect, presents itself as a more striking example which highlights the differences between the black probe method and the extremal probe method \cite{Brandhuber:1998bs, Rey:1998bq}. We also show how the work of \cite{Brandhuber:1998bs, Rey:1998bq} is recovered by taking a non-trivial double scaling limit.

We consider the AdS black hole background with AdS radius $R$ and metric
\beq \label{m1}
ds^2=\frac{R^2}{z^2}\left(-f(z)dt^2+f(z)^{-1}dz^2+dx_{i}^2\right)+R^2d\Omega_{(5)}^2~~,~~f(z)=1-\gamma\frac{z^4}{z_0^4}~~,
\eeq
where $i=1,2,3$, $d\Omega_{(5)}^2$ is the metric on the five-sphere and $z_0=1/(\pi T)$ is the location of the horizon and $T$ its temperature. The boundary is located at $z=0$. The parameter $\gamma$ controls whether we are in the deconfined phase ($\gamma=1$) or in the confined phase ($\gamma=0$). In the latter case, the metric corresponds to thermal AdS. The Hawking-Page transition occurs when the background temperature exceeds $T_{\text{HP}}\sim1/R$ for which the case $\gamma=1$ is thermodynamically preferred. We have ignored the five-form fluxes present in the background since they do not play a role in the examples we consider.

%%%%%%%%%%%%%%%%%%%%%%%%%%%%%%%%%%%%%%%%%%%%%%%%%%%%%%%%%%%%%%

\subsubsection*{Embedding and solution}

We now follow the prescription of \cite{Grignani:2012iw} and choose a static string embedding ($p=1$, $n=6$) with its endpoints on a boundary spatial direction separated by a length $L$, at the locations $x_1=0$ and $x_1=L$, and stretched into the bulk along the $z$-direction. The embedding map is
\beq
t=\tau~~,~~z=\sigma~~,~~x_1=x(\sigma)~~,~~x_2=x_3=d\Omega_{(5)}=0~~.
\eeq
With this choice the problem becomes symmetric around $x_1=L/2$. The string is stretched from $(x_1,z)=(0,0)$ to $(x_1,z)=(L/2,\sigma_0)$ and back again to $(x_1,z)=(L,0)$. The induced metric on $\mathcal{W}_{2}$ becomes
\beq \label{ind1}
\gamma_{ab}d\sigma^{a}d\sigma^{b}=\frac{R^2}{\sigma^2}\left(-f(\sigma)d\tau^2+(x'(\sigma)^2+f(\sigma)^{-1})d\sigma^2\right)~~,~~f(\sigma)=1-\gamma(\pi T\sigma)^4~~,
\eeq
where the prime denotes a derivative with respect to $\sigma$. The norm of the pull-back of the timelike Killing vector field $\partial_t$ onto the world volume yields the redshift factor $R_0=(R/\sigma)\sqrt{f}$ and, since the string is static, we have that $\textbf{k}=R_0$.

Evaluating the determinant of \eqref{ind1} and introducing it in \eqref{freeenergy} leads to the free energy
\beq \label{f1}
\mathcal{F}[x(\sigma)]=-2A\left(\frac{3}{2\pi T}\right)^{6}\int_{0}^{{\sigma_0}}d\sigma\sqrt{1+f(\sigma)x'(\sigma)^2}\thinspace G(\sigma)~~,~~G(\sigma)=\frac{R^8}{\sigma^8}f(\sigma)^3\frac{1+6\sinh^2\alpha}{\cosh^6\alpha}~~,
\eeq
where we have defined $A=\Omega_{(7)}/(16\pi G)$ and used Eqs.~\eqref{t1}-\eqref{t3} as well as \eqref{tT}. The overall factor of 2 accounts for the fact that we are gluing two symmetric patches of string around $x_1=L/2$. The equation of motion that follows from varying \eqref{f1} is
\beq \label{eom1}
\left(\frac{f(\sigma)x'(\sigma)}{\sqrt{1+f(\sigma)x'(\sigma)^2}}G(\sigma)\right)'=0~~.
\eeq
We consider solutions for which $x'(\sigma)>0$ and which satisfy the boundary conditions $x(0)=0$ and $x'(\sigma)\to\infty$ for $\sigma\to\sigma_0$. This yields the solution
\beq \label{s1}
x'(\sigma)=\left(\frac{f(\sigma)^2G(\sigma)^2}{f(\sigma_0)G(\sigma_0)^2}-f(\sigma)\right)^{-\frac{1}{2}}~~.
\eeq
In the case $\gamma=1$, this agrees with the result found in \cite{Grignani:2012iw} while in the case $\gamma=0$, and hence $f(\sigma)\to1$, we obtain
\beq
x'(\sigma)|_{\gamma=0}=\left(\frac{G(\sigma)^2}{G(\sigma_0)^2}-1\right)^{-\frac{1}{2}}~~,~~G(\sigma)|_{\gamma=0}=\frac{R^8}{\sigma^8}\frac{1+6\sinh^2\alpha}{\cosh^6\alpha}~~.
\eeq
We further note that the boundary conditions \eqref{b1} are ensured with the choices we made. In order to see this explicitly we note that the world volume of the string $\mathcal{W}_{2}$ composed of the two symmetric patches only has a boundary at $\sigma=0$. The unit normalised, orthogonal covector to the world volume  boundary is 
\beq 
\eta_bd\sigma^{b}=\frac{R}{\sigma}\frac{\sqrt{1+f(\sigma)x'(\sigma)^2}}{\sqrt{f(\sigma)}}d\sigma~~.
\eeq
With this we evaluate \eqref{b1} leading to
\beq \label{bc1}
T^{\sigma\sigma}\eta_\sigma|_{\sigma=0}=-Q_1\frac{(1+6\sinh^2\alpha)}{6\sinh\alpha\cosh\alpha}\frac{\sigma}{R}=0~~,~~J^{\tau\sigma}\eta_\sigma|_{\sigma=0}=Q_1\frac{\sigma}{R}=0~~,
\eeq
where we have used Eqs.~\eqref{t1}-\eqref{t3} and the fact that $f(\sigma)\to1$ and $G(\sigma)\to\infty$ when $\sigma\to0$.

Finally, we note that the equation of motion \eqref{eom1} also admits the solution $x'(\sigma)=0$ corresponding to two straight strings stretched from the boundary to the black hole horizon ($\sigma=z_0$) in the case $\gamma=1$ or to the origin of thermal AdS ($\sigma=\infty$) in the case $\gamma=0$. In terms of the dual gauge theory, this represents the Polyakov loop. In this case, the geometry also satisfies the boundary conditions \eqref{b1}, in particular, at $\sigma=0$ we obtain the same result as in \eqref{bc1} valid for any choice of $\gamma$.

%%%%%%%%%%%%%%%%%%%%%%%%%%%%%%%%%%%%%%%%%%%%%%%%%%%%%%%%%%%%%%

\subsubsection*{Solution space}
Here we wish to analyse the allowed region of solution space. As in \cite{Grignani:2012iw}, we first introduce a dimensionless coordinate $\hat\sigma=\pi T \sigma$ and define a parameter $\kappa$ using the Eqs.\eqref{t3} and \eqref{tT} according to
\beq \label{kappa}
\kappa\equiv \frac{2^5Q_1}{3^{7}AR^6}=\frac{f(\hat\sigma)^3}{\hat\sigma^6}\frac{\sinh\alpha(\hat\sigma)}{\cosh^5\alpha(\hat\sigma)}~~,
\eeq
where $f(\hat\sigma)=1-\gamma\hat\sigma^4$. The parameter $\kappa$ as we will see below distinguishes between the extremal probe $\kappa=0$ and the non-extremal probe $\kappa\ne0$.\footnote{This distinction only makes sense if the charge $Q_{1}$ is non-vanishing. If the charge vanishes there is no extremal limit since uncharged supergravity/M-theory black branes do not admit an extremal limit. Around Eq.~\eqref{kappa1} we will explain how the extremal limit $\kappa=0$ with $Q_1$ non-vanishing can be obtained.} Therefore we seek to understand the effect of corrections proportional to $\kappa$ in these configurations. Because of the fact that the ratio $\sinh\alpha/\cosh^5\alpha$ is bounded from above by the value $2^{4}/5^{\frac{5}{2}}$, the black string cannot be stretched all the way to the horizon for any value of $\kappa$. The maximum distance into the bulk it can attain is given by the critical distance
\beq \label{sc1}
\hat\sigma_c^2=\frac{1}{\gamma}\left(\sqrt{\gamma+\frac{5^{\frac{5}{3}}}{2^{14/3}}\kappa^{2/3}}-\frac{5^{5/6}}{2^{7/3}}\kappa^{1/3}\right)~~.
\eeq
In the case $\gamma=1$ this agrees with the result obtained in \cite{Grignani:2012iw} and attains its maximum value $\hat\sigma_c=1$, where it touches the horizon, when $\kappa=0$ and decreases for increasing $\kappa$ meaning that when $\kappa>0$ it cannot reach all the way to the horizon. In the confined phase for which $\gamma=0$ we instead get
\beq \label{sc0}
\hat\sigma_c^2|_{\gamma=0}=\frac{2^{4/3}}{5^{5/6}}\kappa^{-\frac{1}{3}}~~,
\eeq
which means that for $\kappa=0$ the string can stretch all the way to the origin of thermal AdS but for $\kappa>0$ it cannot be stretched more than a finite amount. The fact that the distance \eqref{sc1} exists is due to the fact that the local temperature of the black probe $\mathcal{T}$ has a maximum value for a given value of the charge $Q_{1}$. By looking at \eqref{sc1} one may tempted to say that in the case $\gamma=0$ the string can be stretched more into the bulk than when $\gamma=1$. This however is only apparent as by comparing the invariant spacetime lengths along the $z$-direction one finds that
\beq
\hat\sigma_c|_{\gamma=0}=\frac{\hat\sigma_c|_{\gamma=1}}{f(\hat\sigma_c|_{\gamma=1})^{\frac{1}{2}}}~~,
\eeq
and therefore the strings can be stretched exactly the same distance in both cases for a given value of $\kappa$.

\begin{figure}[H] 
\centering
  \includegraphics[width=0.6\linewidth]{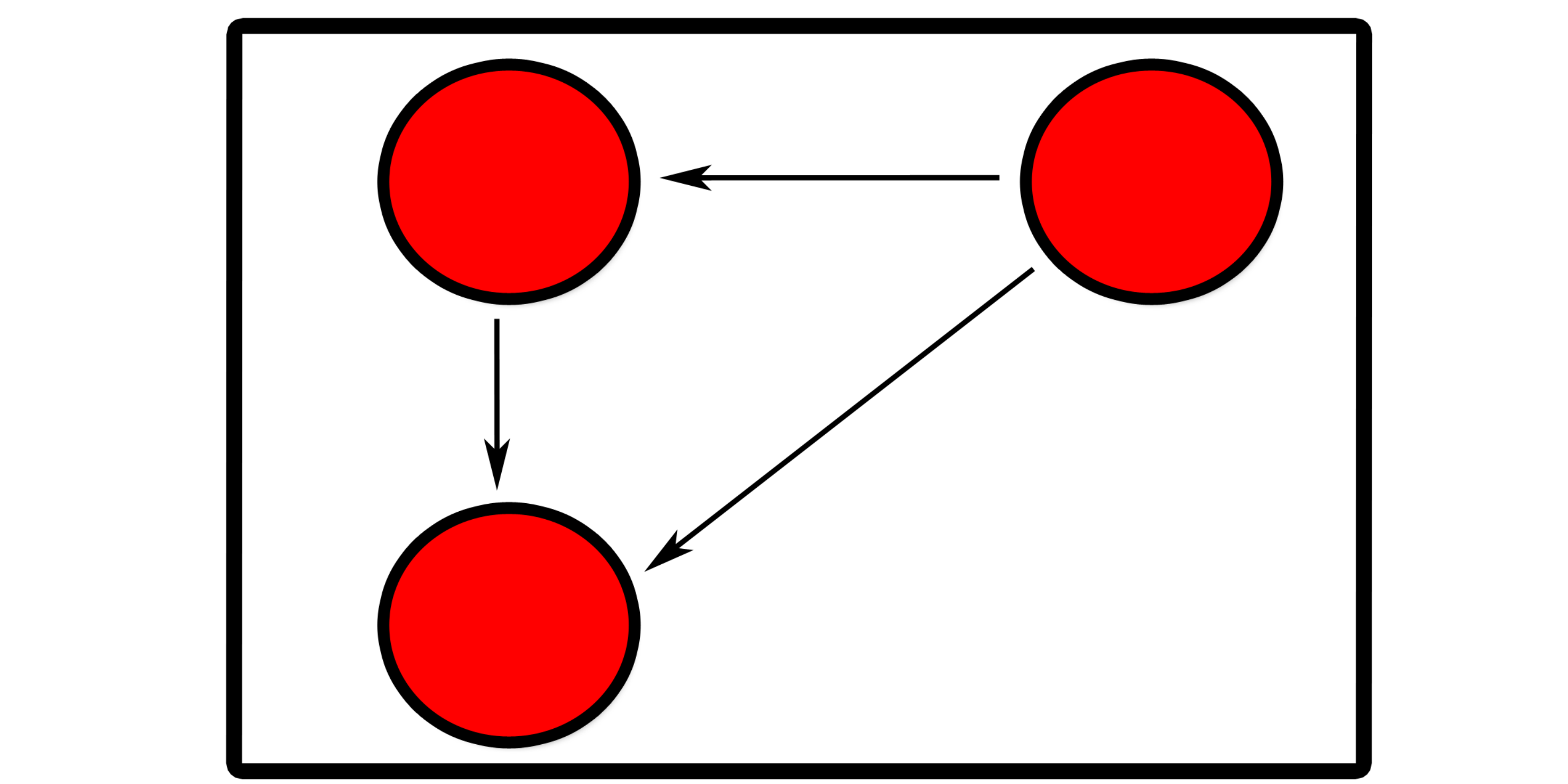}
  \begin{picture}(0,0)(0,0)
\put(-97,113){ $ T, \kappa  $}
\put(-227,113){ $ T,\kappa\!=\!0 $}
\put(-223,27){ $ T\!=\!0 $}
\put(-160,123){ $\kappa\to0 $}
\put(-250,70){ $T\to0 $}
\put(-140,60){ $T\to0 $}
\end{picture} 
\caption{Diagram depicting the solution space and its limits. The solutions at $T=0$ for $\kappa=0$ and $\kappa\ne0$ are equivalent. This diagram is generic for all the configurations constructed in this paper.} \label{diagram}
\end{figure}
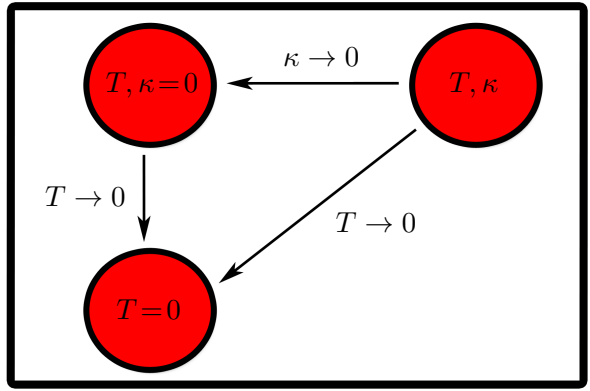

We now comment on how the different regions of parameter space are connected to each other and how the work of \cite{Maldacena:1998im, Rey:1998ik, Brandhuber:1998bs, Rey:1998bq} is recovered. For a given configuration with finite $T,\kappa$, as indicated in Fig.~\ref{diagram}, there are two ways of making the black probe extremal. According to \eqref{tT} we have that
\begin{itemize}
\item $T\to0$ and $\mathcal{T}\to0$ with $\textbf{k}\thinspace\mathcal{T}\to0$.
\item $\mathcal{T}\to0$ and $\textbf{k}\thinspace\mathcal{T}$ remains non-zero in such a way that $T$ remains finite and non-zero.
\end{itemize}
In the first case the background temperature is set to zero and we recover the solution at zero-temperature found in \cite{Maldacena:1998im, Rey:1998ik} regardless of the value of $\kappa$. In the second case, the black brane becomes extremal and is appropriately locally redshifted such that $\textbf{k}\thinspace\mathcal{T}$ remains non-zero. This can only happen if we send $\kappa\to0$ for finite $Q_1$, according to Eq.~\eqref{kappa}.\footnote{As it will be explained below, this implies taking the strict limit $N\to\infty$.} This is in fact the first example of a black hole solution where the near-horizon geometry is locally composed of extremal, zero-temperature, black branes but for which the global geometry has a finite temperature $T$. In this case the results of \cite{Brandhuber:1998bs, Rey:1998bq} are reproduced. If we would now turn off the temperature we would find the results of \cite{Maldacena:1998im, Rey:1998ik}. In more physical terms, we can rewrite $\kappa$ using the AdS/CFT dictionary for the radius $R^4=\lambda l_s^4$, the string coupling $4\pi g_s=\lambda/N$ as well as the relation for the fundamental string tension $T_{F_1}=(2\pi l_s^2)^{-1}$, where $\lambda$ is the 't Hooft coupling of the $\mathcal{N}=4$ SYM theory, $N$ is the number of black D3-branes whose near-horizon limit gave rise to \eqref{m1} and $l_s$ is the string length. Newton's constant can be written in terms of the Planck length $l_p$ according to $16\pi G=(2\pi)^{D-3}l_p^{D-2}$. In type IIB supergravity we have that $l_p^8=g_s^2 l_s^8$. Using this, the parameter $\kappa$ can be written as \cite{Grignani:2012iw}
\beq \label{kappa1}
\kappa=\frac{2^{7}}{3^{6}}\frac{k\thinspace\sqrt{\lambda}}{N^2}~~.
\eeq
Therefore, we see that the work of \cite{Brandhuber:1998bs, Rey:1998bq} is recovered in the strict limit $N\to\infty$ while still satisfying local thermodynamic equilibrium with the black probe in accordance with Eq.~\eqref{tT}.\footnote{The relation $R^4=\lambda l_s^4$ implies that if we send $\kappa\to0$ while keeping $Q_1$ non-zero then we must have that $N\to\infty$ such that $\sqrt{\lambda}/N^2\to0$.}

There is also another region of solution space which has not been previously explored, though of less holographic interest, as it does not connect with the part of solution space scanned by the extremal probe. This region is attained by sending $\kappa\to0$ while keeping $T\ne0$ but sending $Q_{1}\to0$. This branch of solutions does not admit an extremal limit since uncharged supergravity/M-theory black branes do not admit an extremal limit. However, this region is beyond the regime of validity of our method, as we will see below.

%%%%%%%%%%%%%%%%%%%%%%%%%%%%%%%%%%%%%%%%%%%%%%%%%%%%%%%%%%%%%%

\subsubsection*{Regime of validity}
We now wish to consider the regime of validity of the solutions. We consider the regime connected to the extremal probe ($\alpha\to\infty$) for which the characteristic length of the black fundamental string is its charge radius $r_c=r_0\sinh^2\alpha$. 
\begin{itemize}
\item For the solutions to be valid, one needs to require that quantum corrections are sub-leading, i.e., $r_c\gg l_s$ which implies that $k g_s^2\gg1$.  Furthermore, for the near-horizon geometry to be considered locally flat one must require $r_c\ll R$ implying that $k\ll N$, given that $g_s\ll1$. With this one concludes that we must require, 
\beq \label{sa}
1\ll k\ll N~~.
\eeq
The charge radius can also be written in terms of the parameter $\kappa$, which when using \eqref{kappa} reads $r_c=R\kappa^{1/6}$. Therefore, requiring the near-horizon geometry to be locally flat when compared to the two background scales $r_c\ll R$ and $r_c\ll 1/T$ one obtains
\beq \label{val}
\kappa\ll 1~~,~~RT\ll \kappa^{-\frac{1}{6}}~~.
\eeq 
The latter condition, imposing a very loose upper bound on the temperature of the black hole, is only present in the case $\gamma=1$. In terms of the 't Hooft coupling the former condition simply implies that $\lambda\ll N^2$. These conditions had already been derived in \cite{Grignani:2012iw}. 

\item One needs to further require that variations in the local temperature along the string are small compared to the background temperature, i.e.,
\beq
r_c\frac{\mathcal{T}'}{T}=\frac{r_c}{R\sqrt{f}}+2\gamma\frac{(\pi T\sigma)^4}{Rf^{3/2}}\ll1~~.
\eeq
In the case $\gamma=0$ this condition reduces to $r_c\ll R$, which has already been ensured by taking $\kappa\ll1$. In the case $\gamma=1$ this is also satisfied when the distances are such that $\sigma\ll z_0$, otherwise, for small $\kappa$ and for distances $\sigma\sim z_0$ one finds $z_0-\sigma\gg\kappa^{1/9}z_0$. In particular, this implies that for $\gamma=1$ the approximation breaks down for distances close to $\sigma\sim\sigma_c$. 

\item One should also require that the black string is thin compared to the local extrinsic length scale. For this one evaluates the scale of the extrinsic curvature via\footnote{In evaluating \eqref{l1} we have ignored derivatives of the metric since we have taken into account the variations of the background. One can explicitly check that the regimes of validity do not change if one considers these variations in \eqref{l1}.} 
\beq\label{l1}
L_{\text{ext}}(\sigma)=|K^{\rho}N_{\rho}|^{-1}=\frac{R}{\sigma}\frac{(1+fx'(\sigma)^2)^{3/2}}{f|x''(\sigma)|}~~,
\eeq
where $K^{\rho}=\gamma^{ab}{K_{ab}}^{\rho}$ is the mean extrinsic curvature and $N_{\rho}$ is a unit normalised covector orthogonal to $\mathcal{W}_2$. The length scale $L_{\text{ext}}(\sigma)$ is minimal when $\sigma=\sigma_0$. In the confined phase we find that for any $\sigma_0$, including $\sigma_c$, imposing $r_c\ll L_{\text{ext}}(\sigma)|_{\gamma=0}$ implies again $r_c\ll R$, which has already been ensured. In the deconfined phase, for distances $\sigma\ll z_0$ one again finds $r_c\ll R$ but for distances $\sigma\sim z_0$ one finds that $L_{ext}(\sigma_c)\sim r_c$ and hence the approximation breaks down. One should then require $z_0-\sigma\gg\kappa^{1/9}z_0$ as above.\footnote{In general, besides the length scale of variations of the extrinsic curvature \eqref{l1}, one should also consider the length scale of variations associated with the intrinsic curvature of the world volume \eqref{ind1}. It is a straightforward exercise to check that the intrinsic curvature scale is of the same order as \eqref{l1}.} 
\end{itemize}

We note that in the confined phase, in contrast with the deconfined phase, the approximation does not break down near $\sigma\sim\sigma_c$ and is valid all the way up to $\sigma_c$. The reason for this is simple. In the deconfined phase, the presence of the black hole introduces large gradients in the curvature near its horizon which induces large gradients in the extrinsic curvature and local temperature of the black string. On the other hand, in the confined phase, there is no horizon and large gradients are not present as one approaches the centre of AdS.

The straight string with $x'(\sigma)=0$ leads to a divergent $L_{ext}(\sigma)$ and hence also satisfies the validity requirements. In the confined phase, the Polyakov loop can be extended all the way to $\sigma_c$ but it cannot end there as it would not satisfy the boundary conditions \eqref{b1}. In the deconfined phase the approximation breaks down near $\sigma_c$ but if it did not then the string could at most extend up to $\sigma_c$, which would again not satisfy the boundary conditions. We therefore conclude that, the Polyakov loop at finite temperature cannot be constructed using this method.

Finally, we comment on the branch of solutions connecting to uncharged black strings ($\alpha=0$). In this case the relevant scale characterising the black string is the energy density radius $r_{\epsilon}\sim r_0$. Using Eqs.~\eqref{t1}-\eqref{t3} and imposing $r_{\epsilon}\ll R$ one finds that one should have
\beq
\frac{1}{\sigma T}\sqrt{f(\sigma)}\ll1~~.
\eeq
However, for finite $T$ as it is required for this branch of solutions, one sees that near the boundary $\sigma\sim0$ the radius $r_{\epsilon}$ diverges and the condition above cannot be satisfied. We thus conclude that this branch of solutions is beyond the scope of our method.

%%%%%%%%%%%%%%%%%%%%%%%%%%%%%%%%%%%%%%%%%%%%%%%%%%%%%%%%%%%%%%

\subsubsection*{Regularized free energy and large $N$ expansion}
We now take a closer look at the free energy \eqref{f1} and deduce some analytic results in a large $N$ expansion. As noted in \cite{Grignani:2012iw}, the free energy \eqref{f1} can be rewritten, using Eqs.~\eqref{t1}-\eqref{t3} and the AdS/CFT dictionary, as
\beq \label{rf1}
\mathcal{F}[x(\hat\sigma)]=\sqrt{\lambda} k T\int_{0}^{\hat\sigma_0}\frac{d\hat\sigma}{\hat\sigma^2}(1-X)\sqrt{1+fx'(\hat\sigma)^2}~~,~~X=1-\tanh\alpha-\frac{1}{6\sinh\alpha\cosh\alpha}~~,
\eeq
and diverges at $\sigma=0$. This divergence at $\sigma=0$ corresponds to an ultraviolet divergence in the gauge theory and can be dealt with by subtracting a piece to the free energy with the interpretation of being the mass of the W-boson in the gauge theory. The prescription developed in \cite{Grignani:2012iw} consists of introducing the infrared cut-off $\hat\sigma_{\text{cut}}$ and subtracting the piece 
\beq
\mathcal{F}_{\text{sub}}=\sqrt{\lambda}k T\int_{0}^{\hat\sigma_{\text{cut}}}\frac{d\hat\sigma}{\hat\sigma^2}(1-X)~~,
\eeq
such that the free energy difference $\Delta\mathcal{F}$ can be written as 
\beq
\Delta\mathcal{F}=\mathcal{F}-\mathcal{F}_{\text{sub}}=\mathcal{F}_{\text{loop}}-2\mathcal{F}_{\text{P}}~~,
\eeq
where
\beq \label{floop}
\mathcal{F}_{\text{loop}}=\sqrt{\lambda}k T\left(-\frac{1}{\hat\sigma_0}+\int_{0}^{\hat\sigma_0}\frac{d\hat\sigma}{\hat\sigma^2}\left((1-X)\sqrt{1+fx'(\sigma)^2}-1\right)\right)~~,
\eeq
is the regularised free energy associated with the Wilson loop configuration and
\beq \label{fw}
\mathcal{F}_{\text{P}}=-\frac{1}{2}\sqrt{\lambda}k T\left(\frac{1}{\hat\sigma_{\text{cut}}}+\int_{0}^{\hat\sigma_{\text{cut}}}\frac{d\hat\sigma}{\hat\sigma^2}X\right)~~,
\eeq
is the regularised free energy associated with the Polyakov loop.
One then needs a prescription for fixing $\hat\sigma_{\text{cut}}$, which can be done by letting $\hat\sigma_{\text{cut}}$ asymptote to $\hat\sigma_c$. However, one should pay attention to the following facts. As explained previously, Polyakov loops with $\kappa\ne0$ are outside the regime of validity of this approach so one cannot trust the expression \eqref{fw}. On the other hand, in the strict case $\kappa=0$ the expression \eqref{fw} can be trusted since the validity requirements are satisfied. In this case one can expand the function $X$ in powers of $\kappa$ ending up with a result that vanishes when $\kappa\to0$. In the deconfined phase we have that $\hat\sigma_c=1$ when $\kappa=0$ and in the confined phase $\hat\sigma_c\to\infty$ when $\kappa=0$, therefore one finds in general that for $\kappa=0$ one has
\beq \label{fwr}
\mathcal{F}_{\text{P}}=-\frac{1}{2}\gamma\sqrt{\lambda}k T~~,
\eeq
which in the case $\gamma=0$ is in agreement with \cite{Maldacena:1998im} and in the case $\gamma=1$ in agreement with \cite{Grignani:2012iw}. One can then argue that this result is valid for all $\kappa\ll1$ \cite{Grignani:2012iw}. Assuming that the free energy is linear in $T$ for any $\kappa$ (this is an assumption made implicitly in \cite{Grignani:2012iw}), the free energy \eqref{fw} must be of the form $\mathcal{F}_{\text{P}}=\sqrt{\lambda}k Ta(\kappa)$ for some function $a(\kappa)$. Therefore, using Eq.~\eqref{thermo} one finds the entropy $S_{\text{P}}=-\sqrt{\lambda} k a(\kappa)$ which is independent of $T$ since $\kappa$ is independent of $T$. However, at zero-temperature the solution obtained with an extremal probe ($\kappa=0$) and with a non-extremal probe ($\kappa\ne0$) must be equivalent to each other, as illustrated in the diagram of Fig.~\ref{diagram}. Therefore we see that we must have in general $a(\kappa)=a(0)=\gamma$ and hence that \eqref{fwr} is valid for all $\kappa\ll1$. \footnote{However, in general one could consider finite temperature corrections in small $T$ of the form $k\sqrt{\lambda}T (RT)^{m}b_{m}(\kappa)$ with $m>0$, $b_{m}(0)=0$ and $\kappa\ll1$ (since small $T$ means that the probe is near extremal). In order to accommodate these possible corrections and obtain the result \eqref{fw}, one must require that the $(m+1)$'th derivative for all $m$ with respect to $T$ of \eqref{fw} exhibits the same behaviour in the limit $T\to0$ for all $\kappa$, implying that $b_m(\kappa)=0$. The Polyakov loop at finite temperature has a supergravity description in terms of two straight black strings. We see that this requirement implies that the black string entropy is independent of $T$ for small temperatures, which may be a very strict requirement. Even though we use this type of argument throughout this paper, we note that none of the results obtained here are affected by this except the analysis of possible phase transitions from the Wilson loops/surfaces to Polyakov loops/surfaces. }

We now take a look at \eqref{floop}. Since solutions are valid for $\kappa\ll1$ we can analytically expand \eqref{floop} in powers of $\kappa$. An expansion in $\kappa$ means an expansion in large $N$ according to \eqref{kappa1} but also an expansion around $\mathcal{T}=0$, meaning that an expansion in inverse powers of $N$ necessarily requires turning on the thermal degrees of freedom of the black probe. In order to proceed further we note that an expansion around $\mathcal{T}=0$ can be recast as an expansion around $\alpha\to\infty$. Introducing $\phi=\cosh^2\alpha$ in \eqref{kappa} and expanding it around $\phi=\infty$ one finds
\beq \label{exp1}
\frac{\hat\sigma^6}{f(\hat\sigma)^3}\kappa=\frac{1}{\phi^2}-\frac{1}{2\phi^3}-\frac{1}{8\phi^4}+\mathcal{O}(\phi^{-5})~~.
\eeq
Solving this for $\phi$ and expanding it for small $\kappa$ one obtains 
\beq
\cosh^2\alpha=\frac{f(\hat\sigma)^{3/2}}{\hat\sigma^3\sqrt{\kappa}}-\frac{1}{4}-\frac{5}{32}\frac{\hat\sigma^3\sqrt{\kappa}}{f(\hat\sigma)^{3/2}}+\mathcal{O}(\kappa)~~.
\eeq
Using this expansion one can obtain an analytic result for the distance $L$ between the endpoints of the black string for small $\hat\sigma_0$, which in the dual gauge theory is interpreted as the length of the quark-antiquark pair,
\beq \label{L1}
LT=\frac{2}{\pi}\int_{0}^{\hat\sigma_0}\frac{\partial x(\hat\sigma)}{\partial\hat\sigma}=\frac{2\sqrt{2\pi}}{\Gamma(\frac{1}{4})^2}\hat\sigma_0+\left(\frac{\sqrt{2\pi}}{3\Gamma(\frac{1}{4})^2}-\frac{1}{6}\right)\sqrt{\kappa}\thinspace\hat\sigma_0^4-\gamma\frac{2\sqrt{2\pi}}{5\Gamma(\frac{1}{4})^2}\hat\sigma_0^{5}+\mathcal{O}(\hat\sigma_0^7)+\mathcal{O}(\kappa)~~.
\eeq
From this expression we see that for small distances $LT$ one needs a small bulk depth $\hat\sigma_0$. Furthermore, one sees that the effect of the black hole background only enters at order $\hat\sigma_0^5$ and hence the leading order correction to $LT$ comes from the thermal degrees of freedom of the black probe. One can perform a similar expansion in $\mathcal{F}_{\text{loop}}$ and find
\beq \label{F1}
\mathcal{F}_{\text{loop}}=-\sqrt{\lambda}kT\left(\frac{\sqrt{\pi}\Gamma(\frac{3}{4})}{\Gamma(\frac{1}{4})\hat\sigma_0}-\frac{\pi\Gamma(\frac{1}{4})+\sqrt{\pi}\Gamma(\frac{3}{4})}{6\Gamma(\frac{1}{4})}\hat\sigma_0^2\sqrt{\kappa}+\frac{2\sqrt{\pi}\Gamma(\frac{3}{4})}{\Gamma(\frac{1}{4})}\gamma \hat \sigma_0^{3}+\mathcal{O}(\kappa)+\mathcal{O}(\hat\sigma_0^4)\right)~~.
\eeq
Again we see here that the leading order correction comes from the thermal degrees of freedom and not from the background. We now invert \eqref{L1} and introduce it in \eqref{F1} in order to find
\beq \label{fbeta}
\mathcal{F}_{\text{loop}}=-\frac{\sqrt{\lambda}k}{L}\left(\frac{4\pi^2}{\Gamma(\frac{1}{4})^4}+\frac{\Gamma(\frac{1}{4})^4}{96}\sqrt{\kappa}(LT)^3+\gamma\frac{3\Gamma(\frac{1}{4})^4}{160}(LT)^4+...\right)~~.
\eeq
First we note that the combination $\mathcal{F}_{\text{loop}}L$ only depends on the scale invariant combination $LT$, exhibiting the conformal symmetry of the dual gauge theory. Secondly, the leading order correction in $LT$ is due to the properties of the probe and not of the background. This means that this correction term is also present in thermal AdS. A comparison between the last two terms present in \eqref{fbeta} tells us that the term proportional to $\gamma$ will be sub-leading if \cite{Grignani:2012iw}
\beq
LT|_{\gamma=1}\ll\frac{\sqrt{k}\lambda^{1/4}}{N}~~,
\eeq
which requires small enough temperatures. One also sees that in the strict limit $N\to\infty$ the correction due to the background is dominant. From \eqref{fbeta} we can also compute the black string entropy and mass using \eqref{thermo},
\beq 
S_{\text{loop}}=\frac{\sqrt{\lambda}k}{TL}\left(\frac{\Gamma(\frac{1}{4})^4}{32}\sqrt{\kappa}(LT)^3+\gamma\frac{3\Gamma(\frac{1}{4})^4}{40}(LT)^4+...\right)~~,
\eeq
\beq 
M_{\text{loop}}=-\frac{\sqrt{\lambda}k}{L}\left(\frac{4\pi^2}{\Gamma(\frac{1}{4})^4}+\frac{\Gamma(\frac{1}{4})^4}{24}\sqrt{\kappa}(LT)^3+\gamma\frac{3\Gamma(\frac{1}{4})^4}{32}(LT)^4+...\right)~~.
\eeq

%%%%%%%%%%%%%%%%%%%%%%%%%%%%%%%%%%%%%%%%%%%%%%%%%%%%%%%%%%%%%%

\subsubsection*{Comparison between phases}
We now compare the distance between the quark-antiquark pair and the free energy in both phases. We begin by solving for the distance $LT$ numerically for small values of $\kappa$. This is shown in the figures below. The dashed black line represents the value of $LT$ for $\kappa=0$. This can be obtained exactly analytically
\beq
LT|_{\kappa=0}=\frac{2\sqrt{2\pi}}{\Gamma(\frac{1}{4})^2}\hat\sigma_0\sqrt{1-\gamma \hat\sigma_0^4}\thinspace\thinspace \tensor*[_2]{F}{}_1\left(\frac{1}{2},\frac{3}{4},\frac{5}{4};\gamma\hat\sigma_0^4\right)~~.
\eeq
From the figures below one can see that in both phases, there is a critical distance $\hat\sigma_c$ beyond which the solution terminates. In the confined phase, this difference is more pronounced when compared to the $\kappa=0$ case since in this case the extremal probe method tells us that the black string can extend all the way to the origin of AdS.
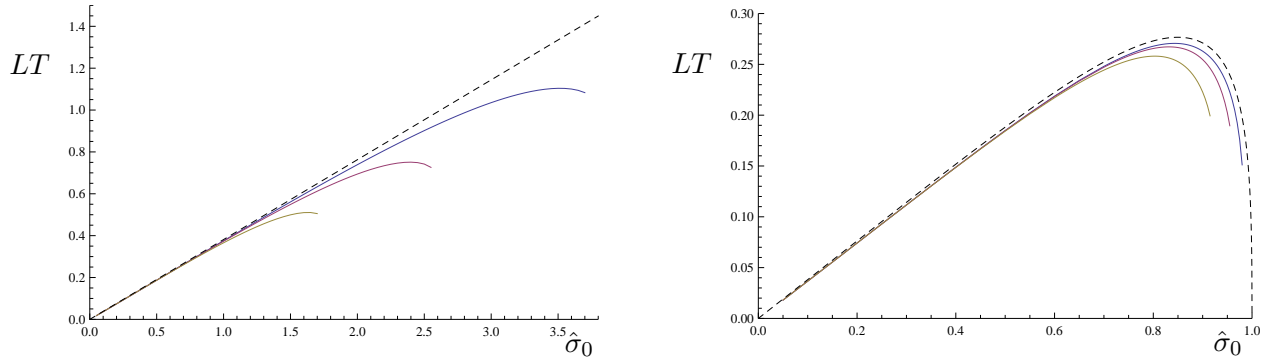
\begin{figure}[H]
\centering
\begin{subfigure}{.5\textwidth}
  \centering
  \includegraphics[width=.8\linewidth]{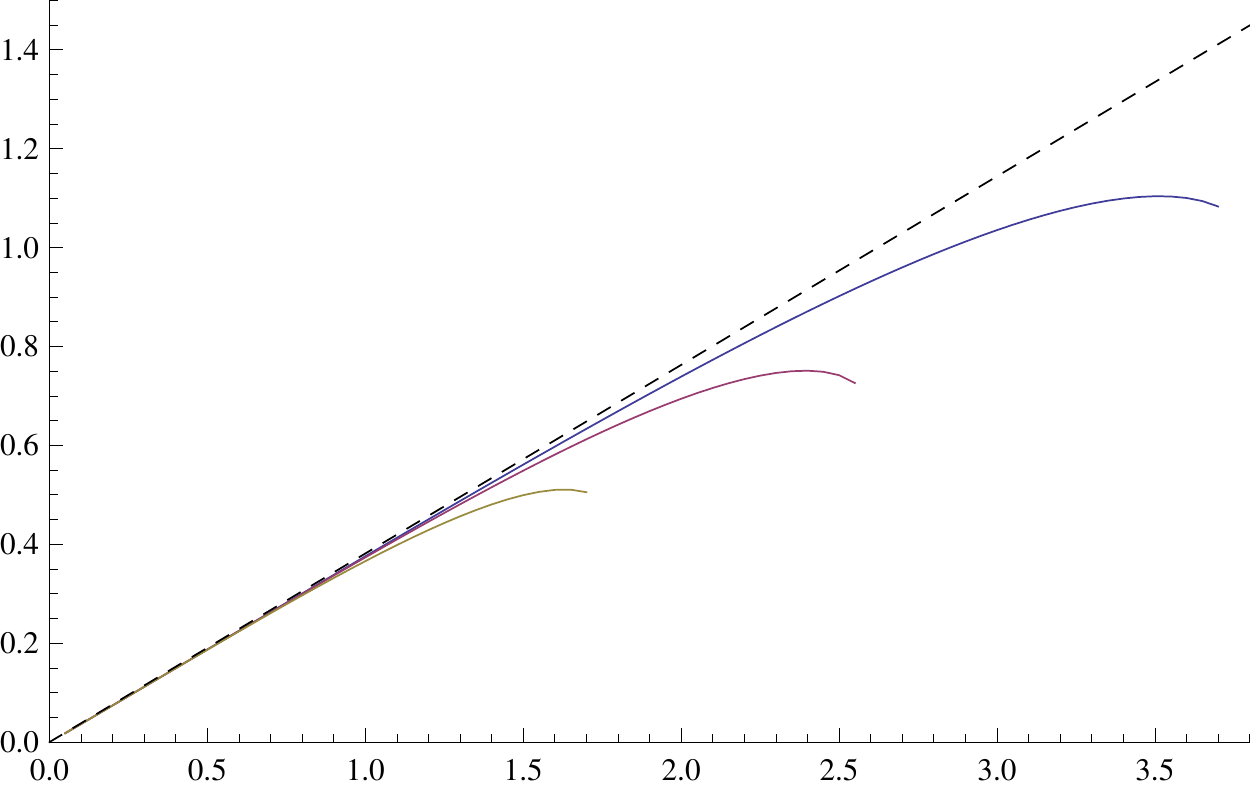}
  \begin{picture}(0,0)(0,0)
\put(-230,100){ $ LT  $}
\put(-20,-5){ $ \hat \sigma_0 $}
\end{picture}	
\end{subfigure}%
\begin{subfigure}{.5\textwidth}
  \centering
  \includegraphics[width=.8\linewidth]{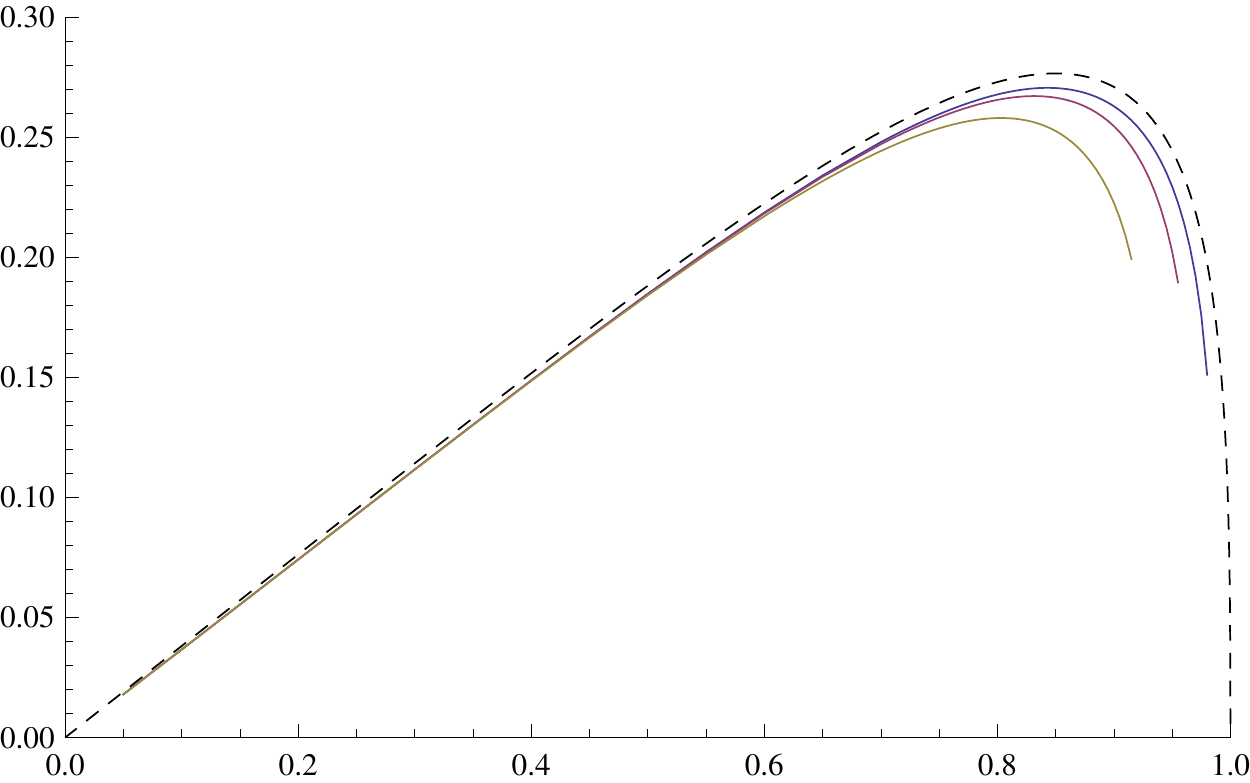}  
  \begin{picture}(0,0)(0,0)
  \put(-230,100){ $ LT  $}
  \put(-25,-5){ $ \hat \sigma_0 $}
  \end{picture}	
\end{subfigure}
\caption{On the left is the plot of the distance $LT$ as a function of $\hat\sigma_0$ for $\kappa=0.01$ (yellow line), $\kappa=0.001$ (red line) and $\kappa=0.0001$ (blue line) in the confined phase. The dashed black line is the result obtained for an extremal probe ($\kappa=0$). On the right there is a plot for the same values of $\kappa$ in the deconfined phase, previously obtained in \cite{Grignani:2012iw}.} \label{adsL}
\end{figure}
We also analyse the free energy difference $\Delta \mathcal{F}$ in both phases. This is depicted in the figures below, in which we have represented with a dashed line the result obtained for $\kappa=0$ which can be found analytically,
\beq
\Delta \mathcal{F}|_{\kappa=0}=\sqrt{\lambda}kT\left(\gamma-\frac{\sqrt{2}\pi^{3/2}(1-\gamma\hat\sigma_0^4)}{\hat\sigma_0\Gamma(\frac{1}{4})^2}\thinspace\thinspace \tensor*[_2]{F}{}_1\left(\frac{1}{2},\frac{3}{4},\frac{1}{4};\gamma\hat\sigma_0^4\right)\right)~~.
\eeq
\begin{figure}[H]
\centering
\begin{subfigure}{.5\textwidth}
  \centering
  \includegraphics[width=.8\linewidth]{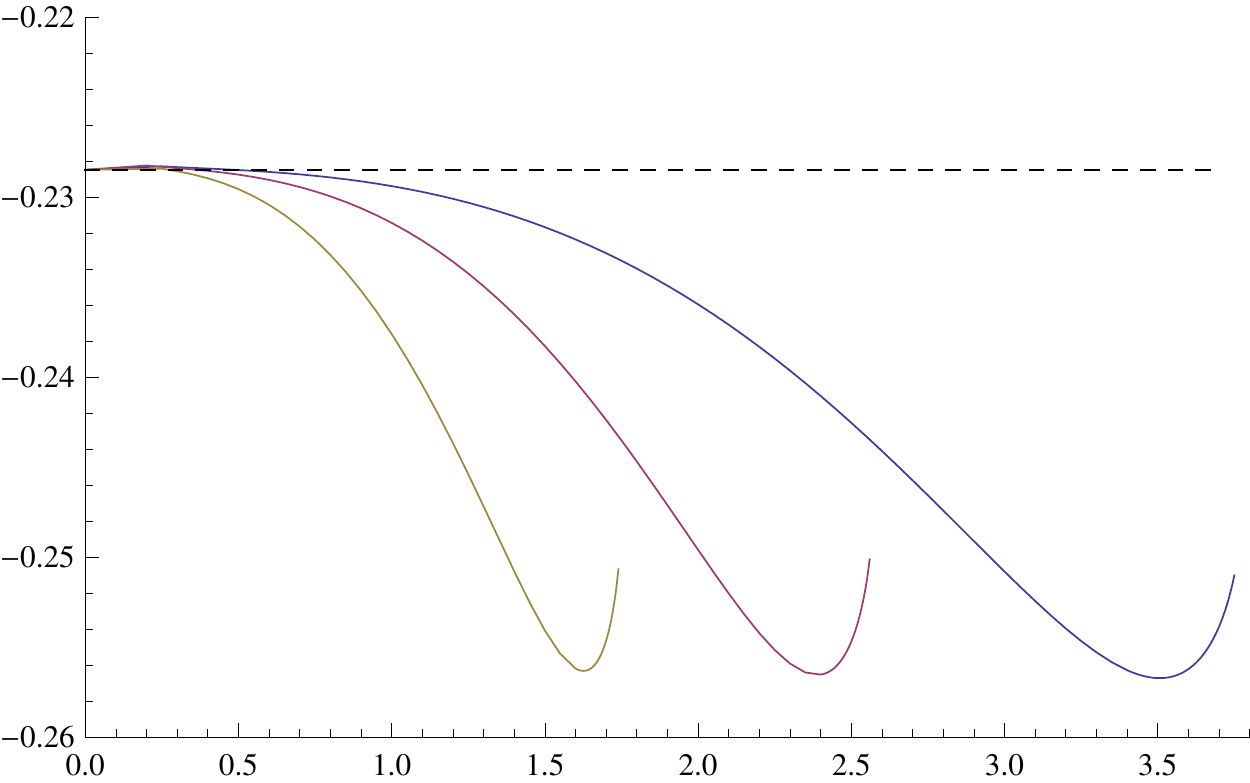}
  \begin{picture}(0,0)(0,0)
\put(-230,100){ $ \frac{L\Delta\mathcal{F}}{\sqrt{\lambda}k}  $}
\put(-20,-5){ $ \hat\sigma_0 $}
\end{picture}	
\end{subfigure}%
\begin{subfigure}{.5\textwidth}
  \centering
  \includegraphics[width=.8\linewidth]{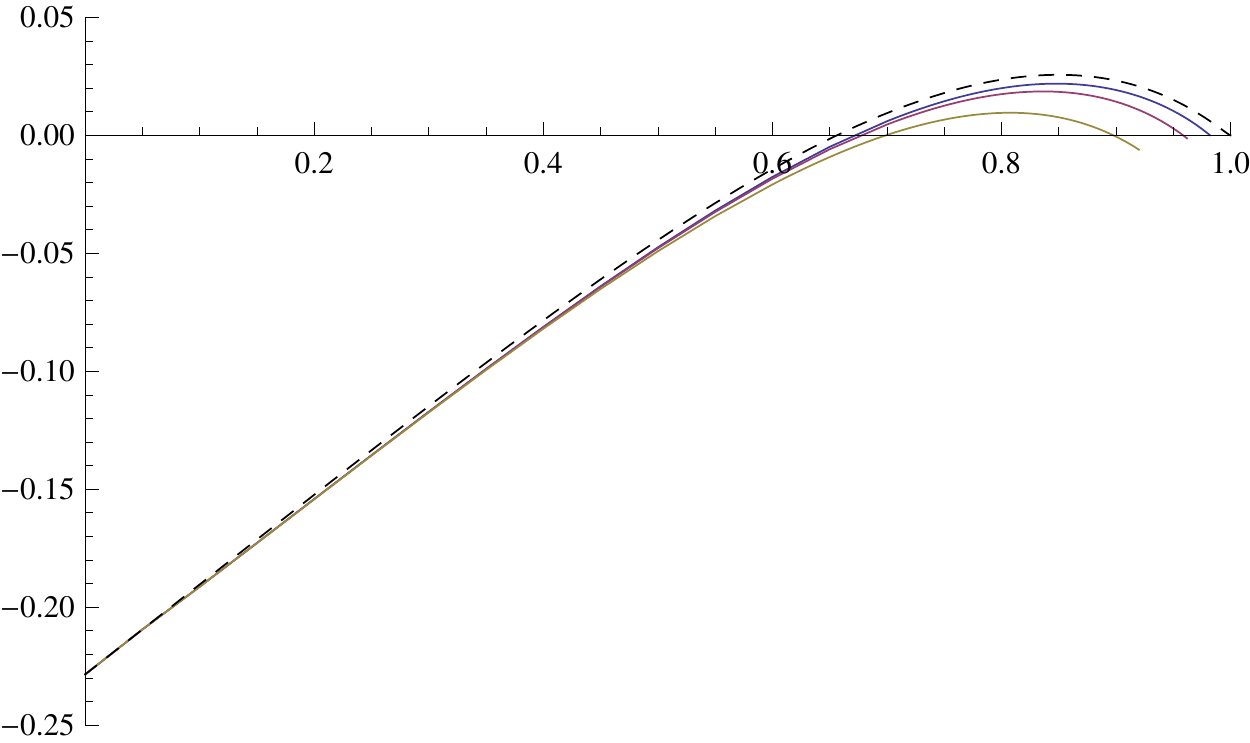}  
  \begin{picture}(0,0)(0,0)
  \put(-230,100){ $ \frac{L\Delta\mathcal{F}}{\sqrt{\lambda}k}  $}
  \put(-25,80){ $ \hat\sigma_0$}
  \end{picture}	
\end{subfigure}
\caption{On the left is the plot of the free energy difference $ \frac{L\Delta\mathcal{F}}{\sqrt{\lambda}k}  $ as a function of $\hat\sigma_0$ obtained numerically. The dashed black line is the result obtained for an extremal probe ($\kappa=0$). On the right there is a plot for the same values of $\kappa$ in the deconfined phase. The colour coding is the same as for the distances $LT$ presented in the previous figures.}
\end{figure}
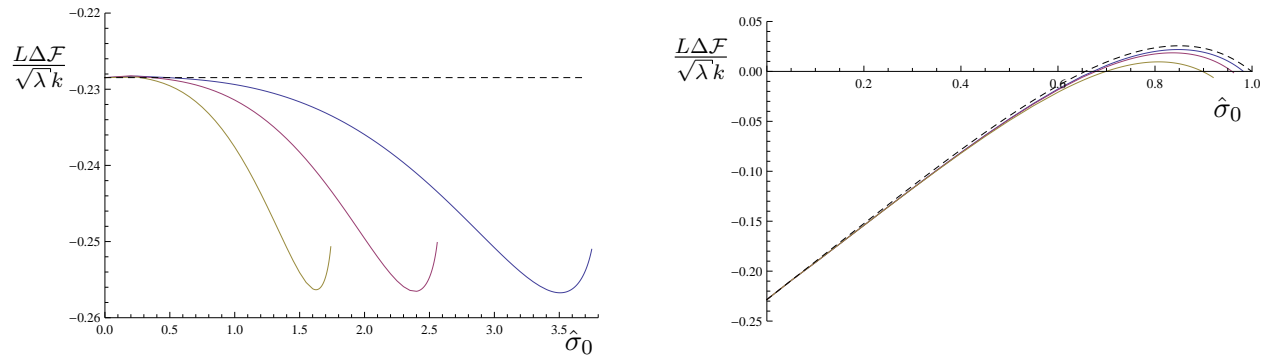
We can see that black strings in thermal AdS do not exhibit any phase transition to the Polyakov loop and hence are always the preferred configuration at a given temperature $T$. On the other hand, in the deconfined phase, a transition occurs around the value $\hat\sigma_0\sim 0.65$. This describes the onset of the Debye screening effect.\footnote{See Ref.~\cite{Grignani:2012iw} for a detailed description of the deconfined phase.}

In the thermal AdS background we see that the differences between the extremal probe method and the black probe method are significant as the free energy varies significantly compared to the constant value obtained for $\kappa=0$. One should note that in thermal AdS, as in the case of a black hole background, there is a double scaling limit ($\kappa\to0$) that makes the probe extremal while still being in thermodynamical equilibrium with the finite temperature background. In thermal AdS, however, this limit leads to the same results as if it were probing AdS at zero-temperature, hence highlighting the qualitative new features between the extremal and black probe methods.  
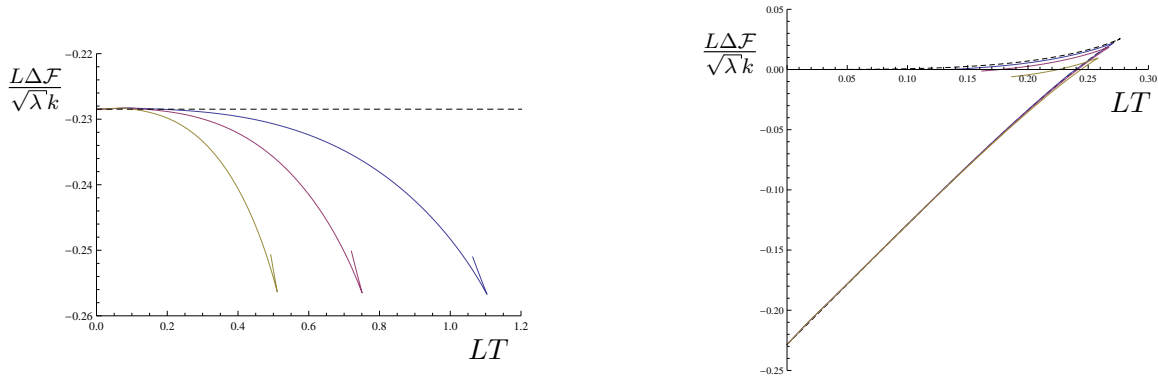
\begin{figure}[H]
\centering
\begin{subfigure}{.5\textwidth}
  \centering
  \includegraphics[width=.7\linewidth]{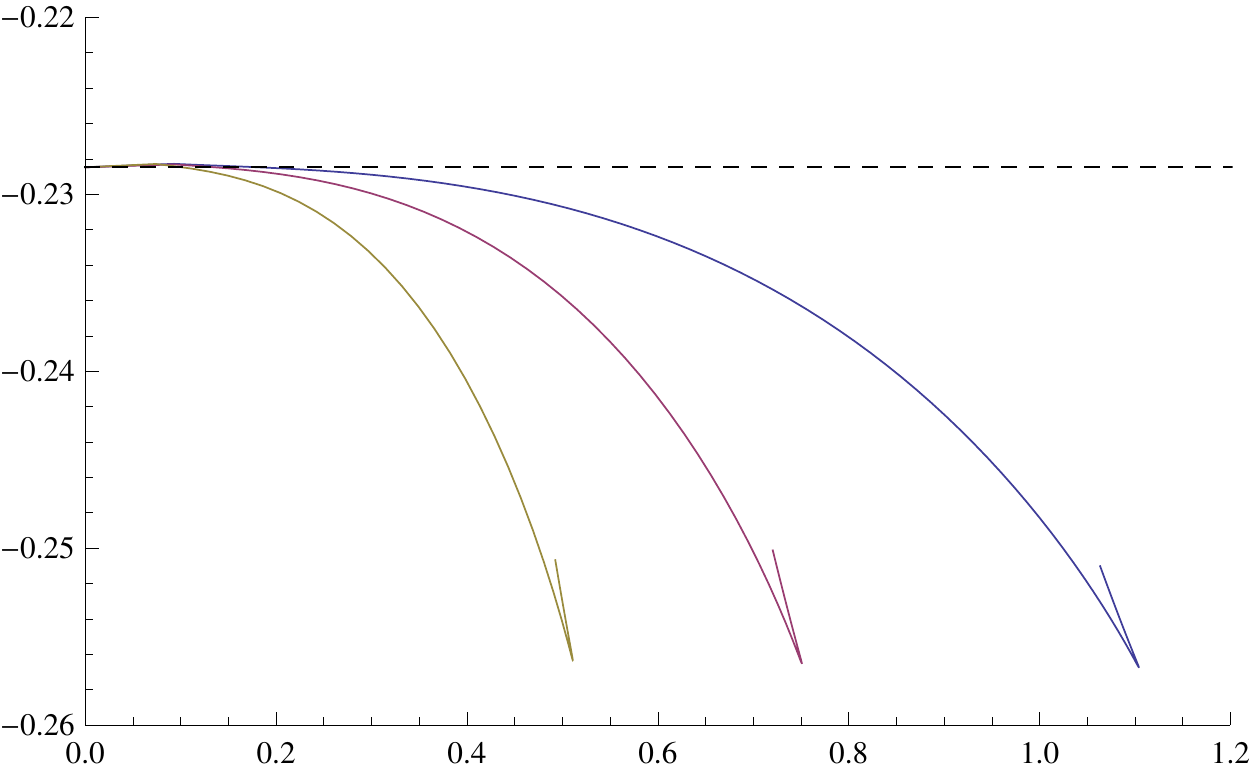}
  \begin{picture}(0,0)(0,0)
\put(-205,90){ $ \frac{L\Delta\mathcal{F}}{\sqrt{\lambda}k}  $}
\put(-30,-10){ $ LT $}
\end{picture}	
\end{subfigure}%
\begin{subfigure}{.5\textwidth}
  \centering
  \includegraphics[width=.6\linewidth]{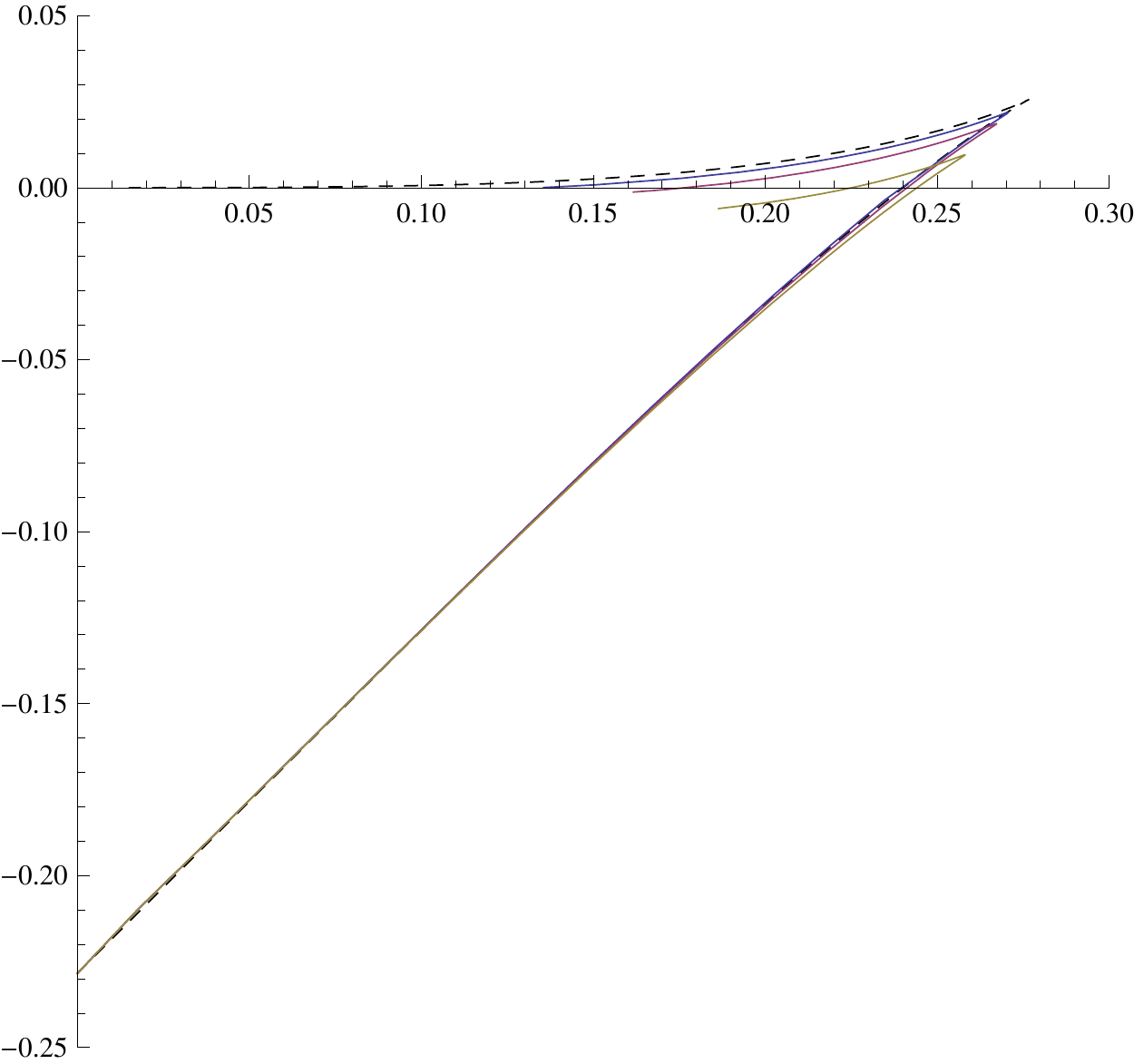}  
  \begin{picture}(0,0)(0,0)
  \put(-180,120){ $ \frac{L\Delta\mathcal{F}}{\sqrt{\lambda}k}  $}
  \put(-25,100){ $ LT$}
  \end{picture}	
\end{subfigure}
\caption{On the left is the plot of the free energy difference $ \frac{L\Delta\mathcal{F}}{\sqrt{\lambda}k}  $ as a function of $LT$ obtained numerically. The dashed black line is the result obtained for an extremal probe ($\kappa=0$). On the right there is a plot for the same values of $\kappa$ in the deconfined phase.}
\end{figure}
One can also plot the free energy differences with respect to $LT$. This is depicted in the figures above.
This again confirms that there is no phase transition occurring in thermal AdS while in the black hole background one finds the Debye screening length \cite{Grignani:2012iw}
\beq
(LT)|_{\gamma=1,\Delta \mathcal{F}=0}\sim0.240038+0.0379706\sqrt{\kappa}~~.
\eeq
The analysis presented here will serve as a means of comparison with our black string constructions in thermal Schr\"{o}dinger in the following section.

%%%%%%%%%%%%%%%%%%%%%%%%%%%%%%%%%%%%%%%%%%%%%%%%%%%%%%%%%%%%%%

\section{Black strings in Schr\"{o}dinger} \label{stringschrodinger}
In this section we construct three different types of black string configurations in thermal Schr\"{o}dinger, two with spatially separated boundary endpoints and another with null separated boundary endpoints. The first configuration interpolates between the black string in AdS constructed in the previous section and a second configuration, which has no AdS counterpart, as the deformation parameter $\ell$ is continuously increased. 

The background we consider has Sch$_{5}\times S^{5}$ asymptotics and is obtained via a null Melvin twist of the extremal D3-brane and subsequently taking the near-horizon limit \cite{Maldacena:2008wh, Adams:2008wt, Herzog:2008wg}. The resulting spacetime is Schr\"{o}dinger with dynamical exponent two\footnote{Unfortunately, dynamical exponents for Schr\"{o}dinger spacetimes are commonly denoted in the literature by $z$, so in this case we have $z=2$.} and metric\footnote{Here we have followed the conventions of \cite{Adams:2008wt}, however we have taken the parameter $\beta$ defined in \cite{Adams:2008wt} and rescaled it such that $\beta\to1/\sqrt{2}\beta$ to avoid keeping track of factors of 2 in the metric.}
\beq \label{dsback}
ds^2=\frac{R^2}{z^2}\left[-\frac{\ell^2}{z^2}dt^2+2dtd\xi+dz^2+dx_i^2\right]+R^2d\Omega^{2}_{(5)}~~,
\eeq
where $\xi$ is the boundary null coordinate and $i=1,2$. The boundary is located at $z=0$ and the origin at $z=\infty$. The parameter $\ell$ is defined as $\ell=\beta R^2$ with $\beta$ having dimensions of inverse length. The choice $\beta=0$ brings us to AdS in null coordinates. Either by a boost in the $(t,\xi)$ plane or by a uniform rescaling of all the Sch$_{5}$ coordinates (these two operations differ by the characteristic anisotropic dilatation symmetry of Sch$_{5}$), the parameter $\ell$ can be scaled to any particular value. When placing a configuration at finite temperature $T$ or with characteristic length $L$ in \eqref{dsback}, this implies that physical quantities are characterised by the dimensionless temperature $\textbf{T}$ and the dimensionless length $\textbf{L}$ defined as
\beq \label{dim}
\textbf{T}\equiv T\ell~~,~~\textbf{L}\equiv \frac{L}{\ell}~~.
\eeq
In this sense, the parameter $\ell$ introduces a new reference scale into the problem, as the black hole horizon introduced a scale via its mass in AdS. The null Melvin twist also generates a $B_{[2]}$ field of the form
\beq
B_{[2]}=2\ell\frac{R^2}{z^2}\left(d\chi+\mathcal{A}\right)\wedge dt~~,~~d\Omega^{2}_{(5)}=ds^2_{\mathbb{P}^2}+\left(d\chi+\mathcal{A}\right)^2~~,
\eeq
where $\mathcal{A}$ is the 1-form potential for the Kahler form on $\mathbb{P}^2$. However, we will always consider configurations at fixed coordinates in the five-sphere and hence couplings to the field $B_{[2]}$ are absent throughout this paper. 

We also note that we will only consider ``thermal Schr\"{o}dinger'', the Schr\"{o}dinger analogue of ``thermal AdS'', and not consider the black hole background with Sch$_{5}\times S^{5}$ asymptotics that one obtains via the null Melvin twist of the non-extremal black D3-brane \cite{Maldacena:2008wh, Adams:2008wt, Herzog:2008wg} because in this case a non-trivial dilaton is generated in the process and how exactly a black probe couples to a background dilaton field is still to be worked out.\footnote{These couplings will be dealt with in \cite{Armasnew}.} In \cite{Hartong:2010ec} it was shown that in Sch$_{5}\times S^{5}$ there is a Schr\"{o}dinger analogue of the Hawking-Page transition to a black hole background (from global Schr\"{o}dinger \cite{Blau:2009gd} to the null Melvin twist (TsT dual) of the (asymptotically plane wave AdS) MMT black hole \cite{Maldacena:2008wh}). As long as we are at sufficiently small temperatures, the preferred phase will be thermal Schr\"{o}dinger. We will then compare our results with those for thermal AdS black strings found in the previous section.

%%%%%%%%%%%%%%%%%%%%%%%%%%%%%%%%%%%%%%%%%%%%%%%%%%%%%%%%%%%%%%
%%%%%%%%%%%%%%%%%%%%%%%%%%%%%%%%%%%%%%%%%%%%%%%%%%%%%%%%%%%%%%

\subsection{Interpolating black strings with spacelike separated boundary endpoints} \label{interpolstring}
In this section we construct the Schr\"{o}dinger geometry for which its AdS counterpart was analysed in the previous section. This geometry interpolates between the AdS result, where $\ell=0$, and a purely Schr\"{o}dinger geometry, which will be studied in detail in the next section, when $\ell$ is very large and the spacetime is highly deformed compared to AdS. In order to do so we introduce a new set of coordinates $t',\xi'$ such that
\beq \label{newc}
t'=\frac{1}{\sqrt{2}}(t-\xi)~~,~~\xi'=\frac{1}{\sqrt{2}}(t+\xi)~~.
\eeq
The metric \eqref{dsback} then takes the form
\beq
ds^2=\frac{R^2}{z^2}\left[-g(z)dt'^2-\frac{\ell^2}{z^2}dt'd\xi'+\left(1-\frac{\ell^2}{2z^2}\right)d\xi'^2+dz^2+dx_i^2\right]+R^2d\Omega^{2}_{(5)}~~,~~g(z)=1+\frac{\ell^2}{2z^2}~~.
\eeq
We see that if we take $\ell\to0$ we obtain the metric \eqref{m1} with $\gamma=0$ where $t'$ is identified with $t$ and $\xi'$ with $x_3$. We now want to consider the analogous configuration as in the previous section where the string is stretched along the $z$-direction from the boundary to the bulk at $z=\sigma_0$ and back to the boundary at $z=0$. For this we chose the embedding map
\beq
t'=\tau~~,~~z=\sigma~~,~~x_1=x(\sigma)~~,~~\xi'=0~~,~~x_i=0~,\forall i=2,3~~,~~d\Omega_{(5)}=0~~,
\eeq
which leads to the induced metric
\beq
\gamma_{ab}d\sigma^{a}d\sigma^{b}=\frac{R^2}{\sigma^2}\left(-g(\sigma)d\tau^2+(1+x'(\sigma)^2)d\sigma^2\right)~~,
\eeq
and reduces to \eqref{ind1} when $\ell=0$. Furthermore the pullback of the timelike Killing vector field $\partial_{t'}$ onto the world volume $\mathcal{W}_2$ yields $\textbf{k}=(R/\sigma)\sqrt{g(\sigma)}$. Indeed we see that if $\ell=0$ we get $\textbf{k}=R/\sigma$ which is the result obtained in AdS when $\gamma=0$, while if we take $\ell$ to be very large we get $\textbf{k}\sim R\ell/(\sigma^2)$ which gives rise to a qualitatively different behaviour. This latter case deserves special attention and will be analysed in detail in the next section.

Using the above induced metric in \eqref{freeenergy} we find the free energy
\beq \label{fi}
\mathcal{F}[x(\sigma)]=-2 A\left(\frac{3}{2\pi T}\right)^{6}\int_{0}^{\sigma_0}d\sigma\sqrt{1+x'(\sigma)^2}Z(\sigma)~~,~~Z(\sigma)=\frac{R^8}{\sigma^8}g(\sigma)^{\frac{7}{2}}\frac{1+6\sinh^2\alpha}{\cosh^6\alpha}~~,
\eeq
which when varied and imposing the conditions $x'(\sigma)\to\infty$ as $\sigma\to\sigma_0$ leads to the solution for $x'(\sigma)$
\beq
x'(\sigma)=\left(\frac{Z(\sigma)^2}{Z(\sigma_0)^2}-1\right)^{-\frac{1}{2}}~~.
\eeq
These solutions are valid solutions when certain requirements, which will be derived in the next section, similar to those obtained in Sec.~\ref{adsstring} are imposed. The case at $T=0$ already presents itself as a very interesting one and difficult to track analytically. In that case we have
\beq
A\left(\frac{3}{2\pi T}\right)^{6} Z(\sigma)|_{T\to0}=Q_{1}\frac{R^2}{\sigma^2}\sqrt{g(\sigma)}~~.
\eeq
One may now attempt to obtain analytic expressions for the distance between the boundary endpoints in terms of the bulk depth. However, this is not possible to obtain analytically for all values of the deformation parameter but only in the case $\ell^2=0$ and in the limit $\ell^2\to\infty$. One may also attempt to expand the sought after quantities for small and large $\ell^2$ but it turns out that such series expansions in $\ell^2$ do not converge for the functions that we are dealing with.\footnote{The reason for this is that the spacetimes with $\ell^2>0$ and $\ell^2<0$ are very different.} Therefore, we need to analyse the problem numerically. For this we consider the regularised free energy, obtained by subtracting to \eqref{fi} a piece
\beq
\mathcal{F}_{\text{sub}}=\sqrt{\lambda}\frac{k}{\pi}\int_{0}^{\sigma_c}\frac{d\sigma}{\sigma^2}\sqrt{g(\sigma)}(1-X)~~,
\eeq
where $X$ was defined in \eqref{rf1} and $X\to0$ as $T\to0$. The resulting free energy takes the form
\beq
\mathcal{F}_{\text{loop}}=\sqrt{\lambda}\frac{k}{\pi}\left(h(\sigma_0)+\int_{0}^{\sigma_0}\frac{d\sigma}{\sigma^2}\sqrt{g(\sigma)}\left((1-X)\sqrt{1-x'(\sigma)^2}-1\right)\right)~~,
\eeq
where we have defined the function $h(\sigma_0)$ as
\beq
h(\sigma_0)=-\frac{1}{2}\left(\frac{\sqrt{\ell^2+2\sigma_0^2}}{\sqrt{2}\sigma_0^2}+\sqrt{2}\frac{\text{arcsinh}(\frac{\ell}{\sqrt{2}\sigma_0})}{\ell}\right)~~.
\eeq
The Polyakov loop configuration consisting of two straight strings with $x'(\sigma)=0$, following the same steps as in Sec.~\ref{adsstring}, can be argued to have zero free energy for any value of $\ell,\kappa,T$. We now obtain expressions for the length $L$, computed in a similar manner as in \eqref{L1}, and $\mathcal{F}_{\text{loop}}$ as a function of $\sigma_0$ at $T=0$. This behaviour is depicted in the figures below.
\begin{figure}[H]
\centering
\begin{subfigure}{.5\textwidth}
  \centering
  \includegraphics[width=.8\linewidth]{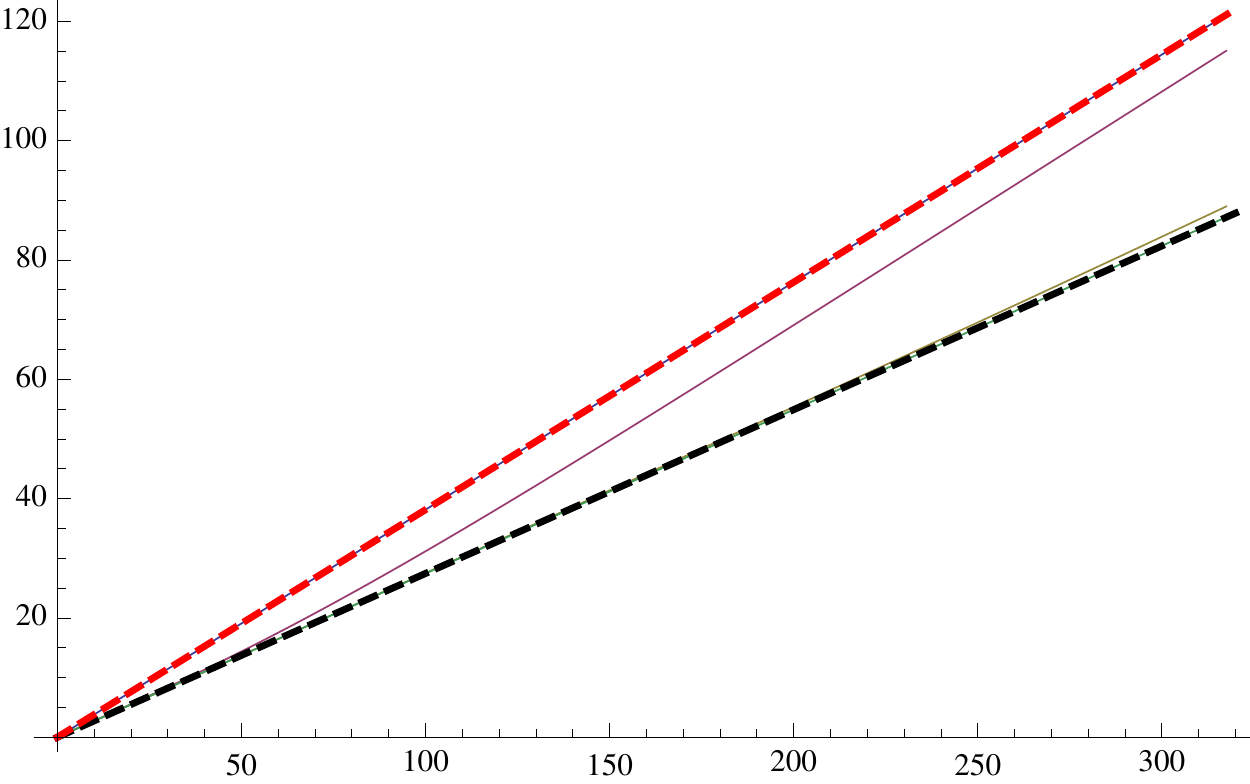}
  \begin{picture}(0,0)(0,0)
\put(-220,110){ $ L  $}
\put(-25,-8){ $ \sigma_0 $}
\end{picture}	
\end{subfigure}%
\begin{subfigure}{.5\textwidth}
  \centering
  \includegraphics[width=.8\linewidth]{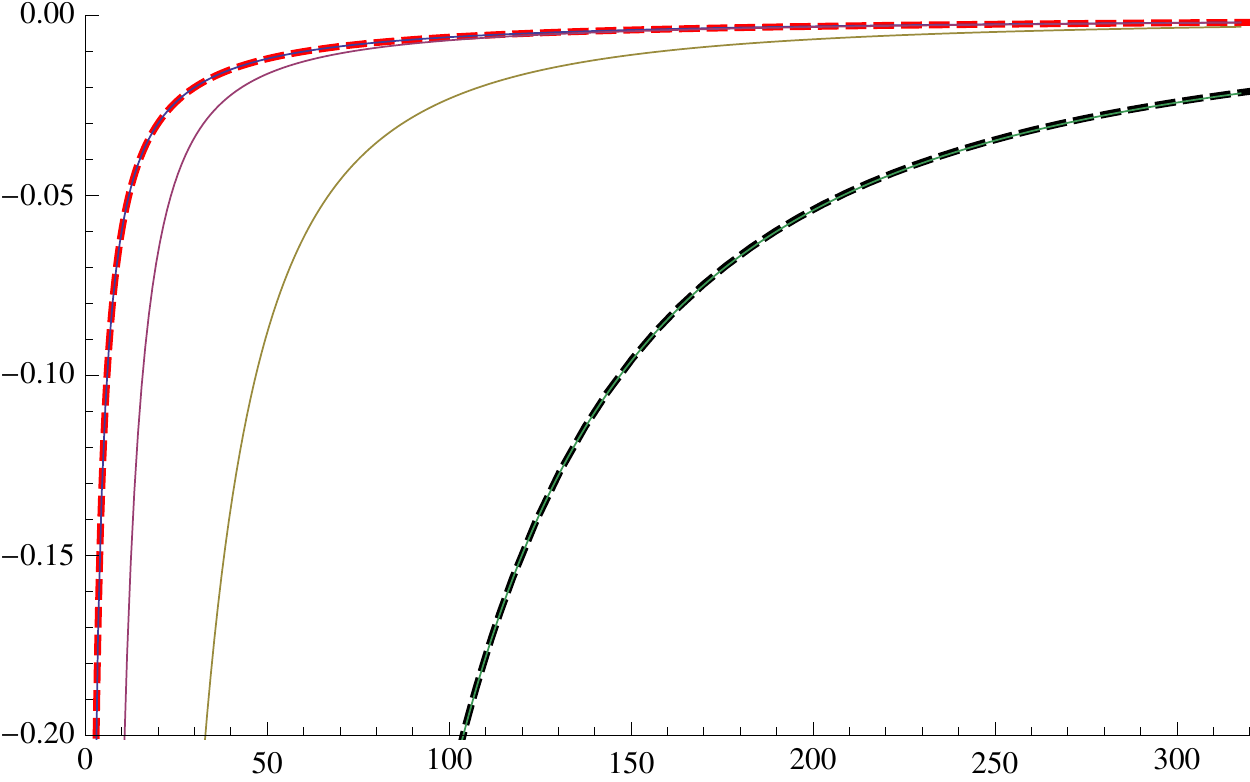}  
  \begin{picture}(0,0)(0,0)
  \put(-225,105){ $\frac{\mathcal{F}_{\text{loop}}\pi}{\sqrt{\lambda}k} $}
  \put(-30,-5){ $\sigma_0$}
  \end{picture}	
\end{subfigure}
\caption{On the left we have the behaviour of the distance $L$ as function of $\sigma_0$ at $T=0$. The dashed solid red line corresponds to the AdS result given by the first term in \eqref{L1} while the dashed solid black line corresponds to the result \eqref{rext}. The blue line represents $(\ell/\sqrt{2})=0.0001$, the magenta line $(\ell/\sqrt{2})=1$, the yellow line $(\ell/\sqrt{2})=1000$ and the green line $(\ell/\sqrt{2})=10000$. On the right we have the behaviour of the free energy as a function of $\sigma_0$ at $T=0$. The dashed solid red line is the first term obtained in \eqref{F1} while the dash solid black line is the result obtained in \eqref{rext}.} \label{ddd}
\end{figure}
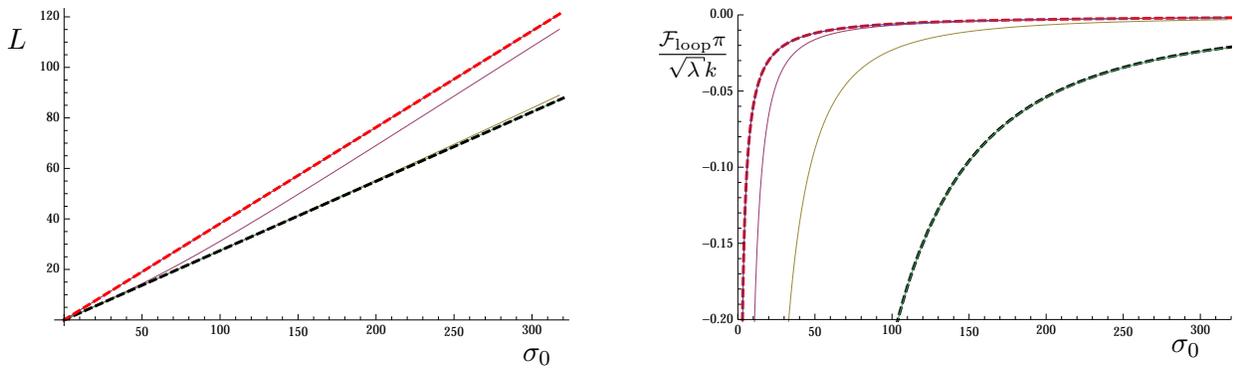
On the l.h.s. of the figure above we have shown the behaviour of the length $L$ as a function of $\sigma_0$ and on the r.h.s. the behaviour of the free energy as a function of $\sigma_0$. The dashed red line on the l.h.s. corresponds to the zero-temperature AdS result for the length $L$ given in \eqref{L1}  while the dashed line on the r.h.s. corresponds to the leading order result for AdS given in \eqref{F1}. The dashed black lines depict the result obtained analytically in the limit $\ell\to\infty$
\beq \label{rext}
L|_{\ell\to\infty}=3\sqrt{\pi}\sigma_0\frac{\Gamma(\frac{5}{3})}{\Gamma(\frac{1}{6})}~~,~~\frac{\mathcal{F}_{\text{loop}}\pi}{\sqrt{\lambda}k}|_{\ell\to\infty}=\frac{\sqrt{\pi}\ell}{6\sigma_0^2}\frac{\Gamma(-\frac{1}{6})}{\Gamma(\frac{1}{6})}
\eeq
which will be derived in the next section. We can see from the plots above that the behaviour for any value of $\ell$ is bounded by the cases $\ell=0$ and $\ell\to\infty$ and hence between an AdS geometry and a purely Schr\"{o}dinger geometry. This means, as we will see explicitly below, that since $\mathcal{F}_{\text{loop}}$ scales with $\sigma_0^{-1}$ for $\ell=0$ and with $\sigma_0^{-2}$ for very large $\ell$ and since the behaviour of the curve $(L,\sigma_0)$ is approximately linear (closer scrutiny reveals that for small values of $\sigma_0$ there is a departure from the linear behaviour exhibited in Fig.~\ref{ddd}), then the free energy is scaling like $L^{-1}$ for $\ell=0$ and like $L^{-2}$ for $\ell\to\infty$. Therefore we see that in terms of the dual theory, the expectation value of these operators is changing substantially for large values of the deformation parameter.

We now consider the finite temperature case and for non-zero $\kappa$. For these purposes we introduce the dimensionless coordinate $\hat\sigma=\pi T \sigma$ and define the parameter $\kappa$ as in \eqref{kappa} but now with $f(\hat\sigma)$ replaced by $g(\hat\sigma)$ on the r.h.s. of \eqref{kappa}. We further introduce the parameter $\hat\ell$ such that
\beq
g(\hat\sigma)=1+\frac{\hat\ell^2}{\hat\sigma^2}~~,~~\hat\ell^2=\frac{\pi^2 T^2\ell^2}{2}~~.
\eeq
The qualitative behaviour of the dimensionless length $LT$ and the free energy at finite temperature is qualitatively similar to that seen in Fig.~\ref{ddd}, the difference being that the strings cannot extend all the way to the bulk but instead have a maximum distance as in \eqref{sc1} given by
\beq
\hat\sigma_c^2=\frac{2^{\frac{1}{3}}}{5^{\frac{5}{6}}}\kappa^{-\frac{1}{3}}\left(1+\sqrt{1+2^{\frac{2}{3}}5^{\frac{5}{6}}\hat\ell^2\kappa^{\frac{1}{3}}}\right)~~.
\eeq
We now solve for the free energy numerically as a function of $\hat\sigma_0$ for several values of $\hat\ell$ and $\kappa=0.001$. The result is depicted in the figure below. One can observe that, as in the case $T=0$ the results for any $\hat\ell$ are bounded between the AdS result and the large $\hat\ell$ result. 
\begin{figure}[H]
\centering
  \includegraphics[width=0.4\linewidth]{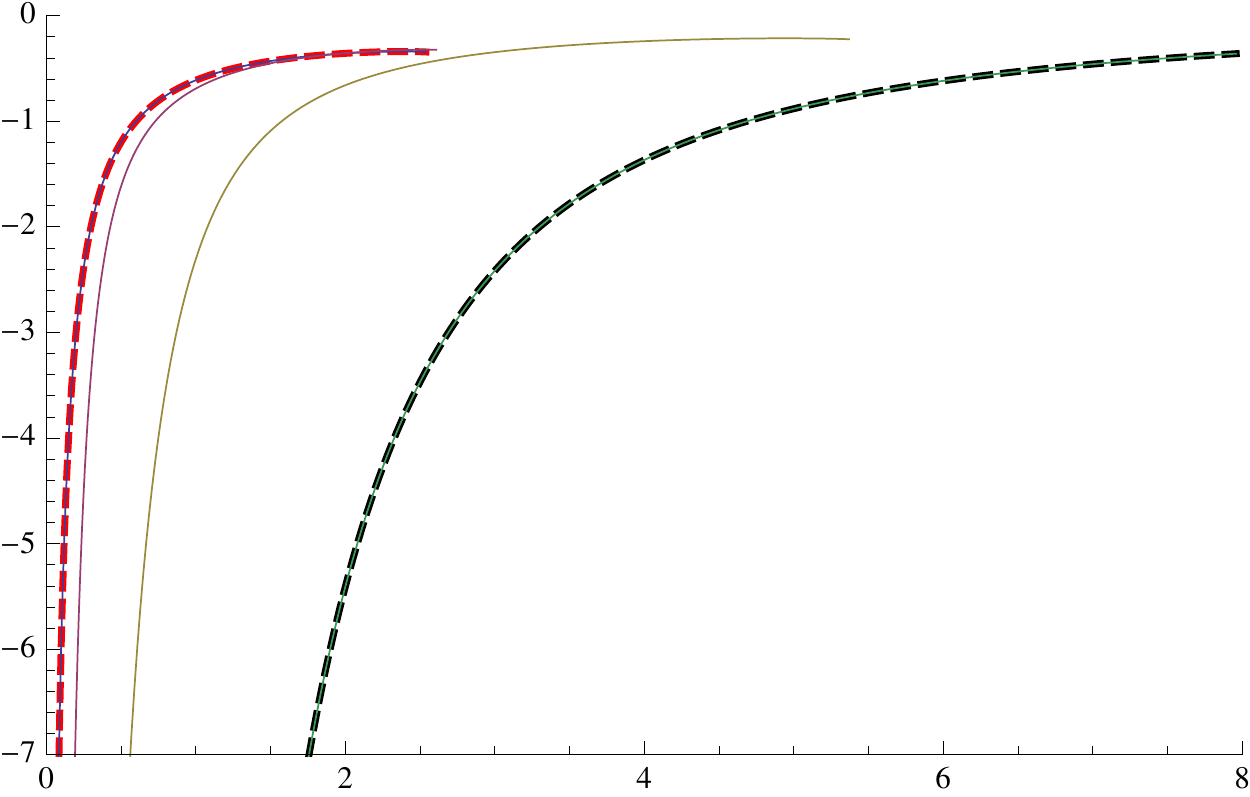}
  \begin{picture}(0,0)(0,0)
\put(-240,110){ $\frac{\mathcal{F}_{\text{loop}}}{\sqrt{\lambda}k T} $}
\put(-20,-5){ $ \tilde \sigma_0 $}
\end{picture}	
\caption{The free energy as a function of $\hat\sigma_0$ for $\kappa=0.001$ obtained numerically. The red dashed line is the result obtained for $\hat\ell=0$ while the dashed black line is the result for $\hat\ell=100$ obtained using the results of the next section. The blue line corresponds to the case $\hat\ell=0.001$, the magenta line to the case $\hat\ell=1$, the yellow line to $\hat\ell=10$ and the green line to $\hat\ell=100$. } \label{Fiv}
\end{figure}
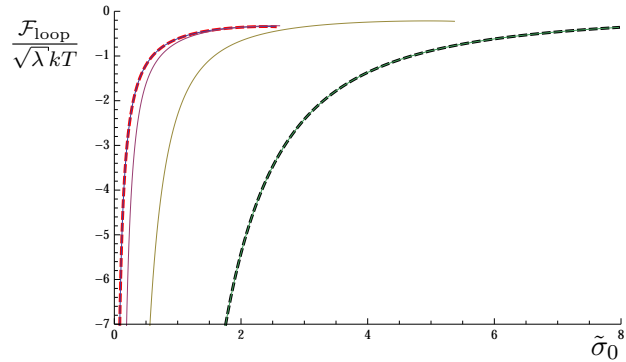
The behaviour exhibited here is generic for any small value of $\kappa$. The behaviour of $LT$ as a function of $\hat\sigma_0$ is also approximately linear, except near $\hat\sigma_c$, and hence the dependence of $\mathcal{F}_{\text{loop}}$ on $LT$ shows approximately the same behaviour as what is depicted below. We will now analyse the case $\hat\ell\to\infty$ in detail.

%%%%%%%%%%%%%%%%%%%%%%%%%%%%%%%%%%%%%%%%%%%%%%%%%%%%%%%%%%%%%%
%%%%%%%%%%%%%%%%%%%%%%%%%%%%%%%%%%%%%%%%%%%%%%%%%%%%%%%%%%%%%%
\subsection{Black strings with spacelike separated boundary endpoints} \label{blackspatial}
In this section we construct another class of black strings with endpoints on a boundary spatial direction. This class of solutions, valid for any value of the deformation parameter $\ell$, meets the class of interpolating black strings studied in the previous section when $\ell$ is taken to be very large. To that end, we go back to the coordinates $(t,\xi)$. As in the previous cases, we fix the endpoints of the string in the $x_1$-direction and stretch it into the bulk along the $z$-direction in the background geometry \eqref{dsback}. This leads to the embedding map
\beq
t=\tau~~,~~z=\sigma~~,~~x_1=x(\sigma)~~,~~\xi=x_2=d\Omega_{(5)}=0~~,
\eeq
and hence the problem is symmetric around $x_1=L/2$. The induced metric on $\mathcal{W}_2$, using \eqref{dsback}, is
\beq \label{ind2}
\gamma_{ab}d\sigma^{a}d\sigma^{b}=\frac{R^2}{\sigma^2}\left(-\frac{\ell^2}{\sigma^2}d\tau^2+(1+x'(\sigma)^2)d\sigma^2\right)~~.
\eeq
The pull-back of the background time-like Killing vector field $\partial_t$ leads to a world volume timelike Killing vector field $\partial_\tau$ with norm $R_0=\textbf{k}=R\ell/\sigma^2$. This different power of $\sigma$ when compared with the case of Sec.~\ref{adsstring} will lead to different local properties of the black string. In particular we see that when $\ell\to0$ we get $\textbf{k}\to0$ and hence these solutions have no AdS counterpart. In terms of the dual DLCQ$_{\beta}$ field theory, these presumably correspond to operators that are well defined in the $\beta$-deformed DLCQ theory but are singular in the conventional DLCQ of the theory (due to the notorious zero mode problem, see e.g. \cite{Hellerman:1997yu,Ganor:1997jx}).

The free energy takes the same form as in \eqref{f1} but with a different function $H(\sigma)$
\beq \label{f2}
\mathcal{F}[x(\sigma)]=-2A\left(\frac{3}{2\pi T}\right)^{6}\int_{0}^{{\sigma_0}}d\sigma\sqrt{1+x'(\sigma)^2}\thinspace H(\sigma)~~,~~H(\sigma)=\frac{R^{8}\ell^7}{\sigma^{15}}\frac{1+6\sinh^2\alpha}{\cosh^6\alpha}~~,
\eeq
where the overall factor of 2 in the free energy is again due to the fact that the string is stretched all the way from $(x_1,z)=(0,0)$ to $(x_1,z)=(L/2,\sigma_0)$ and back again to $(x_1,z)=(L,0)$. The equation of motion that follows from varying \eqref{f2} is
\beq \label{eom2}
\left(\frac{x'(\sigma)}{\sqrt{1+x'(\sigma)^2}H(\sigma)}\right)'=0~~.
\eeq
As in the AdS case, we consider solutions for which $x'(\sigma)>0$ and which satisfy the boundary conditions $x(0)=0$ and $x'(\sigma)\to\infty$ for $\sigma\to\sigma_0$. This gives rise to the solution
\beq \label{s1}
x'(\sigma)=\left(\frac{H(\sigma)^2}{H(\sigma_0)^2}-1\right)^{-\frac{1}{2}}~~.
\eeq
We note that in thermal Schrodinger there is also an analogous solution to the Polyakov loop in AdS, characterised by two straight strings stretched from the boundary to the bulk with $x'(\sigma)=0$, trivially satisfying \eqref{eom2}. The boundary conditions \eqref{b1} must also be satisfied. In this case, the unit normalised, orthogonal covector to the world volume  boundary is 
\beq 
\eta_bd\sigma^{b}=\frac{R}{\sigma}\sqrt{1+x'(\sigma)^2}d\sigma~~.
\eeq
Evaluating \eqref{b1} explicitly at the boundary $\sigma=0$ leads to
\beq \label{bc2}
T^{\sigma\sigma}\eta_\sigma|_{\sigma=0}=-Q_1\frac{(1+6\sinh^2\alpha)}{6\sinh\alpha\cosh\alpha}\frac{\sigma}{R}=0~~,~~J^{\tau\sigma}\eta_\sigma|_{\sigma=0}=Q_1\frac{\sigma^2}{R\ell}=0~~,
\eeq
where we have used the fact that $H(\sigma)\to\infty$ when $\sigma\to0$. The same result holds for the straight string.

\subsubsection*{Critical distance and regime of validity}
As in \eqref{kappa} one can define the parameter $\kappa$ via the relation
\beq \label{kappas}
\kappa=\frac{2^5 Q_1}{3^7 A R^6}=\frac{1}{\tilde\sigma^{12}}\frac{\sinh\alpha}{\cosh^5\alpha}~~,~~\tilde \sigma=\sqrt{\pi\frac{T}{\ell}}\sigma~~.
\eeq
The rescaling of the coordinate $\sigma$ was chosen such that $\kappa$ has no dependence on the temperature and remains the same as in the AdS case. This leads us to the  result for the critical distance $\tilde\sigma_c$
\beq \label{sigmas}
\tilde\sigma_c^2=\frac{2^{\frac{2}{3}}}{5^{\frac{5}{12}}}\kappa^{-\frac{1}{6}}~~.
\eeq
The largest scale characterising the probe is $r_c=R\kappa^{1/6}$ and the validity of the supergravity approximation still requires \eqref{sa}, while the requirement $r_c\ll R$ again yields $\kappa\ll1$. Furthermore, one must have that the local temperature variations compared to the background temperature must be small, i.e.,
\beq
r_c\frac{\mathcal{T'}}{T}=2r_c \frac{\sigma}{R\ell}\ll1~~.
\eeq
Near the boundary this is trivially satisfied, while for $\sigma\sim\sigma_c$ we obtain the requirement
\beq
\kappa\ll\left(\frac{\ell}{T}\right)^{6}~~,
\eeq
which just implies a very mild bound on the ratio $\ell/T$, in particular, for low temperatures this is easily satisfied. One further needs to require that the string is thin compared to the extrinsic curvature length scale $L_{\text{ext}}$ given by
\beq\label{l2}
L_{\text{ext}}(\sigma)=|K^{\rho}N_{\rho}|^{-1}=\frac{R}{\sigma}\frac{(1+x'(\sigma)^2)^{3/2}}{|x''(\sigma)|}~~.
\eeq
The minimum curvature is attained when $\sigma=\sigma_0$ and requiring $r_c\ll L_{\text{ext}}(\sigma_0=\sigma_c)$ just results in $r_c\ll R$ which has already been imposed. We thus conclude that these black strings in thermal Schr\"{o}dinger are valid solutions all the way up to $\sigma=\sigma_c$. We further note that the Polyakov loop in the case $\kappa\ne0$, as in AdS, lies outside the regime of validity of the method we are employing.

%%%%%%%%%%%%%%%%%%%%%%%%%%%%%%%%%%%%%%%%%%%%%%%%%%%%%%%%%%%%%%

\subsubsection*{Regularized free energy and large $N$ expansion}
The free energy \eqref{f2} using \eqref{t3} and the AdS/CFT dictionary can be written as 
\beq
\mathcal{F}[x(\tilde\sigma)]=\sqrt{\lambda} k T\int_{0}^{\tilde\sigma_0}\frac{d\tilde\sigma}{\tilde\sigma^3}(1-X)\sqrt{1+x'(\tilde\sigma)^2}~~,
\eeq
where $X$ was defined in \eqref{rf1}, and diverges at the boundary $\sigma=0$. Applying the same prescription as for AdS, we subtract a piece
\beq \label{fsub2}
\mathcal{F}_{\text{sub}}=\sqrt{\lambda}k T\int_{0}^{\tilde\sigma_{\text{cut}}}\frac{d\tilde\sigma}{\tilde\sigma^3}(1-X)~~,
\eeq
such that the free energy difference $\Delta\mathcal{F}$ can be written as $\Delta\mathcal{F}=\mathcal{F}-\mathcal{F}_{\text{sub}}=\mathcal{F}_{\text{loop}}-2\mathcal{F}_{\text{P}}$, where
\beq \label{floop2}
\mathcal{F}_{\text{loop}}=\sqrt{\lambda}k T\left(-\frac{1}{2\tilde\sigma_0^2}+\int_{0}^{\tilde\sigma_0}\frac{d\tilde\sigma}{\tilde\sigma^3}\left((1-X)\sqrt{1+x'(\tilde\sigma)^2}-1\right)\right)~~,
\eeq
\beq \label{fw1}
\mathcal{F}_{\text{P}}=-\frac{1}{2}\sqrt{\lambda}k T\left(\frac{1}{2\tilde\sigma_{\text{cut}}^2}+\int_{0}^{\tilde\sigma_{\text{cut}}}\frac{d\tilde\sigma}{\tilde\sigma^3}X\right)~~.
\eeq
Using similar arguments as in the AdS case, relying on the independence of $\kappa$ on $T$ and demanding the results at $T=0$ to coincide for both the extremal and black probes, one concludes that $\mathcal{F}_{\text{P}}=0$ for any $\kappa$. In order to obtain analytic results, we now proceed and make a similar expansion as in \eqref{exp1} and solve for $\phi$ obtaining
\beq \label{exp2}
\cosh^2\alpha=\frac{1}{\tilde\sigma^6\sqrt{\kappa}}-\frac{1}{4}-\frac{5}{32}\tilde\sigma^6\sqrt{\kappa}+\mathcal{O}(\kappa)~~.
\eeq
With this in hand we can compute the total distance between the boundary endpoints to order $\mathcal{O}\left(\kappa\right)$
\beq \label{ls1}
\textbf{L}\textbf{T}^{\frac{1}{2}}=\frac{2}{\sqrt{\pi}}\int_{0}^{\tilde\sigma_0}\frac{d x(\tilde\sigma)}{d\tilde\sigma}=3\tilde\sigma_0\frac{\Gamma(\frac{5}{3})}{\Gamma(\frac{1}{6})}\left(1-\tilde\sigma_0^6\sqrt{\kappa}+\mathcal{O}\left(\kappa\right)\right)~~,
\eeq
where the dimensionless parameters $\textbf{L}$ and $\textbf{T}$ were introduced in Eq.~\eqref{dim}. The leading order result in \eqref{ls1}, when written in terms of $\sigma_0$ leads to the result presented on the l.h.s. of \eqref{rext}. Similarly, we can expand \eqref{floop2} in powers of $\kappa$, obtaining 
\beq\label{fs1}
\mathcal{F}_{\text{loop}}=\sqrt{\lambda}kT\frac{\sqrt{\pi}}{6\hat\sigma_0^2}\frac{\Gamma(-\frac{1}{3})}{\Gamma(\frac{1}{6})}\left(1+\frac{4}{3}\hat\sigma_0^6\sqrt{\kappa}+\mathcal{O}\left(\kappa\right)\right)~~,
\eeq
for which the leading order term, when written in terms of $\sigma_0$ yields the result presented on the r.h.s. of \eqref{rext}. Inverting now \eqref{ls1}, introducing in \eqref{fs1} and expanding for small $\kappa$ and small $L$ we find
\beq \label{ff1}
\boldsymbol{\mathcal{F}}_{\text{loop}}=-2\sqrt{\pi}\frac{\sqrt{\lambda}k}{\textbf{L}^2}\left(\frac{\Gamma(-\frac{4}{3})\Gamma(\frac{5}{3})^{2}}{\Gamma(\frac{1}{6})^{3}}-\frac{\Gamma(\frac{1}{6})^3}{96\Gamma(\frac{2}{3})^3}\textbf{L}^{6}\textbf{T}^{3}+\mathcal{O}(\kappa)+\mathcal{O}(\textbf{L}^{10})\right)~~,
\eeq
where we have defined $\boldsymbol{\mathcal{F}}_{\text{loop}}=\mathcal{F}_{\text{loop}} \ell$. We see that the free energy has a different structure when compared with the free energy in AdS \eqref{fbeta}. In particular it scales with $\textbf{L}^{-2}$, as advertised in the previous section, while temperature corrections come in powers of $\textbf{L}\textbf{T}^{\frac{1}{2}}$. The combination $\boldsymbol{\mathcal{F}}_{\text{loop}}\textbf{L}^2$ is scale invariant and also $\textbf{L}\textbf{T}^{\frac{1}{2}}$, therefore we see that the quark-antiquark potential in Sch$_5\times S^5$ is capturing the conformal invariance of the dual non-relativistic CFT.

With \eqref{ff1} in hand, we can evaluate the black string entropy and mass using \eqref{thermo}. Specifically, we find
\beq
S=2\sqrt{\pi}\frac{\sqrt{\lambda}k}{\textbf{T} \textbf{L}^2}\left(-\frac{\Gamma(\frac{1}{6})^3}{16\Gamma(\frac{2}{3})^3}\textbf{L}^{6}\textbf{T}^{3}+\mathcal{O}(\kappa)+\mathcal{O}(\textbf{L}^{10})\right)~~.
\eeq
\beq
\textbf{M}=-2\sqrt{\pi}\frac{\sqrt{\lambda}k }{\textbf{L}^2}\left(\frac{\Gamma(-\frac{4}{3})\Gamma(\frac{5}{3})^{2}}{\Gamma(\frac{1}{6})^{3}}-\frac{7\Gamma(\frac{1}{6})^3}{96\Gamma(\frac{2}{3})^3}\textbf{L}^{6}\textbf{T}^{3}+\mathcal{O}(\kappa)+\mathcal{O}(\textbf{L}^{10})\right)~~
\eeq
where we have defined $\textbf{M}=M\ell$.

We now use a mixture of analytic and numerical methods to obtain the solution space of these black strings. We begin by solving for the distance $\textbf{L}\textbf{T}^{\frac{1}{2}}$ numerically, which is depicted on the l.h.s. of the figure below.
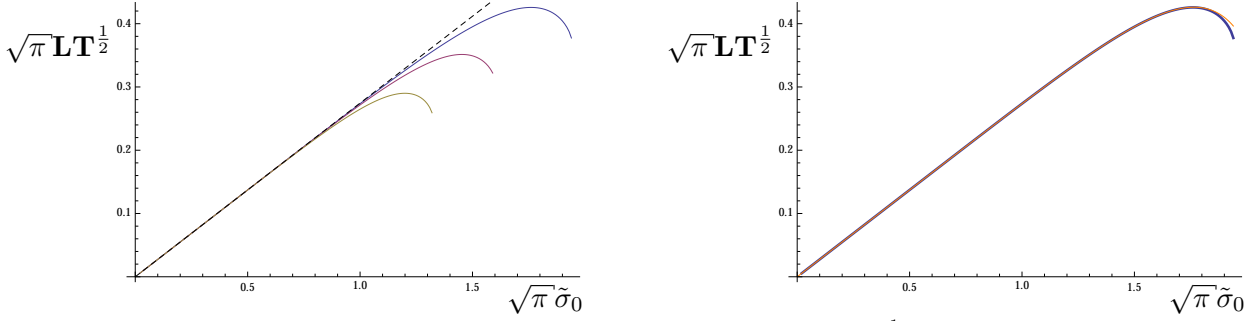
\begin{figure}[H]
\centering
\begin{subfigure}{.5\textwidth}
  \centering
  \includegraphics[width=.7\linewidth]{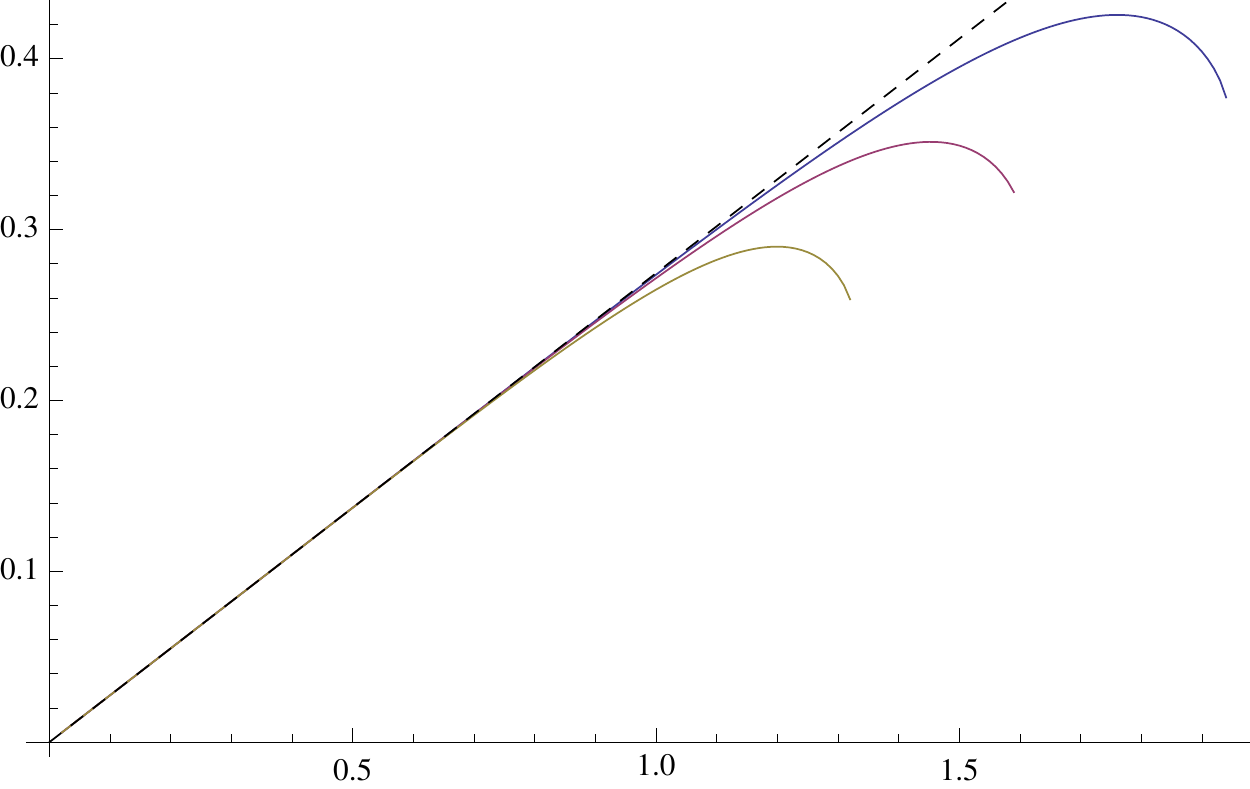}
  \begin{picture}(0,0)(0,0)
\put(-225,90){ $ \sqrt{\pi}\textbf{L}\textbf{T}^{\frac{1}{2}} $}
\put(-35,-5){ $ \sqrt{\pi}\tilde\sigma_0 $}
\end{picture}	
\end{subfigure}%
\begin{subfigure}{.5\textwidth}
  \centering
  \includegraphics[width=.7\linewidth]{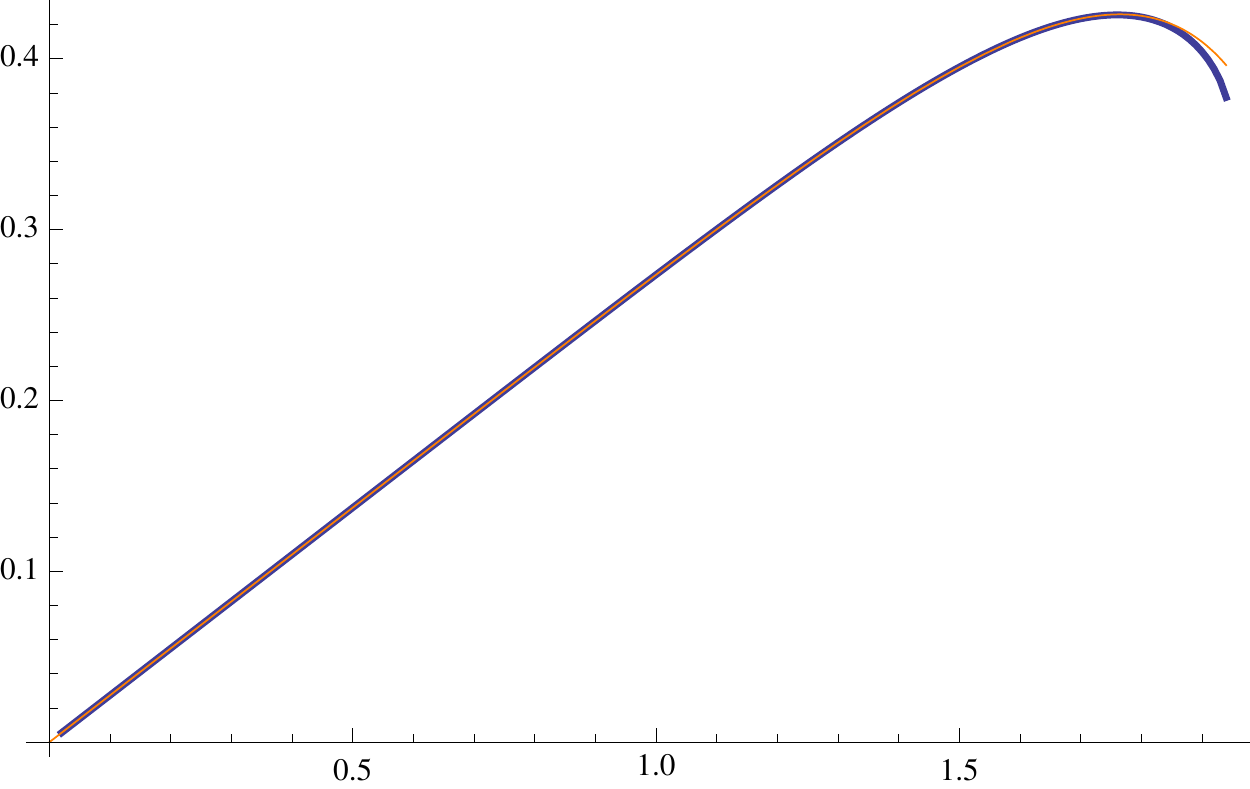}  
  \begin{picture}(0,0)(0,0)
  \put(-225,90){ $\sqrt{\pi}\textbf{L}\textbf{T}^{\frac{1}{2}} $}
  \put(-35,-5){ $ \sqrt{\pi}\tilde\sigma_0 $}
  \end{picture}	
\end{subfigure}
\caption{On the left is the plot of the dimensionless distance $\sqrt{\pi}\textbf{L}\textbf{T}^{\frac{1}{2}}$ as a function of $\sqrt{\pi}\tilde\sigma_0$ for $\kappa=0.01$ (yellow line), $\kappa=0.001$ (red line) and $\kappa=0.0001$ (blue line). The dashed black line is the case $\kappa=0$. On the right hand side we have $\sqrt{\pi}\textbf{L}\textbf{T}^{\frac{1}{2}}$ as a function of $\sqrt{\pi}\tilde\sigma_0$, where the thick blue line represents $\kappa=0.0001$ while thin orange line represents the analytic expansion to order $\mathcal{O}(\kappa^{5/2})$ given in \eqref{ls11}.}
\end{figure}
The dashed black line in the l.h.s. of the figure above represents the result for an extremal probe obtained by setting $\kappa=0$ in \eqref{ls1}. The dependence of the dimensionless distance on the bulk depth $\tilde\sigma_0$ is similar to that seen in thermal AdS in Fig.~\ref{adsL}. In order to describe these curves analytically for small $\kappa$ we can expand the dimensionless distance to $\mathcal{O}(\kappa^{5/2})$ in order to find
\beq \label{ls11}
\textbf{L}\textbf{T}^{\frac{1}{2}}=3\tilde\sigma_0\frac{\Gamma(\frac{5}{3})}{\Gamma(\frac{1}{6})}\left(1-\tilde\sigma_0^6\sqrt{\kappa}-\frac{83}{504}\tilde\sigma_0^{12}\kappa-\frac{10471}{78624}\tilde\sigma_0^{18}\kappa^{3/2}-\frac{820955}{5975424}\tilde\sigma_0^{24}\kappa^2+\mathcal{O}\left(\kappa^{5/2}\right)\right)~~,
\eeq
where we have used the fact that to this order
\beq
\cosh^2\alpha=\frac{1}{\tilde\sigma^6\sqrt{\kappa}}-\frac{1}{4}-\frac{5}{32}\tilde\sigma^6\sqrt{\kappa}-\frac{5}{32}\tilde\sigma^{12}\kappa-\frac{385}{2048}\tilde\sigma^{18}\kappa^{3/2}-\frac{19}{128}\tilde\sigma^{24}\kappa^{2}+\mathcal{O}\left(\kappa^{5/2}\right)~~.
\eeq
Using \eqref{ls11} we have drawn the thin orange curve in the r.h.s. of the figure above. As seen from the plot, the expansion \eqref{ls11} captures all the essentially features accurately up to values $\sigma\sim\sigma_c$. This behaviour is generic for small values of $\kappa$. We can perform a similar expansion in the free energy \eqref{floop2} obtaining
\beq\label{fs11}
\mathcal{F}_{\text{loop}}=\sqrt{\lambda}k T\frac{\sqrt{\pi}}{6\tilde\sigma_0^2}\frac{\Gamma(-\frac{1}{3})}{\Gamma(\frac{1}{6})}\left(1+\frac{4}{3}\tilde\sigma_0^6\sqrt{\kappa}+\frac{25}{84}\tilde\sigma_0^{12}\kappa+\frac{7955}{39312}\tilde\sigma_0^{18}\kappa-\frac{51805}{689472}\tilde\sigma_0^{24}\kappa^2+\mathcal{O}\left(\kappa^{5/2}\right)\right)~~.
\eeq
Below we show the behaviour of $\mathcal{F}_{\text{loop}}$ using \eqref{fs11} and \eqref{ls11} as a function of the bulk depth and as a function of the dimensionless distance between boundary endpoints.
\begin{figure}[H]
\centering
\begin{subfigure}{.5\textwidth}
  \centering
  \includegraphics[width=.8\linewidth]{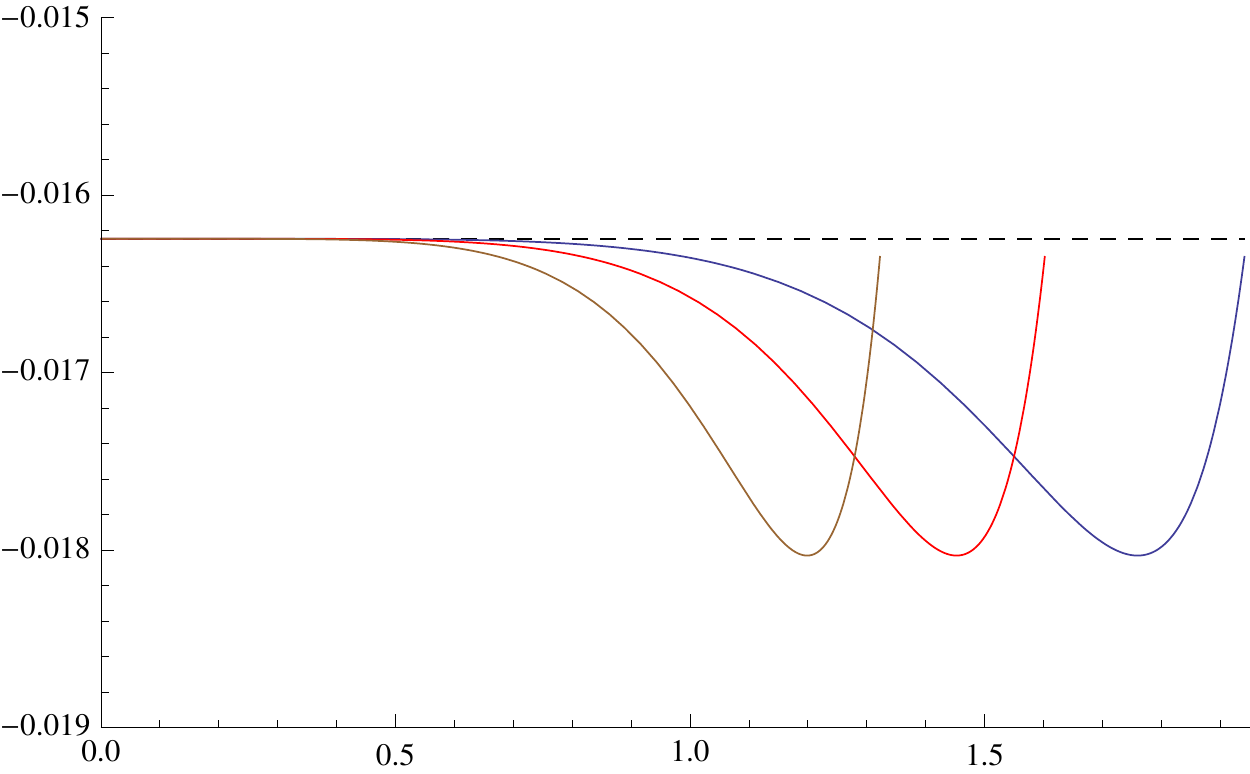}
  \begin{picture}(0,0)(0,0)
\put(-240,100){ $ \frac{\Delta\boldsymbol{\mathcal{F}}\textbf{L}^2}{ \sqrt{\lambda}k} $}
\put(-30,-5){ $ \tilde\sigma_0 $}
\end{picture}	
\end{subfigure}%
\begin{subfigure}{.5\textwidth}
  \centering
  \includegraphics[width=.8\linewidth]{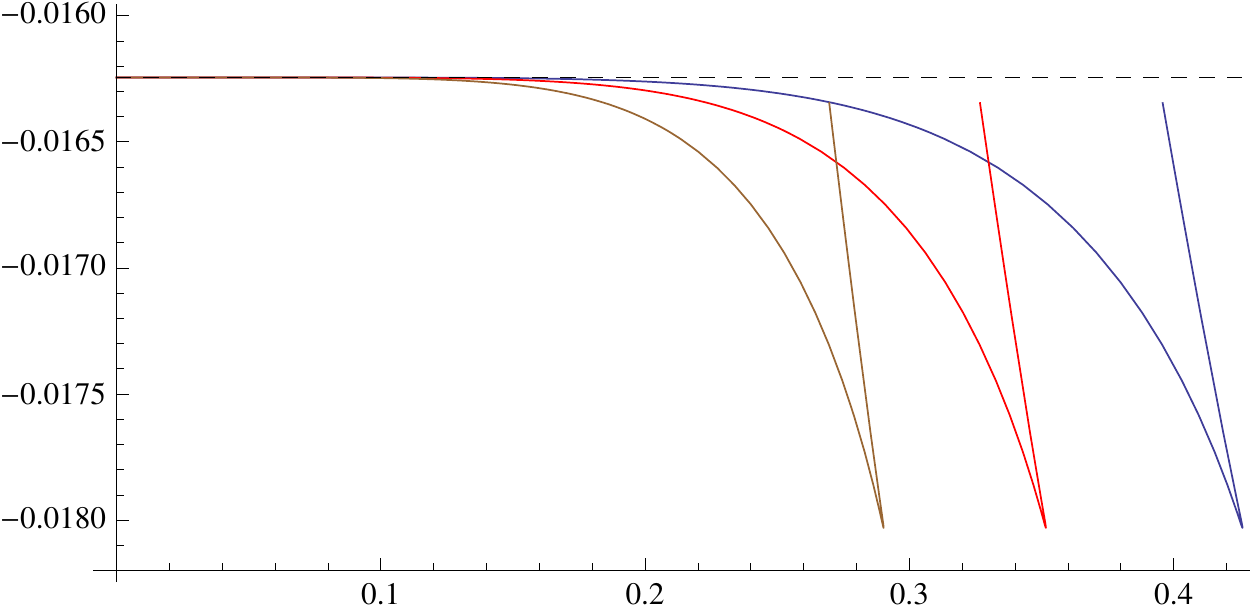}  
  \begin{picture}(0,0)(0,0)
  \put(-240,80){ $ \frac{\Delta\boldsymbol{\mathcal{F}}\textbf{L}^2}{ \sqrt{\lambda}k}  $}
  \put(-30,-10){ $\textbf{L}\textbf{T}^{\frac{1}{2}}$}
  \end{picture}	
\end{subfigure}
\caption{On the left is the plot of $\frac{\Delta\boldsymbol{\mathcal{F}}\textbf{L}^2}{ \sqrt{\lambda}k} $, where $\Delta\boldsymbol{\mathcal{F}}\equiv \boldsymbol{\mathcal{F}}_{\text{loop}}-\mathcal{F}_{\text{W}} \ell$, as function of $\tilde\sigma_0$ and as a function of $\textbf{L}\textbf{T}^{\frac{1}{2}}$ on the right. The colour coding is the same as on the l.h.s. of the previous figure. The dashed line is the result for $\kappa=0$.}
\end{figure}
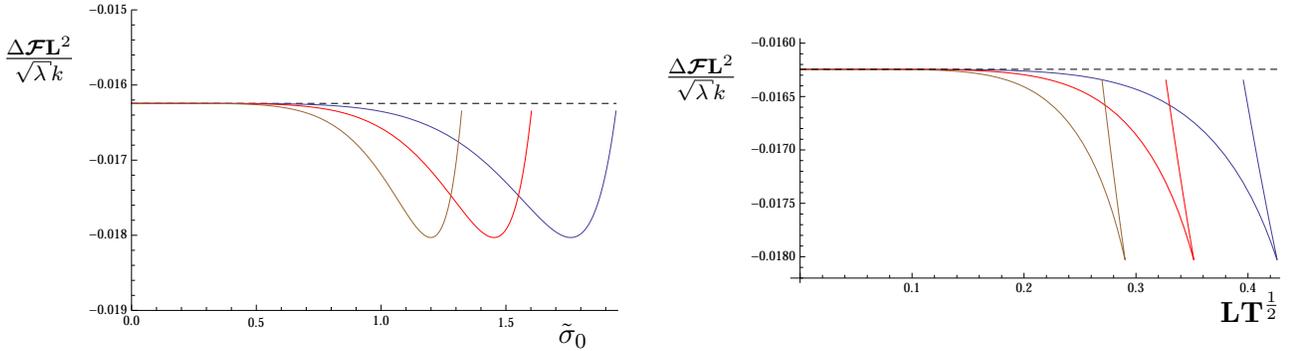
The black dashed line depicted above is the result for $\kappa=0$ which using Eq.~\eqref{fs11} is given by
\beq
\frac{\Delta\boldsymbol{\mathcal{F}}\textbf{L}^2}{ \sqrt{\lambda}k}|_{\kappa=0}=\frac{3\sqrt{\pi}}{2\tilde\sigma_0^2}\frac{\Gamma(-\frac{1}{3})\Gamma(\frac{5}{3})^2}{\Gamma(\frac{1}{6})^3}
\eeq
From the above figures, it is clear that, as in thermal AdS, there is no phase transition in thermal Schr\"{o}dinger and hence the black string configuration is always preferred compared to the Polyakov loop. When working in thermal Schr\"{o}dinger spacetimes we will use the term ``phase transition'' in this sense, i.e. we analyse possible phase transitions representing the onset of Debye screening effects, by which we mean transitions from Wilson loops to Polyakov loops.

%%%%%%%%%%%%%%%%%%%%%%%%%%%%%%%%%%%%%%%%%%%%%%%%%%%%%%%%%%%%%%
\subsection{Black strings with lightlike separated boundary endpoints} \label{nullstring}
In this section we consider black strings with lightlike separated boundary endpoints which have no AdS counterpart. In this case we fix the endpoints of the string in the null $\xi$-direction of the metric \eqref{dsback} and stretch it into the bulk along the z-direction. The embedding map is now
\beq
t=\tau~~,~~z=\sigma~~,~~\xi=\xi(\sigma)~~,~~x_1=x_2=d\Omega_{(5)}=0~~,
\eeq
rendering the problem symmetric around $\xi=L/2$. With this we find the induced metric which is now stationary
\beq
\gamma_{ab}d\sigma^{a}d\sigma^b=\frac{R^2}{\sigma^2}\left(-\frac{\ell^2}{\sigma^2}d\tau^2+2\xi'(\sigma)d\tau d\sigma+d\sigma^2\right)~~,
\eeq
and the corresponding free energy
\beq \label{f3}
\mathcal{F}[\xi(\sigma)]=-2A\left(\frac{3}{2\pi T}\right)^{6}\int_{0}^{{\sigma_0}}d\sigma\sqrt{1+\frac{\sigma^2}{\ell^2}\xi'(\sigma)^2}\thinspace H(\sigma)~~,
\eeq
where $H(\sigma)$ was given in \eqref{f2}. Here, we have used the fact that the norm of the Killing vector field $\partial_t$ on the world volume is $\textbf{k}=R\ell/\sigma^2$. The equations of motion that follow from \eqref{f3} are
\beq
\left(\frac{\sigma^2\xi'(\sigma)}{\ell^2\sqrt{1+\frac{\sigma^2}{\ell^2}\xi'(\sigma)^2}}H(\sigma)\right)'=0~~.
\eeq
We look for configurations that stretch from the boundary to $\sigma_0$ and back again to the boundary and that satisfy $\xi'(\sigma)>0$ as well as $\xi(\sigma)\to\infty$ as $\sigma\to\sigma_0$. In this case we obtain the solution
\beq
\xi'(\sigma)=\left(\frac{\sigma^4H^2(\sigma)}{\ell^2\sigma_0^2H^2(\sigma_0)}-\frac{\sigma^2}{\ell^2}\right)^{-\frac{1}{2}}~~.
\eeq
We note that the equation of motion also admits a solution of a stretched string $\xi'(\sigma)=0$ analogous to the Polyakov loop.

The boundary conditions \eqref{b1} are also satisfied. In order to see this explicitly, we construct the normalised orthogonal covector to the world volume boundary
\beq
\eta_{b}d\sigma^{b}=\frac{R}{\sigma}\sqrt{1+\frac{\sigma^2}{\ell^2}\xi'(\sigma)^2}d\sigma~~.
\eeq
Evaluating the boundary conditions \eqref{b1} explicitly leads to the same results as in \eqref{bc2}, however in this case we have an extra component of the stress-energy tensor that must also be considered
\beq
T^{\tau\sigma}\eta_\sigma|_{\sigma=0}=-Q_{1}\frac{(1+6\sinh^2\alpha)}{6\sinh\alpha\cosh\alpha}\frac{\sigma^3}{R\ell^2}\xi'(\sigma)=0~~,
\eeq
where we have used the fact that $\xi'(\sigma)\to0$ when $\sigma\to0$. The same result holds for the straight string.

We have established the existence of this type of black strings in Sch$_5\times S^5$ but we need to make sure that they lie within the regime of validity of our approximation. This analysis is in fact the same as in Sec.~\ref{blackspatial}, in particular the critical distance $\tilde\sigma_c$ remains unaltered. The only difference resides in the specific details of the extrinsic curvature length scale. In this case we find
\beq
L_{\text{ext}}(\sigma)=R\frac{(1+\xi'(\sigma)^2)^{3/2}}{|\xi'(\sigma)+\xi'(\sigma)^3-\sigma\xi''(\sigma)|}~~.
\eeq
The minimum curvature scale is found at $\sigma=\sigma_0$. It is easy to check that for small $\kappa$ either for $\sigma_0\sim0$ or $\sigma_0\sim\sigma_c$ we find that $L_{\text{ext}}(\sigma_0)\sim R$. This requires that $\kappa\ll1$, which has already been imposed. We conclude that these black strings are valid solutions all the way up to $\sigma_c$.

%%%%%%%%%%%%%%%%%%%%%%%%%%%%%%%%%%%%%%%%%%%%%%%%%%%%%%%%%%%%%%
\subsubsection*{Regularized free energy and large $N$ expansion}
As in the previous cases, the free energy \eqref{f3} can be rewritten as 
\beq
\mathcal{F}[\xi(\tilde\sigma)]=\sqrt{\lambda} k T\int_{0}^{\tilde\sigma_0}\frac{d\tilde\sigma}{\tilde\sigma^3}(1-X)\sqrt{1+\tilde\sigma^2\xi'(\tilde\sigma)^2}~~,
\eeq
and diverges at $\tilde\sigma=0$. In order to deal with this divergence we subtract the piece \eqref{fsub2} obtaining the regularised free energy
\beq \label{floop3}
\mathcal{F}_{\text{loop}}=\sqrt{\lambda}k T\left(-\frac{1}{2\tilde\sigma_0^2}+\int_{0}^{\tilde\sigma_0}\frac{d\tilde\sigma}{\tilde\sigma^3}\left((1-X)\sqrt{1+\tilde\sigma^2 \xi'(\tilde\sigma)^2}-1\right)\right)~~,
\eeq
while $\mathcal{F}_W$ in \eqref{fw1} still vanishes using the same type of analysis. We can now use the expansion \eqref{exp2} in order to evaluate the distance between the string endpoints
\beq \label{lnn}
\textbf{L}=2\int_{0}^{\tilde\sigma_0}d\tilde\sigma\frac{\partial\xi(\tilde\sigma)}{\partial\tilde\sigma}=\frac{\pi}{2}-\frac{2\hat\sigma_0^6}{3}\sqrt{\kappa}+\mathcal{O}(\kappa)~~.
\eeq
This expression is significantly different than what we have encountered before. First of all, in the case of an extremal probe ($\kappa=0$) the distance between the endpoints does not depend on how much the string is stretched into the bulk. Secondly, the distance decreases for increasing bulk depth $\tilde\sigma_0$ and non-zero $\kappa$. Thirdly, it does not depend on the temperature $T$. Performing a similar expansion in \eqref{floop3}, inverting \eqref{lnn} and introducing in \eqref{floop3} leads to
\beq\label{ff2}
\boldsymbol{\mathcal{F}}_{\text{loop}}=-\sqrt{\lambda}k \textbf{T}\left(\frac{9(\pi-2\textbf{L})^2}{2^7}\right)^{\frac{1}{3}}\kappa^{\frac{1}{6}}+\mathcal{O}(\kappa)~~,
\eeq
which is again qualitatively different than in the previous cases, namely, it is temperature dependent and the first correction is proportional to $\kappa^{\frac{1}{6}}$ and vanishes for zero $\kappa$. This configuration is thus more sensitive to corrections due to the internal degrees of freedom of the probe. This indicates that black probes can be used to distinguish Schr\"{o}dinger spacetimes from AdS in null coordinates.\footnote{The form of the free energy \eqref{ff2} is of a rather different nature when compared with, e.g., \eqref{ff1}. This suggest that perhaps the correct interpretation of $\textbf{L}$ in \eqref{ff2} is not as length but instead as a conserved particle number (see Ref.~\cite{Hartong:2013cba} for a discussion of this matter). }

We can also evaluate the black string entropy and mass using \eqref{ff2} and \eqref{thermo},
\beq
S=\sqrt{\lambda}k \left(\frac{9(\pi-2\textbf{L})^2}{2^7}\right)^{\frac{1}{3}}\kappa^{\frac{1}{6}}+\mathcal{O}(\kappa)~~,~~\textbf{M}=-\sqrt{\lambda}k \textbf{T}\left(\frac{9(\pi-2\textbf{L})^2}{2^6}\right)^{\frac{1}{3}}\kappa^{\frac{1}{6}}+\mathcal{O}(\kappa)~~,
\eeq
yielding an entropy which is finite, independent of $T$ and vanishing when $\kappa=0$.

In order to depict the solution space, we solve numerically for the length $\textbf{L}$ obtaining the figure below, which is accurately reproduced up to values $\sigma\sim\sigma_c$ and for small $\kappa$ by the following expansion
\beq \label{tL}
\begin{split}
\textbf{L}=&\frac{\pi}{2}-\frac{2}{3}\tilde\sigma_0^6\sqrt{\kappa}+\frac{1}{384}(-256+55\pi)\tilde\sigma_0^{12}\kappa-\frac{5(7856-2079\pi)}{36288}\tilde\sigma_0^{18}\kappa^{3/2} \\
&+\frac{5 (2063061 \pi -7178240)\tilde\sigma_0^{24} }{9289728 \pi }\kappa ^2 +\mathcal{O}\left(\kappa^{5/2}\right)~~.
\end{split}
\eeq
\begin{figure}[H]
\centering
  \includegraphics[width=0.4\linewidth]{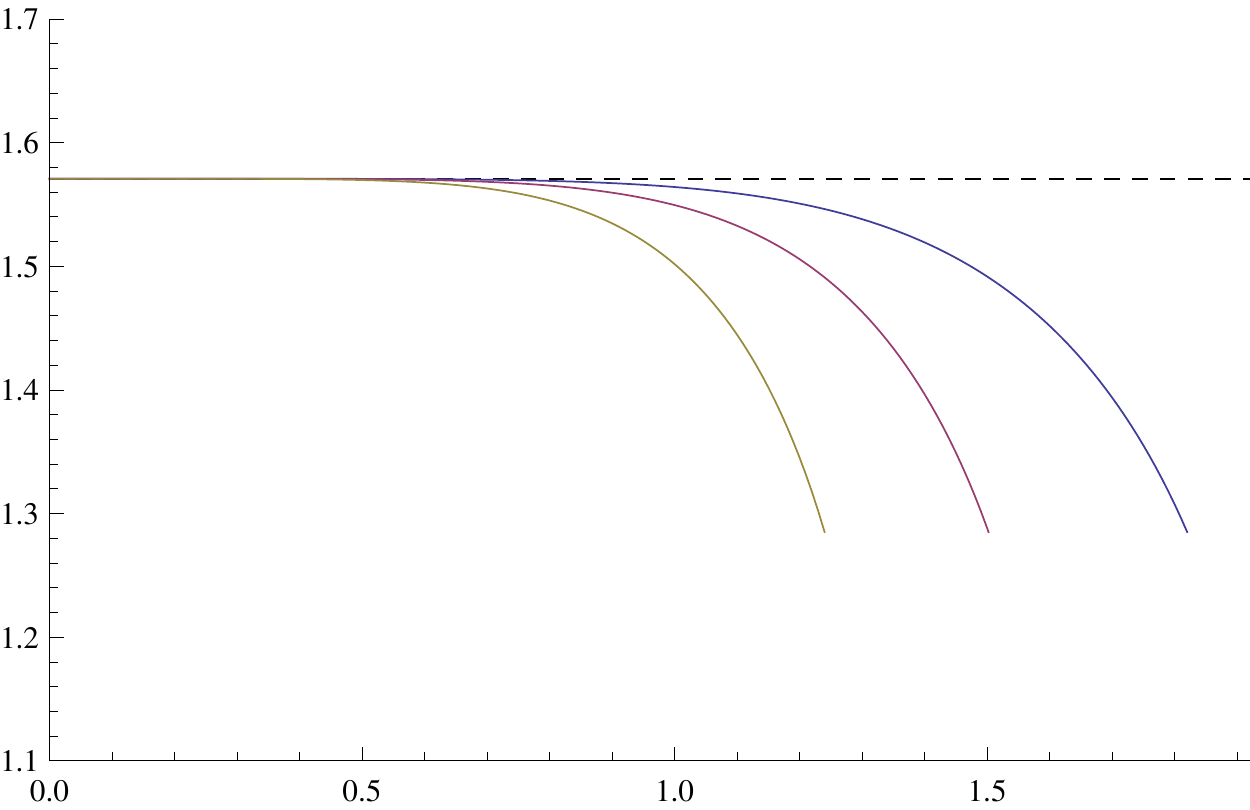}
  \begin{picture}(0,0)(0,0)
\put(-220,100){ $ \textbf{L}  $}
\put(-20,-5){ $ \tilde \sigma_0 $}
\end{picture}	
\caption{The dimensionless distance $\textbf{L}$ numerically obtained as a function of $\tilde\sigma_0$ for $\kappa=0.01$ (yellow line), $\kappa=0.001$ (red line) and $\kappa=0.00001$ (blue line). The dashed line represents the result for $\kappa=0$.}
\end{figure}
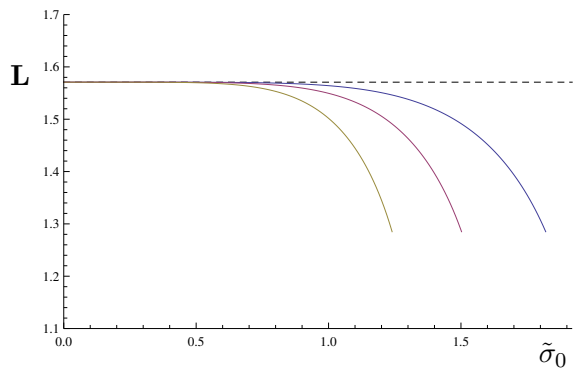
As we can see from the figure above, the maximum length is achieved when $\kappa=0$ for which we have from \eqref{tL} that $\textbf{L}=\pi/2$ and then decreases for increasing $\tilde\sigma_0$. We can also expand \eqref{floop3} to higher orders and obtain
\beq \label{fe3}
\begin{split}
\frac{\boldsymbol{\mathcal{F}}_{\text{loop}}}{\sqrt{\lambda}k \textbf{T}}=&-\frac{1}{2}\tilde\sigma_0^4\sqrt{\kappa}+\frac{1}{384}(-128+33\pi)\tilde\sigma_0^{10}\kappa-\frac{1}{8064}(4112-1155)\tilde\sigma_0^{16}\kappa^{3/2} \\
&+\frac{(483328-99315 \pi )\tilde\sigma_0^{22}}{5308416}\kappa ^2 +\mathcal{O}\left(\kappa^{5/2}\right)~~,
\end{split}
\eeq
which we can use to study phase transitions when comparing it to the Polyakov loop with vanishing free energy. This is depicted in the figures below.
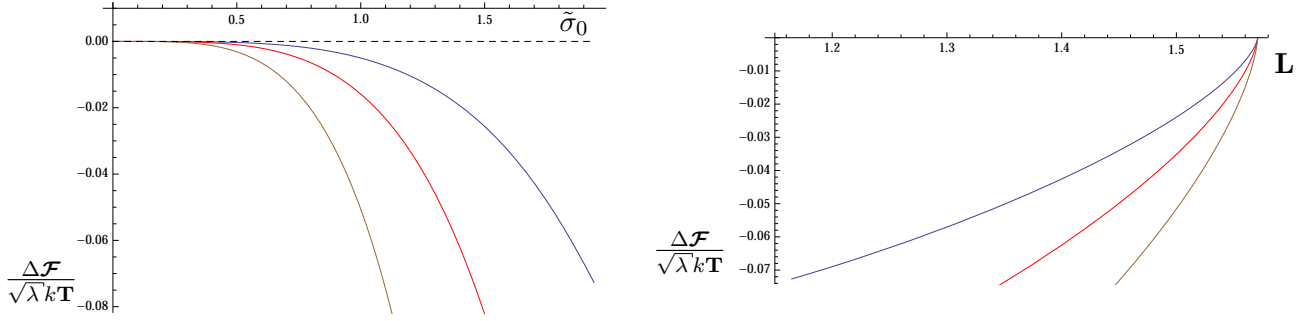
\begin{figure}[H]
\centering
\begin{subfigure}{.5\textwidth}
  \centering
  \includegraphics[width=.8\linewidth]{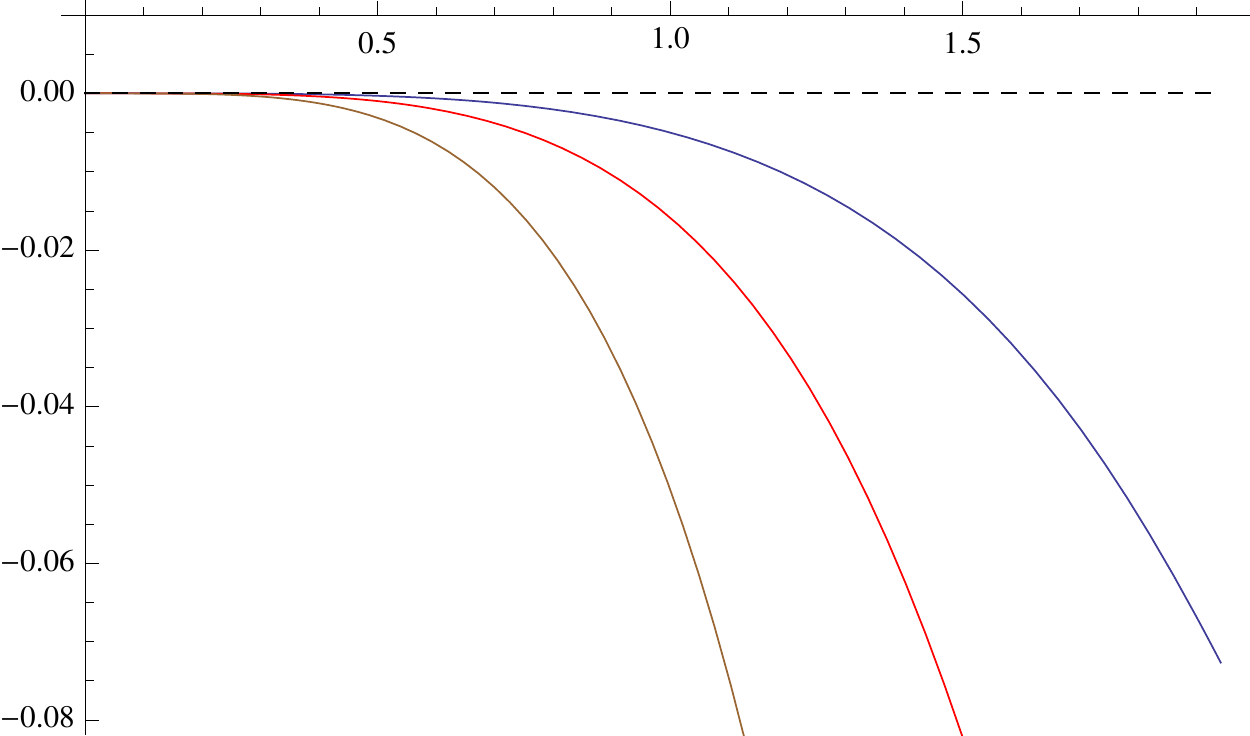}
  \begin{picture}(0,0)(0,0)
\put(-235,10){ $ \frac{\Delta\boldsymbol{\mathcal{F}}}{ \sqrt{\lambda}k \textbf{T}} $}
\put(-25,107){ $ \tilde\sigma_0 $}
\end{picture}	
\end{subfigure}%
\begin{subfigure}{.5\textwidth}
  \centering
  \includegraphics[width=.8\linewidth]{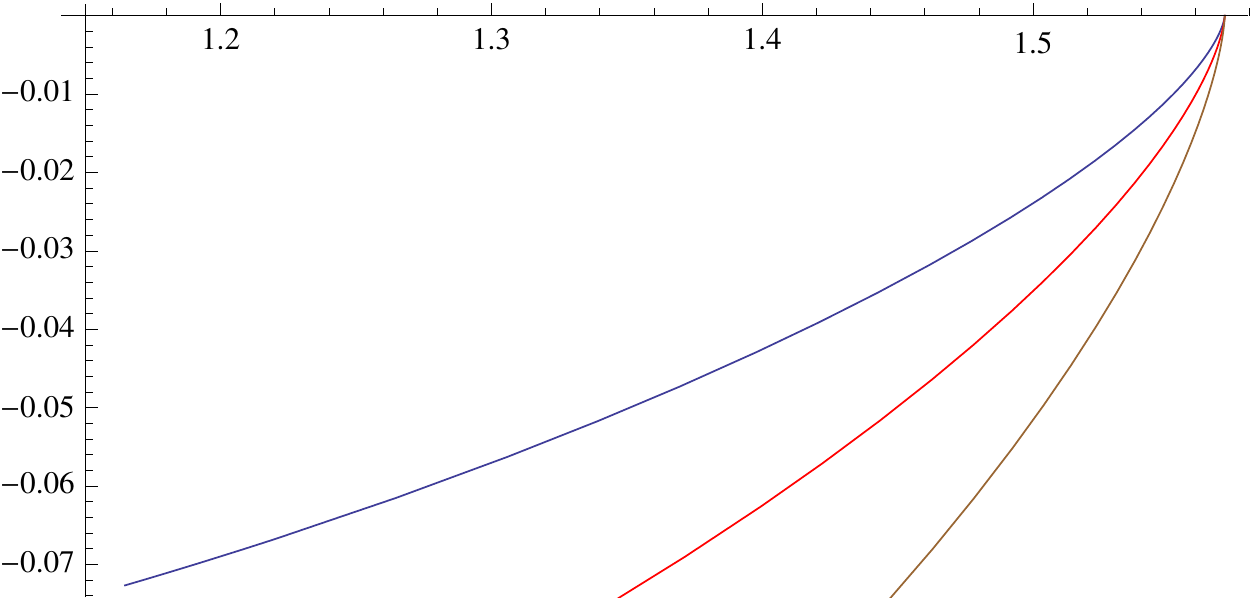}  
  \begin{picture}(0,0)(0,0)
  \put(-240,10){ $   \frac{\Delta\boldsymbol{\mathcal{F}}}{ \sqrt{\lambda}k \textbf{T}}  $}
  \put(-5,80){ $ \textbf{L}$}
  \end{picture}	
\end{subfigure}
\caption{On the left is the plot of $ \frac{\Delta\boldsymbol{\mathcal{F}}}{ \sqrt{\lambda}k \textbf{T}}$ as function of $\tilde\sigma_0$ and as a function of $\textbf{L}$ on the right obtained using \eqref{fe3} and \eqref{tL}. The colour coding is the same as in previous figure. The dashed line is the result for $\kappa=0$.}
\end{figure}
We can see from the figures below that $\Delta\boldsymbol{\mathcal{F}}\le0$. Therefore there is no phase transition. As a function of $\textbf{L}$, the free energy decreases for decreasing $\textbf{L}$ which is in stark contrast with the previous cases. Since we now have three competing configurations given by \eqref{fs11}, \eqref{fw1} and \eqref{fe3} regularised int the same way, one may wonder which is the preferred configuration for a fixed temperature and $\tilde\sigma_0$. Indeed, as seen in the figures below, the preferred configuration is always the black string with boundary endpoints spacelike separated. On the l.h.s. we have plotted the free energies \eqref{fs11} and \eqref{fe3} as a function of the bulk depth $\tilde\sigma_0$ while on the r.h.s. we have plotted the difference between \eqref{fs11} and \eqref{fe3} which we have denoted by $\Delta\boldsymbol{\mathcal{F}}_{(12)}$. 
\begin{figure}[H]
\centering
\begin{subfigure}{.5\textwidth}
  \centering
  \includegraphics[width=.7\linewidth]{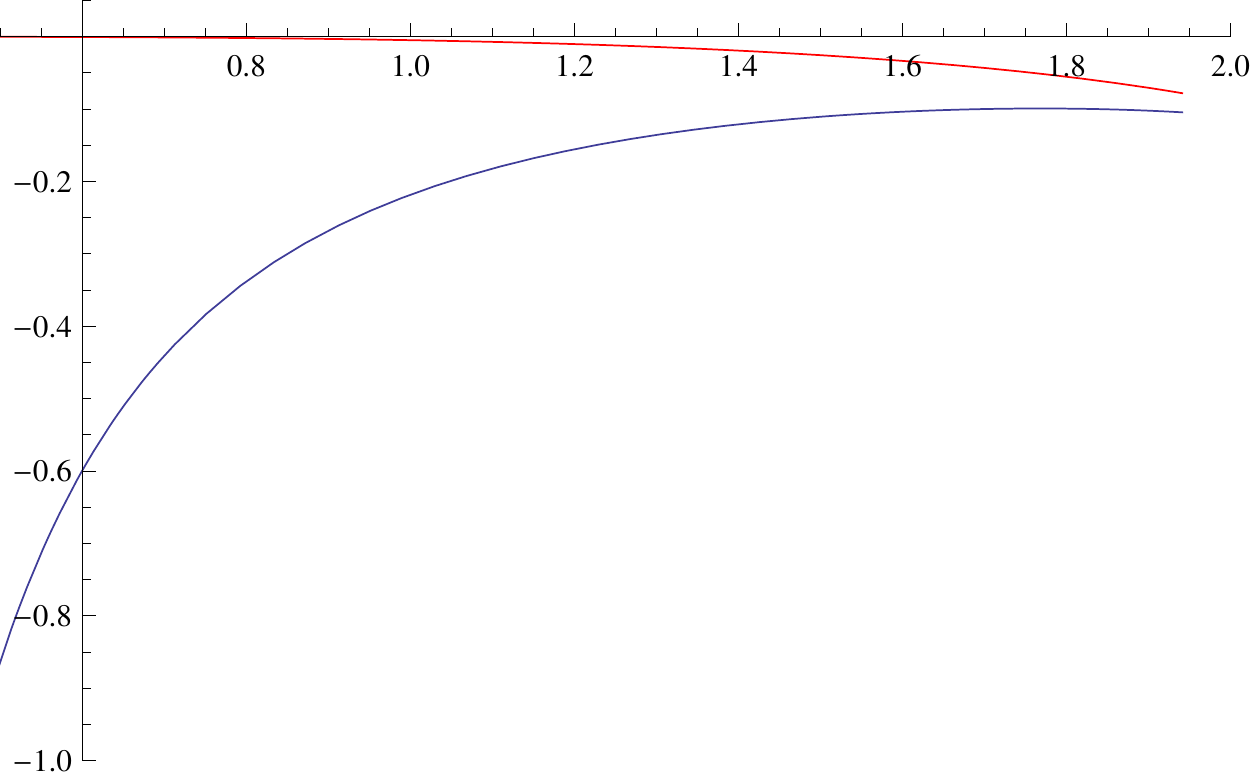}
  \begin{picture}(0,0)(0,0)
\put(-215,5){ $ \frac{\boldsymbol{\mathcal{F}}_{\text{loop}}}{\sqrt{\lambda}k \textbf{T}} $}
\put(-25,110){ $ \tilde\sigma_0 $}
\end{picture}	
\end{subfigure}%
\begin{subfigure}{.5\textwidth}
  \centering
  \includegraphics[width=.7\linewidth]{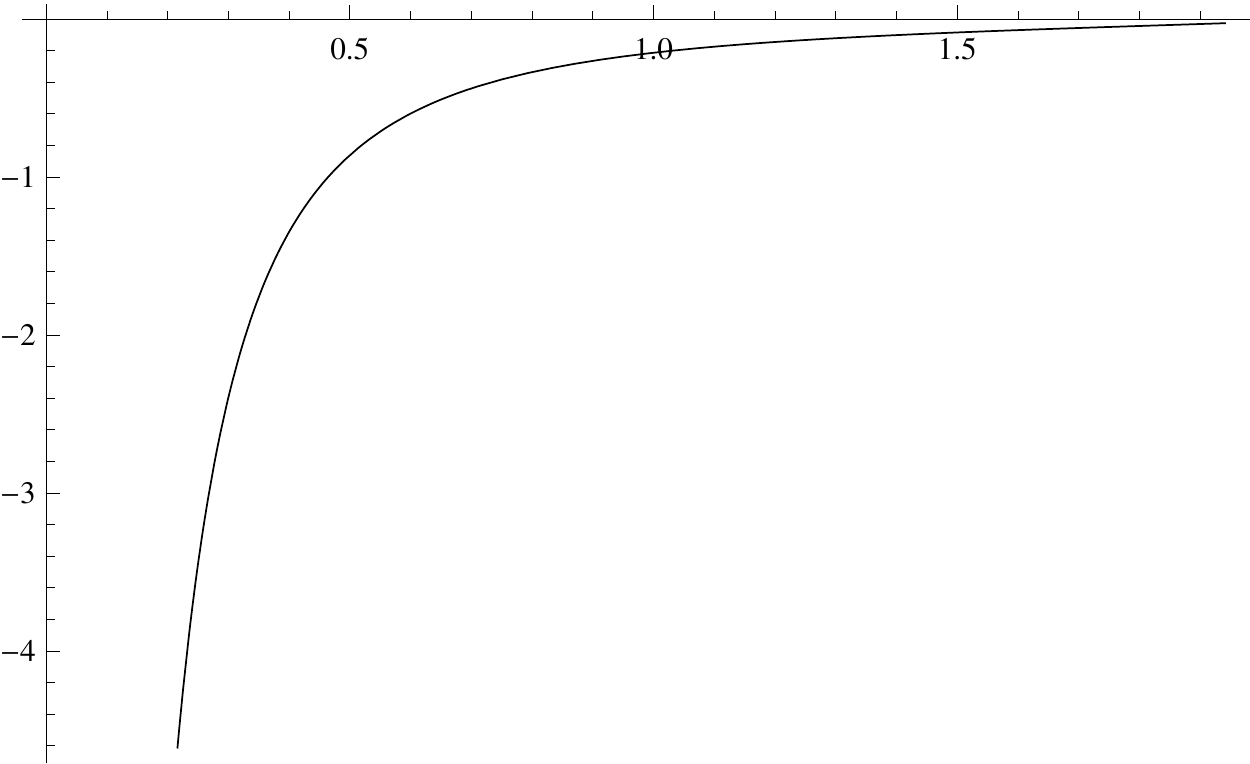}  
  \begin{picture}(0,0)(0,0)
  \put(-215,5){ $ \frac{\Delta\boldsymbol{\mathcal{F}}_{(12)}}{\sqrt{\lambda}k\textbf{T}} $}
  \put(-25,110){ $ \tilde\sigma_0$}
  \end{picture}	
\end{subfigure}
\caption{On the left is the plot of $\boldsymbol{\mathcal{F}}_{\text{loop}}/(\sqrt{\lambda}k\textbf{T})$ as a function of $\tilde\sigma_0$ for $\kappa=0.0001$ for black strings with free energies \eqref{fs11} (blue line) and with free energies \eqref{fe3} (red line). The qualitative features of this plot are the same for any small value of $\kappa$. On the right is the plot of the difference between both, denoted by $\Delta\boldsymbol{\mathcal{F}}_{12}$.}\end{figure}
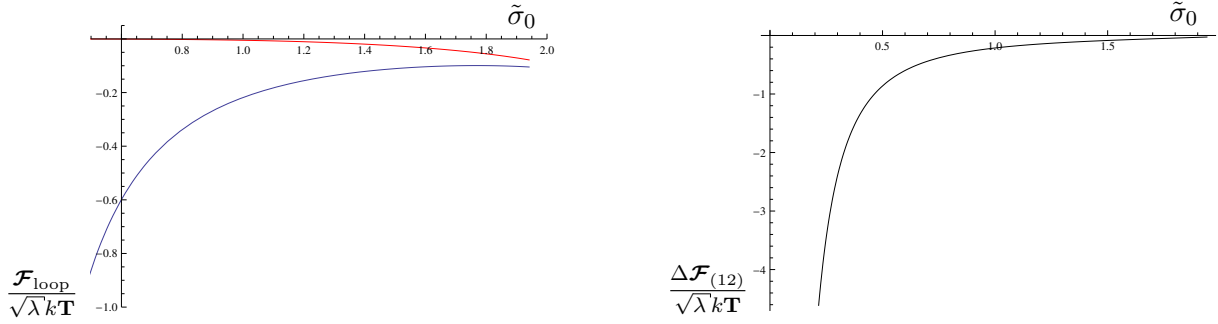
From the graph on the l.h.s. we see that the two configurations occupy two different regions of solution space without meeting each other and, in particular, the black strings with spacelike separated boundary endpoints always have lower free energies. From the plot on the r.h.s. we see that the difference $\Delta\boldsymbol{\mathcal{F}}_{12}$ is always negative, meaning that black strings with free energies \eqref{fs11} are always the preferred configurations among the two.

%%%%%%%%%%%%%%%%%%%%%%%%%%%%%%%%%%%%%%%%%%%%%%%%%%%%%%%%%%%%%%

\section{Black membranes in M-theory} \label{Mtheory}
In this section we consider black membranes stretched into the bulk of the M5-brane background ending on one-dimensional strings. At zero temperature, these configurations, known as Wilson surfaces, have been constructed in \cite{Maldacena:1998im} and their dual interpretation is that of the phase factor associated to the propagation of a very heavy string on the brane. At finite temperature, these solutions have been constructed in \cite{Chen:2008ds} using the extremal probe method. Here we will begin by constructing such Wilson surfaces in AdS$_7\times S^{4}$ using the black probe method, both in the presence of a black hole and without it, and then proceed to construct analogous configurations in thermal Sch$_7\times S^{4}$.

%%%%%%%%%%%%%%%%%%%%%%%%%%%%%%%%%%%%%%%%%%%%%%%%%%%%%%%%%%%%%%

\subsection{Black membranes in Anti-de Sitter}
We consider the M-theory background of $N$ coincident black M5-branes with metric
\beq \label{mads}
ds^2=\frac{R^2}{z^2}\left(-f(z)dt^2+f(z)^{-1}dz^2+dx_i^2\right)+\frac{R^2}{4}d\Omega^2_{(4)}~~,~~f(z)=1-\gamma \frac{z^6}{z_0^6}~~,
\eeq
where $i=1,...,5$ and the horizon radius $z_0$ is related to the temperature $T$ via $z_0=3/(2\pi T)$. The boundary of AdS is located at $z=0$. The parameter $\gamma$ was introduced as in Sec.~\ref{adsstring} to distinguish between thermal AdS ($\gamma=0$) and the AdS black hole ($\gamma=1$). We now wish to construct a configuration of a stretched black membrane along the $z$-direction and whose boundary endpoints are one-dimensional strings placed along the $x_2$ direction with length $\mathbb{L}$. In terms of the world volume coordinates $(\tau,\sigma,\sigma_2)$, the embedding map is given by
\beq
t=\tau~~,~~z=\sigma~~,~~x_1=x(\sigma)~~,~~x_2=\sigma_2~~,~~x_i=0~\forall~i=3,4,5~~,~~d\Omega_{(4)}=0~~.
\eeq
The membrane is stretched from the boundary at $(x_1,z)=(0,0)$ to the bulk until $(x_1,z)=(L/2,\sigma_0)$ and back again to the boundary at $(x_1,z)=(L,0)$. The problem is symmetric around $x_1=L/2$ and also translational invariant along $x_2$ so this direction can be integrated out and plays no role in the problem. The induced metric becomes
\beq \label{indM}
\gamma_{ab}d\sigma^{a}d\sigma^{b}=\frac{R^2}{\sigma^2}\left(-f(\sigma)d\tau^2+(f(\sigma)^{-1}+x'(\sigma)^2)d\sigma^2+d\sigma_2^2\right)~~,
\eeq
while the norm of the timelike Killing vector field on the world volume is $\textbf{k}=R\sqrt{f}/\sigma$. The probe black membrane being stretched carries a $Q_{2}$ charge, which in $D=11$ requires $n=6$ in Eqs.\eqref{t1}-\eqref{t3}. The free energy \eqref{freeenergy} can then be written as
\beq \label{fm1}
\mathcal{F}[x(\sigma)]=-2A\mathbb{L}\left(\frac{3}{2\pi T}\right)^{6}\int_{0}^{{\sigma_0}}d\sigma\sqrt{1+x'(\sigma)^2}\thinspace F(\sigma)~~,~~F(\sigma)=\frac{R^{9}}{\sigma^{9}}f(\sigma)^{3}\frac{1+6\sinh^2\alpha}{\cosh^6\alpha}~~,
\eeq
where we have integrated along the direction $\sigma_2$ yielding the overall factor of $\mathbb{L}$. The solution to the equations of motion that follow from \eqref{fm1} is identical to \eqref{s1} but with the function $G(\sigma)$ replaced by $F(\sigma)$, i.e.,
\beq \label{s2}
x'(\sigma)=\left(\frac{f(\sigma)^2F(\sigma)^2}{f(\sigma_0)F(\sigma_0)^2}-f(\sigma)\right)^{-\frac{1}{2}}~~,
\eeq
where we have imposed $x'(\sigma)\to\infty$ as $\sigma\to\sigma_0$. The boundary conditions \eqref{b1} are also satisfied. We also note that the equations of motion admit a solution of the type $x'(\sigma)=0$ corresponding to a straight membrane stretched from the boundary to the bulk, analogous to the Polyakov loop in Sec.~\ref{adsstring}.

\subsubsection*{Critical distance and regime of validity}
Introducing the dimensionless coordinate $\hat\sigma=2\pi T \sigma/3$, one can define a parameter $\bar\kappa$ as in \eqref{kappa} via
\beq \label{barkappa}
\bar\kappa\equiv\frac{Q_{2}}{6AR^6}=\frac{f(\hat\sigma)^3}{\hat\sigma^6}\frac{\sinh\alpha(\hat\sigma)}{\cosh^5\alpha(\hat\sigma)}~~,
\eeq
where $f(\hat\sigma)=1-\gamma\hat\sigma^6$. The fact that the function $f(\hat\sigma)$ depends on a different power of $\hat\sigma$ implies that the critical distance will change. In particular, we have that
\beq
\hat\sigma_c^2|_{\gamma=1}=\frac{\left(18+\sqrt{324+75\sqrt{5}\bar\kappa}\right)^{2/3}-3^{1/3}5^{5/6}\bar\kappa^{1/3}}{6^{2/3}\left(18+\sqrt{324+75\sqrt{5}\bar\kappa}\right)^{1/3}}~~,
\eeq
while for $\gamma=0$ we obtain again \eqref{sc0}. We see that the membranes in the case $\gamma=1$ cannot stretch all the way to the horizon for non-zero $\bar\kappa$, i.e., $\hat\sigma_c\le1$. These solutions, however, are not valid all the way to $\hat\sigma_c$ as we will see below.

The validity of the approximation requires that one can ignore quantum corrections. The largest scale characterising the membrane is $r_c=R\bar\kappa^{1/6}$. We therefore need to require $r_c\gg l_p$ which implies that $N_{(2)}\gg1$ where $N_{(2)}$ stands for the number of M2 black probe membranes and where we have used the fact that the brane tension can be written as $T_{(p)}=((2\pi)^{p}l_p^{p+1})^{-1}$. Requiring the radius $r_c$ to be much smaller than the curvature of the background $r_c\ll R$ yields the condition $N_{(2)}\ll N$ where $N$ should now be seen as the number of M5-branes that compose the background geometry.\footnote{Here we have used the AdS/CFT dictionary $R^3=2N/(\pi T_{(2)})$.} Therefore we must have that
\beq
1\ll N_{(2)}\ll N~~.
\eeq
Note that in terms of these quantities, we can write $\bar\kappa$, defined in \eqref{barkappa}, as $\bar\kappa= N_{(2)}/(2N^2)$. The requirement $r_c\ll R$ and $r_c\ll 1/T$ lead again to \eqref{val}. Furthermore, we should have small variations of the local temperature compared to the background temperature,
\beq
r_c\frac{\mathcal{T}'}{T}=\frac{r_c}{R\sqrt{f}}+3\gamma r_c\frac{(2\pi T\sigma/3)^6}{Rf^{3/2}}\ll1~~.
\eeq
Near the boundary this is satisfied as well as in general for the case $\gamma=0$ given that $\bar\kappa\ll1$. In the case $\gamma=1$ and near the critical distance we find that the l.h.s. above is proportional to $\bar\kappa^{-1/3}$ and hence the approximation breaks down near $\sigma_c$. For distances close to $z_0$ we need to require $z_0-\sigma\gg\bar\kappa^{1/9}z_0$. The extrinsic curvature length takes the same form as in \eqref{l1} and the requirement $r_c\ll L_{\text{ext}}(\sigma)$ is satisfied provided $z_0-\sigma\gg\bar\kappa^{1/9}z_0$.

\subsubsection*{Regularized free energy and large $N$ expansion}
The free energy \eqref{fm1} can be rewritten as
\beq 
\mathcal{F}[x(\hat\sigma)]=\frac{4^2}{3^2}\pi \mathbb{L} N_{(2)}^{3/2}\sqrt{\lambda_M}T^2\int_{0}^{\hat\sigma_0}\frac{d\hat\sigma}{\hat\sigma^3}(1-X)\sqrt{1+f(\hat\sigma)x'(\hat\sigma)^2}~~,
\eeq
where we have introduced the t' Hooft-like coupling $\lambda_M=N^2/N_{(2)}$ found in \cite{Niarchos:2012cy} in the context of the M2-M5 intersection. The characteristic $N_{(2)}^{3/2}$ dependence of M2-branes is also found in the context of giant gravitons in this background \cite{Armas:2013ota}. The free energy diverges at $\hat\sigma=0$ so, as in the previous cases, we must subtract a piece of the form
\beq
\mathcal{F}_{\text{sub}}=\frac{4^2}{3^2}\pi \mathbb{L} N_{(2)}^{3/2}\sqrt{\lambda_M}T^2\int_{0}^{\hat\sigma_{cut}}\frac{d\hat\sigma}{\hat\sigma^3}(1-X)~~,
\eeq
such that $\Delta \mathcal{F}=\mathcal{F}-\mathcal{F}_{\text{sub}}=\mathcal{F}_{\text{surf}}-2\mathcal{F}_{\text{P}_s}$, where
\beq \label{fsurf}
\mathcal{F}_{\text{surf}}=\frac{4^2}{3^2}\pi \mathbb{L} N_{(2)}^{3/2}\sqrt{\lambda_M}T^2\left(-\frac{1}{2\hat\sigma_0^2}+\int_{0}^{\hat\sigma_0}\frac{d\hat\sigma}{\hat\sigma^3}\left((1-X)\sqrt{1+fx'(\sigma)^2}-1\right)\right)~~,
\eeq
\beq \label{fw55}
\mathcal{F}_{\text{P}_s}=-\frac{1}{2}\frac{4^2}{3^2}\pi \mathbb{L}N_{(2)}^{3/2}\sqrt{\lambda_M}T^2\left(\frac{1}{2\hat\sigma_{\text{cut}}^2}+\int_{0}^{\hat\sigma_{\text{cut}}}\frac{d\hat\sigma}{\hat\sigma^3}X\right)~~.
\eeq
Here $\mathcal{F}_{\text{surf}}$ is the regularised free energy of the stretched black membrane while $\mathcal{F}_{\text{P}_s}$ is the regularised free energy of the straight black membrane. By similar arguments as those used for the Polyakov loop, but now requiring configurations at $T=0$ for any $\bar\kappa$ to have the same value of $\partial S/\partial T$, one finds that $\mathcal{F}_{\text{P}_s}=-(4^2/3^2)\gamma \pi \mathbb{L} N_{(2)}^{3/2}\sqrt{\lambda_M}T^2/4$. We can now proceed and obtain analytic results in a small $\bar\kappa$ expansion. We find that the length along the direction $x_1$ for small $\bar\kappa$ and small $\hat\sigma_0$ is given by
\beq \label{ltm}
LT=\frac{3\Gamma(\frac{2}{3})}{\sqrt{\pi}\Gamma(\frac{1}{6})}\hat\sigma_0+\frac{1}{3\sqrt{\pi}}\left(\frac{\Gamma(\frac{2}{3})}{\Gamma(\frac{1}{6})}+\frac{\Gamma(\frac{7}{6})}{\Gamma(-\frac{1}{3})}\right)\hat\sigma_0^4\sqrt{\bar\kappa}-\gamma\frac{27\Gamma(\frac{5}{3})}{28\sqrt{\pi}\Gamma(\frac{1}{6})}\hat\sigma_0^7+\mathcal{O}(\bar\kappa)+\mathcal{O}(\hat\sigma_0^{8})~~.
\eeq
Again, we see that the leading order correction only depends on the properties of the probe and not of the background. The background contribution when $\gamma=1$ had already been obtained in \cite{Chen:2008ds}. Expanding the free energy \eqref{fsurf} for small $\kappa$ and $L$ and inverting \eqref{ltm} we find
\beq \label{fm1}
\begin{split}
\mathcal{F}_{\text{surf}}=-\frac{4^2\pi }{3^2}\frac{\mathbb{L}}{L^2} N_{(2)}^{3/2}\sqrt{\lambda_M}\Big(\frac{3^2}{2\sqrt{\pi}}\frac{\Gamma(\frac{2}{3})^3}{\Gamma(\frac{1}{6})^3}-&\frac{2\Gamma(-\frac{2}{3})\Gamma(\frac{7}{6})^2}{3\sqrt{3}\Gamma(\frac{2}{3})}\sqrt{\bar\kappa}(TL)^3\\
&+\gamma\frac{2\pi^{5/2}}{567}\frac{\Gamma(\frac{1}{6})^3}{\Gamma(\frac{2}{3})^3}(TL)^6+\mathcal{O}(\bar\kappa)+\mathcal{O}(L^8)\Big)~~.
\end{split}
\eeq 
The leading order result agrees with the result presented in \cite{Maldacena:1998im} and the coefficient proportional to $\gamma$ agrees with the one found in \cite{Chen:2008ds} when $\gamma=1$. Again, the corrections due to the probe are the leading order corrections and they enter at the same order as in the case of the Wilson loops of Sec.~\ref{adsstring}. On the other hand the contribution due to the background enters at higher order when compared to the Wilson loops of Sec.~\ref{adsstring}. For the black probe corrections to be leading it is sufficient to require
\beq
(TL)^3\ll \frac{1}{\lambda_{M}}~~,
\eeq
where we have used the fact that, as noted above, $\bar\kappa=N_{(2)}/(2N^2)$. This is always satisfied for small enough temperatures, and more easily satisfied for small temperatures than in the Wilson loop case due to the cubic factor in $TL$. The combination $\mathcal{F}_{\text{surf}}L^2/\mathbb{L}$ is scale invariant as well as the combination $LT$, therefore we see the underlying conformal symmetry of the dual gauge theory. 

\subsubsection*{Numerical analysis}
We now analyse the solution space in detail by solving numerically for the various quantities. We begin by solving for the distance between the strings on the boundary for both values of $\gamma$. This is depicted in the figures below. 
\begin{figure}[H]
\centering
\begin{subfigure}{.5\textwidth}
  \centering
  \includegraphics[width=0.8\linewidth]{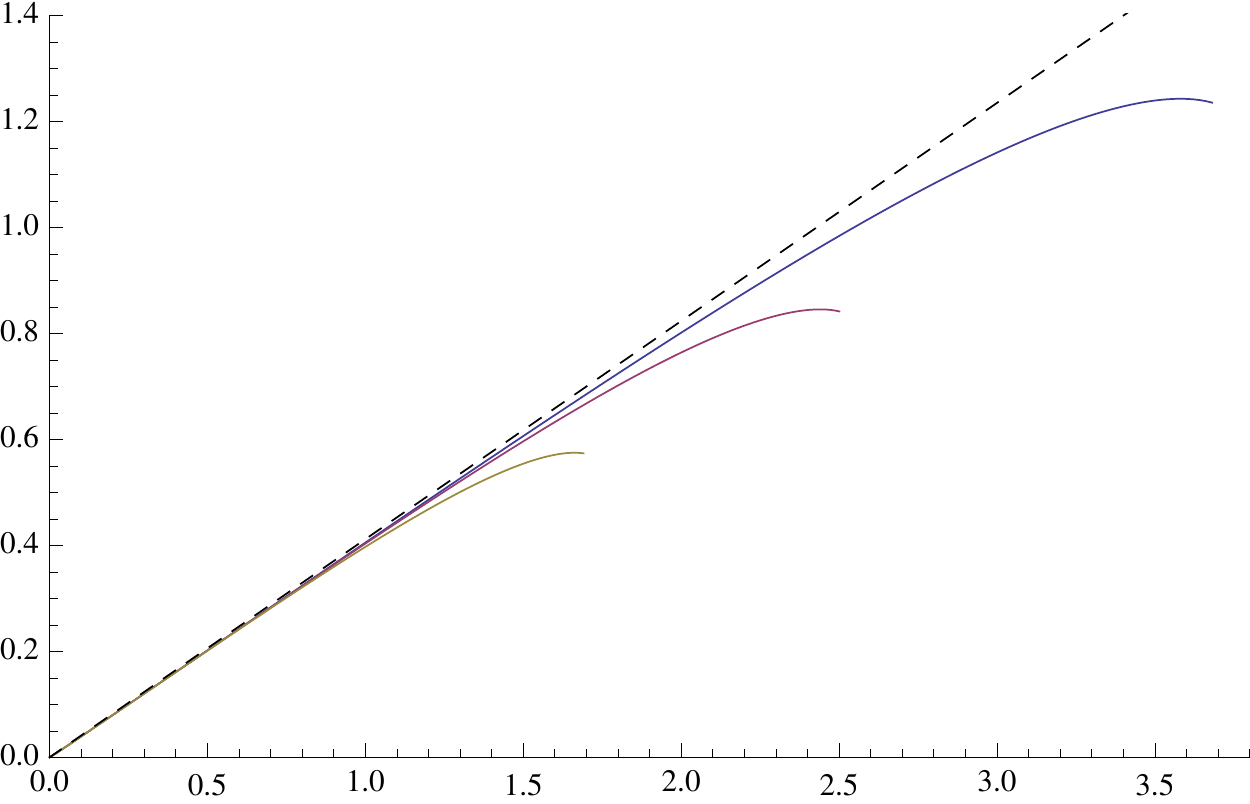}
  \begin{picture}(0,0)(0,0)
\put(-230,120){ $ LT  $}
\put(-20,-5){ $ \hat\sigma_0 $}
\end{picture}	
\end{subfigure}%
\begin{subfigure}{.5\textwidth}
  \centering
  \includegraphics[height=0.47\linewidth,width=0.8\linewidth]{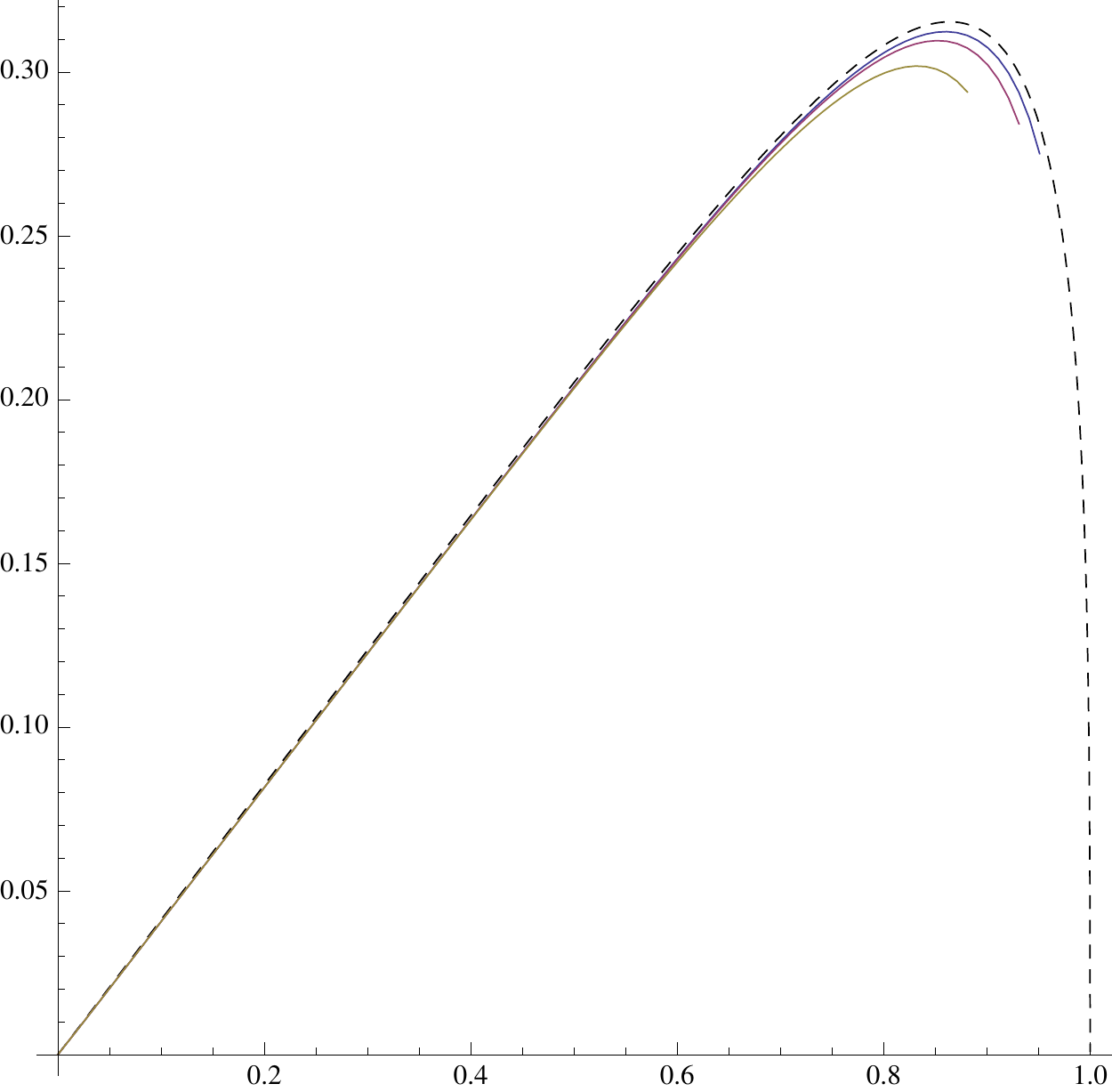}  
  \begin{picture}(0,0)(0,0)
  \put(-230,110){ $ LT $}
  \put(-20,-10){ $ \hat\sigma_0 $}
  \end{picture}	
\end{subfigure}
\caption{On the left we have the numerical results for the dimensionless length $LT$ as a function of $\hat\sigma_0$ in thermal AdS for $\bar\kappa=0.01$ (yellow curve), $\bar\kappa=0.001$ (red curve) and $\bar\kappa=0.0001$ (blue curve). The dashed line is the result for $\bar\kappa=0$. On the right we have the numerical results in the presence of a black hole.}
\end{figure}
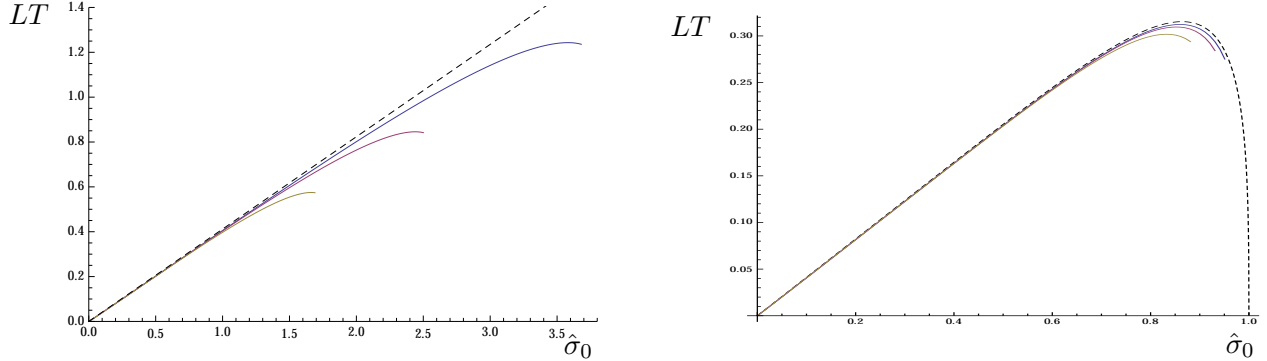
The qualitative features are the same as for the Wilson loop of Sec.~\ref{adsstring}. In particular we observe the existence of the critical distance $\hat\sigma_c$ beyond which solutions cease to exist. In the presence of the black hole, this means that the black membrane cannot stretch all the way to the horizon. The dashed curve representing the case $\kappa=0$ can be obtained exactly,
\beq
LT|_{\kappa=0}=\frac{3\Gamma(\frac{2}{3})}{\sqrt{\pi}\Gamma(\frac{1}{6})}\hat\sigma_0\sqrt{1-\gamma \hat\sigma_0^6}\thinspace\thinspace \tensor*[_2]{F}{}_1\left(\frac{1}{2},\frac{2}{3},\frac{7}{6};\gamma\hat\sigma_0^6\right)~~,
\eeq
and agrees with the result of \cite{Chen:2008ds} when $\gamma=1$. We now present the free energy difference between the two configurations. On the l.h.s. of the plot below we have the dependence of the free energy difference as a function of $\hat\sigma_0$ and on the r.h.s. as a function of $LT$ in the case $\gamma=1$. We observe that $\Delta\mathcal{F}$ becomes positive for large values of $LT$, which implies a similar effect as the Debye screening effect for the quark-antiquark pair. The exact onset can be obtained numerically and reads
\beq
LT|_{\Delta\mathcal{F}=0}\sim0.2779+0.1131\sqrt{\bar\kappa}~~.
\eeq
\begin{figure}[H]
\centering
\begin{subfigure}{.5\textwidth}
  \centering
  \includegraphics[width=0.75\linewidth]{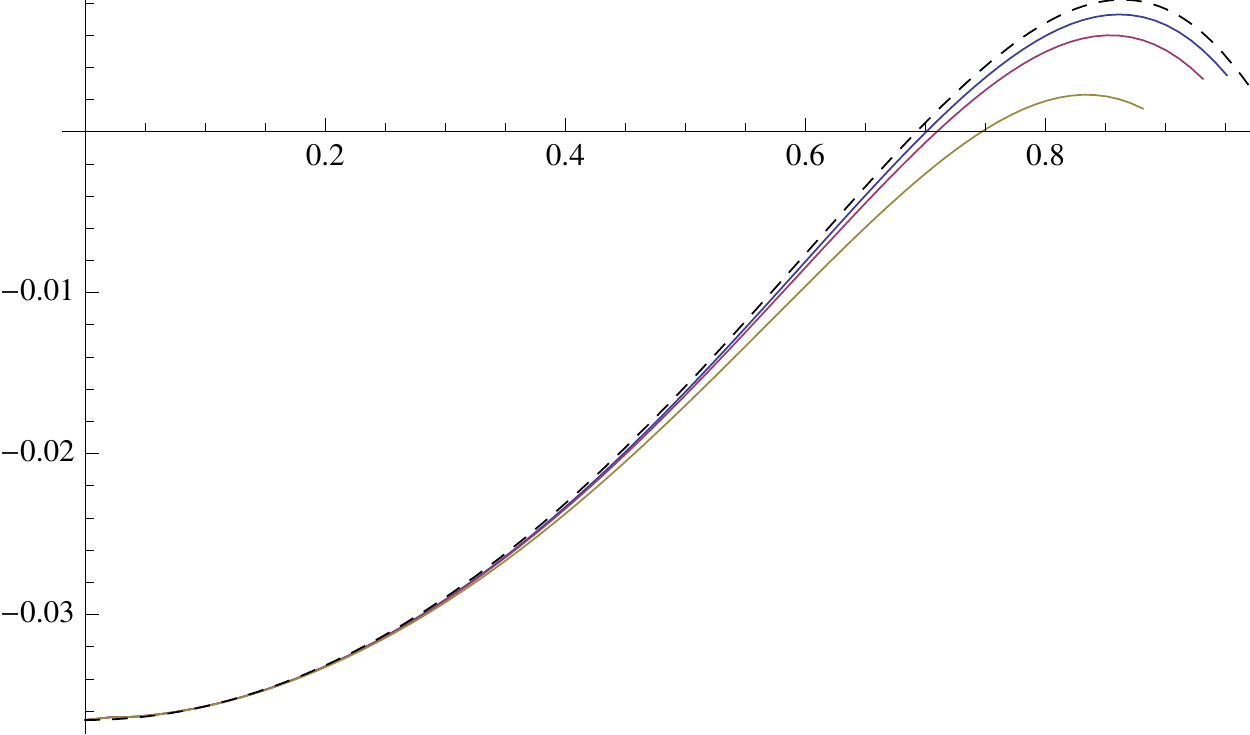}
  \begin{picture}(0,0)(0,0)
\put(-250,95){ $ \frac{3^2\Delta\mathcal{F}L^2}{4^2\pi\mathbb{L} N_{2}^{3/2}\sqrt{\lambda_M}} $}
\put(-20,80){ $ \hat\sigma_0 $}
\end{picture}	
\end{subfigure}%
\begin{subfigure}{.5\textwidth}
  \centering
  \includegraphics[width=0.65\linewidth,height=0.45\linewidth]{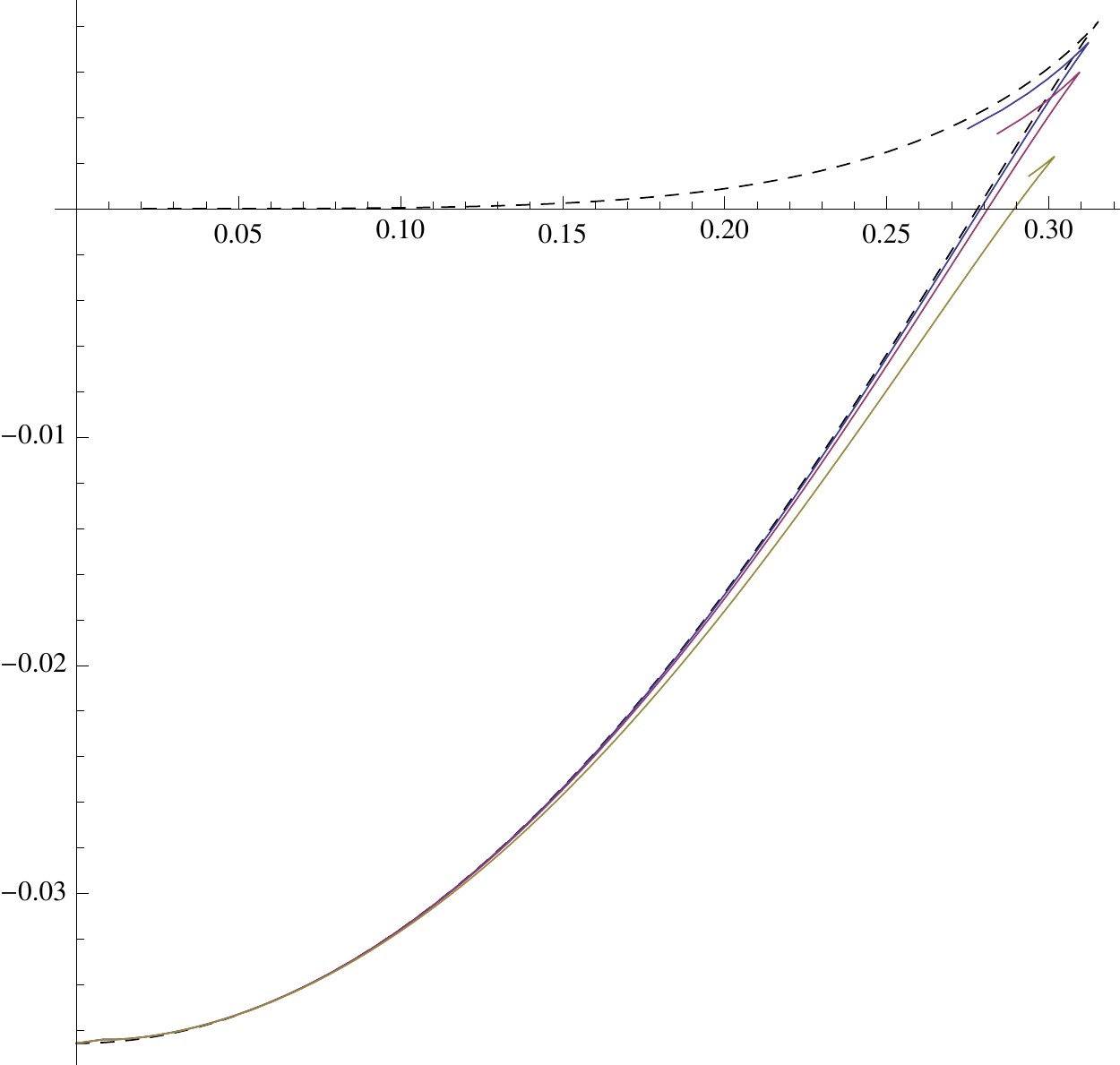}  
  \begin{picture}(0,0)(0,0)
  \put(-225,100){ $\frac{3^2\Delta\mathcal{F}L^2}{4^2\pi\mathbb{L} N_{2}^{3/2}\sqrt{\lambda_M}} $ }
  \put(-20,75){ $ LT $}
  \end{picture}	
\end{subfigure}
\caption{On the left we have the free energy difference as a function of $\hat\sigma_0$ and on the right as a function of $LT$ in the case $\gamma=1$. The dashed line represents $\bar\kappa=0$ and has been obtained in \cite{Chen:2008ds}.}
\end{figure}
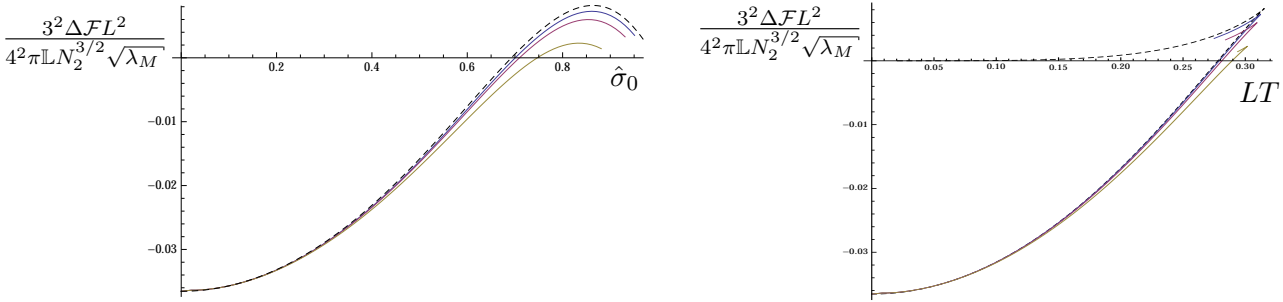
In the case $\gamma=0$ the results are also similar to the Wilson loop as one can see in the figures below, namely, there is no phase transition.
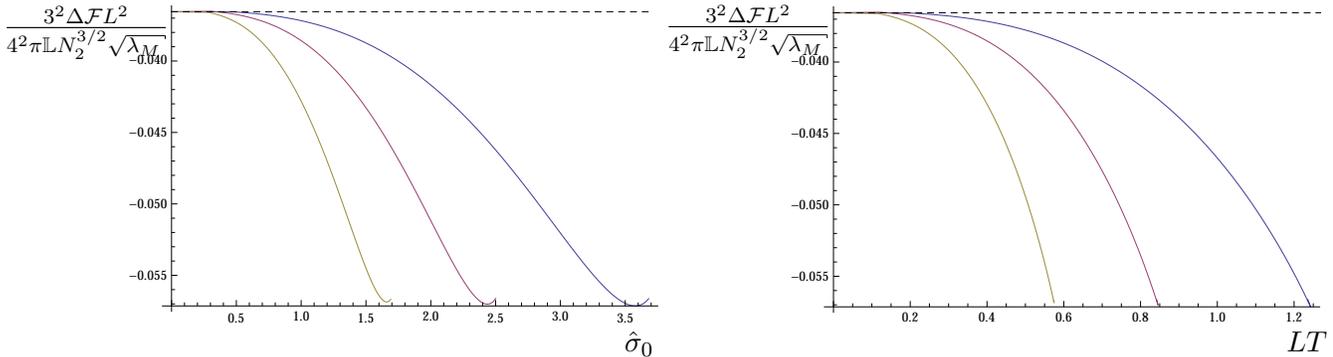
\begin{figure}[H]
\centering
\begin{subfigure}{.5\textwidth}
  \centering
  \includegraphics[width=0.8\linewidth]{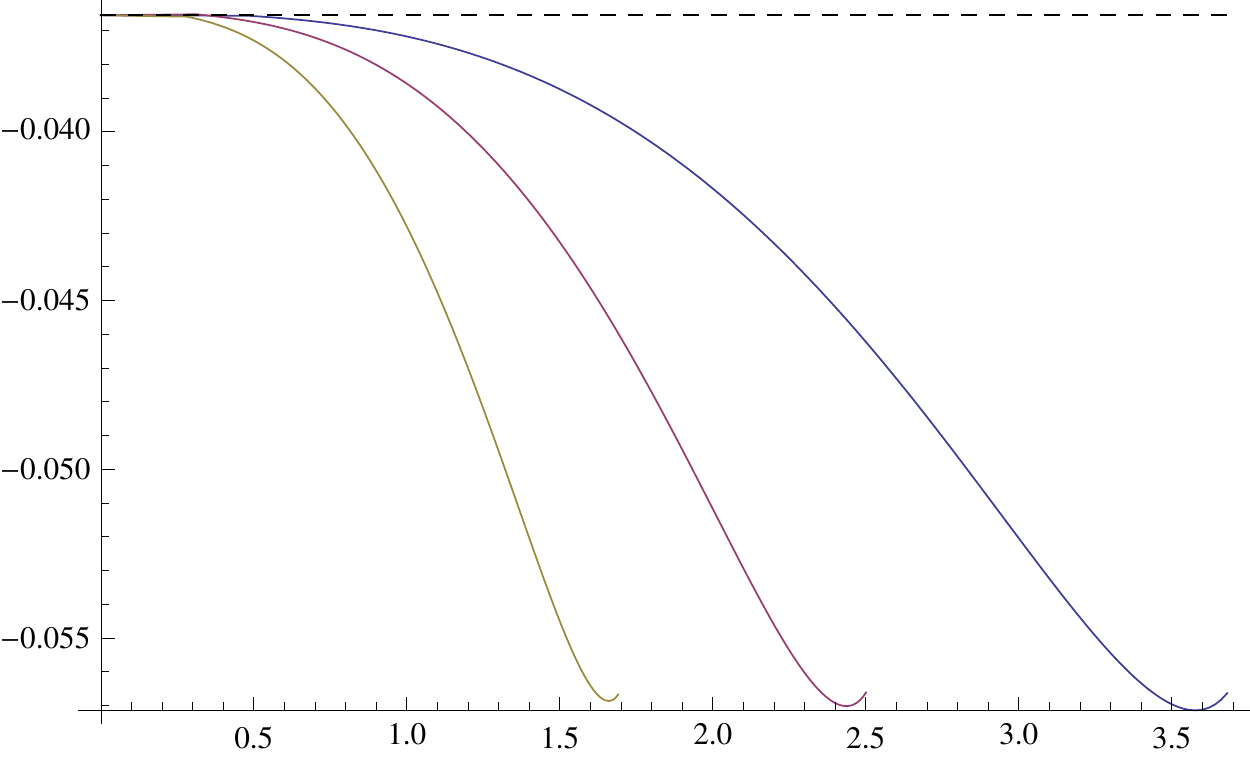}
  \begin{picture}(0,0)(0,0)
\put(-255,110){ $ \frac{3^2\Delta\mathcal{F}L^2}{4^2\pi\mathbb{L} N_{2}^{3/2}\sqrt{\lambda_M}} $}
\put(-20,-10){ $ \hat\sigma_0 $}
\end{picture}	
\end{subfigure}%
\begin{subfigure}{.5\textwidth}
  \centering
  \includegraphics[width=0.8\linewidth]{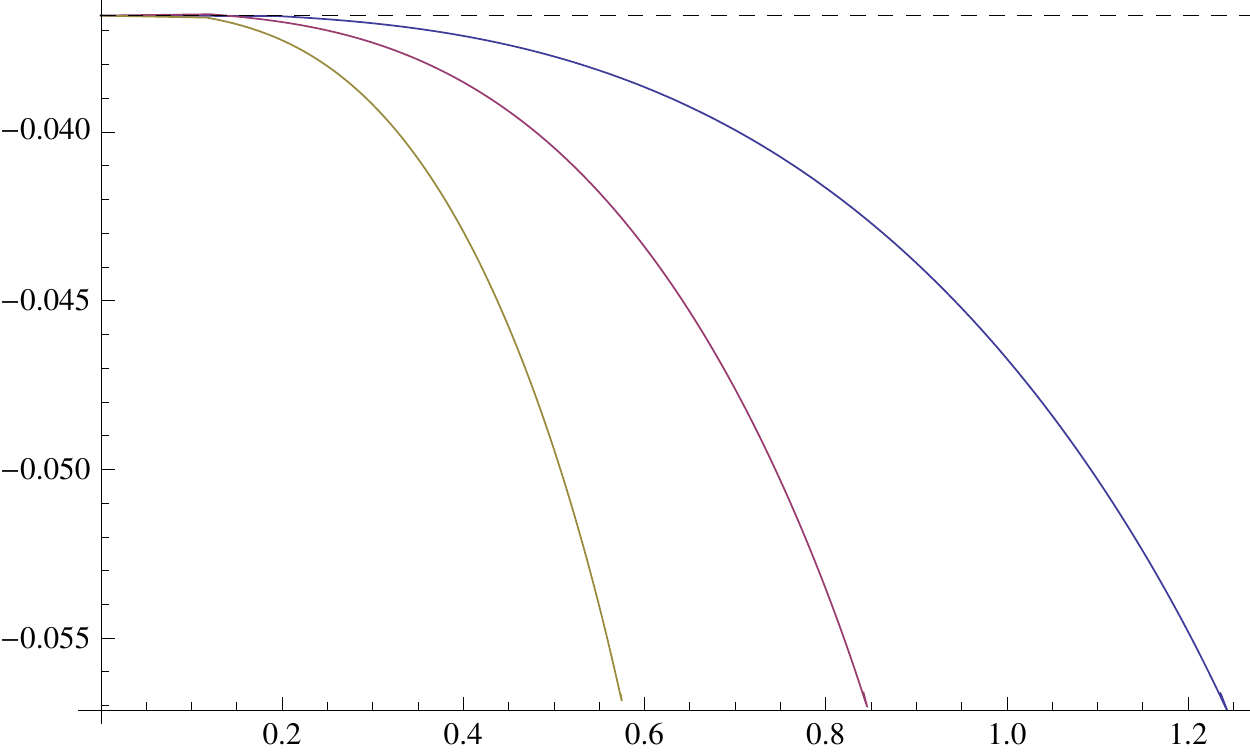}  
  \begin{picture}(0,0)(0,0)
  \put(-255,110){ $\frac{3^2\Delta\mathcal{F}L^2}{4^2\pi\mathbb{L} N_{2}^{3/2}\sqrt{\lambda_M}} $ }
  \put(-20,-10){ $ LT $}
  \end{picture}	
\end{subfigure}
\caption{On the left we have the free energy difference as a function of $\hat\sigma_0$ and on the right as a function of $LT$ for $\gamma=0$. The colour coding is the same as in the previous figure.}
\end{figure}

%%%%%%%%%%%%%%%%%%%%%%%%%%%%%%%%%%%%%%%%%%%%%%%%%%%%%%%%%%%%%%
\subsection{Black membranes in Schr\"{o}dinger} \label{smembranes}
In this section we construct several membrane geometries in Sch$_7\times S^{4}$, obtained via the null Melvin twist of the extremal M5-brane. This spacetime geometry can be written as \cite{Mazzucato:2008tr}
\beq \label{sds}
ds^2=\frac{R^2}{z^2}\left(-\frac{\ell^4}{z^4}dt^2+2dtd\xi+dz^2+dx_i^2\right)+\frac{R^2}{4}d\Omega^2_{(4)}~~,
\eeq
where $i=1,..,4$ and $\ell^2=\beta R^3$ measures the deformation of null-AdS into Schr\"{o}dinger.\footnote{Note that our parameter $\beta$ has been rescaled when compared with the parameter $\beta$ used in \cite{Mazzucato:2008tr}, namely, we have taken the metric in \cite{Mazzucato:2008tr} and performed the rescaling $\beta\to\beta\sqrt{2}$.} As in the case of the Sch$_5\times S^{5}$ introduced in \eqref{dsback}, boosts in the $(t,\xi)$ plane or rescalings of the Sch$_7$ coordinates imply that the physical quantities describing configurations at finite temperature and with characteristic lengths $L$  and $\mathbb{L}$ are given in terms of the dimensionless quantities
\beq \label{dim2}
\textbf{T}=T \ell~~,~\textbf{L}\equiv \frac{L}{\ell}~~,~~\tilde{\mathbb{L}}\equiv \frac{\mathbb{L}}{\ell}.
\eeq
In this spacetime we will construct four different types of stretched black membranes. The first configuration we construct is the analogue of the one presented in Sec.~\ref{interpolstring} and consists of an interpolating membrane between the bulk dual to the AdS Wilson surface presented in the previous section and a new Schr\"{o}dinger configuration. The boundary endpoints of these configurations are one-dimensional strings with length $\mathbb{L}$ and the remaining three configurations are differentiated depending on how these strings are placed: (i) spatially extended and spacelike separated, (ii) null extended and spacelike separated, and (iii) spatially extended and lightlike separated. 

Before analysing each configuration individually, we note that all these last three configurations share the same form of $\textbf{k}=R\ell^2/z^3$, therefore allowing us to make general statements about their validity. In particular, we can define the parameter $\bar\kappa$ as in \eqref{barkappa} via
\beq
\bar\kappa=\frac{1}{\tilde\sigma^{18}}\frac{\sinh\alpha(\tilde\sigma)}{\cosh^5\alpha(\tilde\sigma)}~~,~~\tilde\sigma=\left(\frac{2\pi}{3}\frac{T}{\ell^2}\right)^{\frac{1}{3}}\sigma~~,
\eeq 
where the rescaling of the coordinate $\tilde\sigma$ ensures that $\bar\kappa$ is the same both in AdS and in Schr\"{o}dinger. With this we can obtain the critical distance $\tilde\sigma_c$ beyond which these three black membranes in this spacetime cannot extend
\beq
\tilde\sigma_c^2=\frac{2^{\frac{4}{9}}}{5^{\frac{5}{18}}}\bar\kappa^{-\frac{1}{9}}~~.
\eeq
The remaining validity analysis of these configurations is similar to the case $\gamma=0$ studied in the previous section. One should, however, pay attention to the variation of the local temperature $r_c\mathcal{T}'/T=3r_c\sigma^2/(R\ell^2)\ll1$ which implies near the critical distance that
\beq
\kappa\ll\left(\frac{\ell}{T}\right)^{12}~~.
\eeq
Again, this requirement is generic for all the three types of black membranes mentioned above and imposes a mild bound on the ratio $\ell/T$. Furthermore, a careful analysis of the extrinsic curvature length $L_{\text{ext}}(\sigma)$, which can be evaluated explicitly using the information provided below for each black membrane, tell us that no new requirements appear. This indicates that these configurations are valid solutions all the way up to $\sigma_c$. We will now analyse each of these three cases and also the case of the interpolating membrane one-by-one but we will be succinct in the details since the method we follow has been repeatedly used in previous sections.

%%%%%%%%%%%%%%%%%%%%%%%%%%%%%%%%%%%%%%%%%%%%%%%%%%%%%%%%%%%%%%

\subsubsection*{Interpolating membranes with spatially extended and spacelike separated boundary strings}
In this section we construct a membrane configuration which interpolates between the Wilson surface analysed in the previous section, with $\gamma=0$, and a new Schr\"{o}dinger configuration when $\ell$ is large, which we will analyse separately in the next section. These are the analogue of the interpolating black strings studied in Sec.~\ref{interpolstring}. We begin by introducing the coordinates $(t',\xi')$ as in \eqref{newc} which brings the metric \eqref{sds} to the form.
\beq
ds^2=\frac{R^2}{z^2}\left[-y(z)dt'^2-\frac{\ell^4}{z^4}dt'd\xi'+\left(1-\frac{\ell^4}{2z^4}\right)d\xi'^2+dz^2+dx_i^2\right]+R^2d\Omega^{2}_{(4)}~~,~~g(z)=1+\frac{\ell^4}{2z^4}~~.
\eeq
By taking $\ell\to0$ we obtain the metric \eqref{mads} with $\gamma=0$ where $t'$ is identified with $t$ and $\xi'$ with $x_3$. We now choose an embedding map such that
\beq
t'=\tau~~,~~z=\sigma~~,~~x_1=x(\sigma)~~,~~x_2=\sigma_2~~,~~\xi'=0~~,~~x_i=0~,\forall i=3,4~~,~~d\Omega_{(4)}=0~~,
\eeq
which results in an induced metric of the form
\beq
\gamma_{ab}d\sigma^{a}d\sigma^{b}=\frac{R^2}{\sigma^2}\left(-y(\sigma)d\tau^2+(1+x'(\sigma)^2)d\sigma^2+d\sigma_2^2\right)~~.
\eeq
This metric reduces to \eqref{indM} when $\ell=0$. The pullback of the timelike Killing vector field $\partial_{t'}$ onto the world volume $\mathcal{W}_3$ gives rise to $\textbf{k}=(R/\sigma)\sqrt{y(\sigma)}$. Hence, if $\ell=0$ we obtain $\textbf{k}=R/\sigma$ which is the result obtained in AdS when $\gamma=0$, while if we take $\ell\to\infty$ we get $\textbf{k}\sim R\ell^2/\sigma^3$.

Solving now the equation of motion that arises from varying \eqref{freeenergy} leads to the familiar solution
\beq
x'(\sigma)=\left(\frac{U(\sigma)^2}{U(\sigma_0)^2}-1\right)^{-\frac{1}{2}}~~,~~U(\sigma)=\frac{R^9}{\sigma^9}y(\sigma)^{\frac{7}{2}}\frac{1+6\sinh^2\alpha}{\cosh^6\alpha}~~.
\eeq
We now focus on the zero-temperature case with $\kappa=0$. In order to do so we consider the regularised free energy which takes the form
\beq
\mathcal{F}_{\text{surf}}=\frac{4\mathbb{L}N_{(2)}^{3/2}}{\pi}\sqrt{\lambda_M}\left(u(\sigma_0)+\int_{0}^{\sigma_0}\frac{d\sigma}{\sigma^3}\sqrt{y(\sigma)}\left((1-X)\sqrt{1-x'(\sigma)^2}-1\right)\right)~~,
\eeq
where we have defined the function $u(\sigma_0)$ via
\beq
u(\sigma_0)=-\frac{1}{4}\left(\frac{\sqrt{\ell^4+2\sigma_0^4}}{\sqrt{2}\sigma_0^4}+\sqrt{2}\frac{\text{arcsinh}(\frac{\ell^2}{\sqrt{2}\sigma_0^2})}{\ell^2}\right)~~.
\eeq
The Polyakov surface can be argued to have zero free energy for any value of $\ell,\kappa,T$ with similar arguments as in the previous cases. We now obtain expressions for the length $L$ and $\mathcal{F}_{\text{surf}}$ as a function of $\sigma_0$ at $T=0$. This behaviour is shown in the plots below. 
\begin{figure}[H]
\centering
\begin{subfigure}{.5\textwidth}
  \centering
  \includegraphics[width=.8\linewidth]{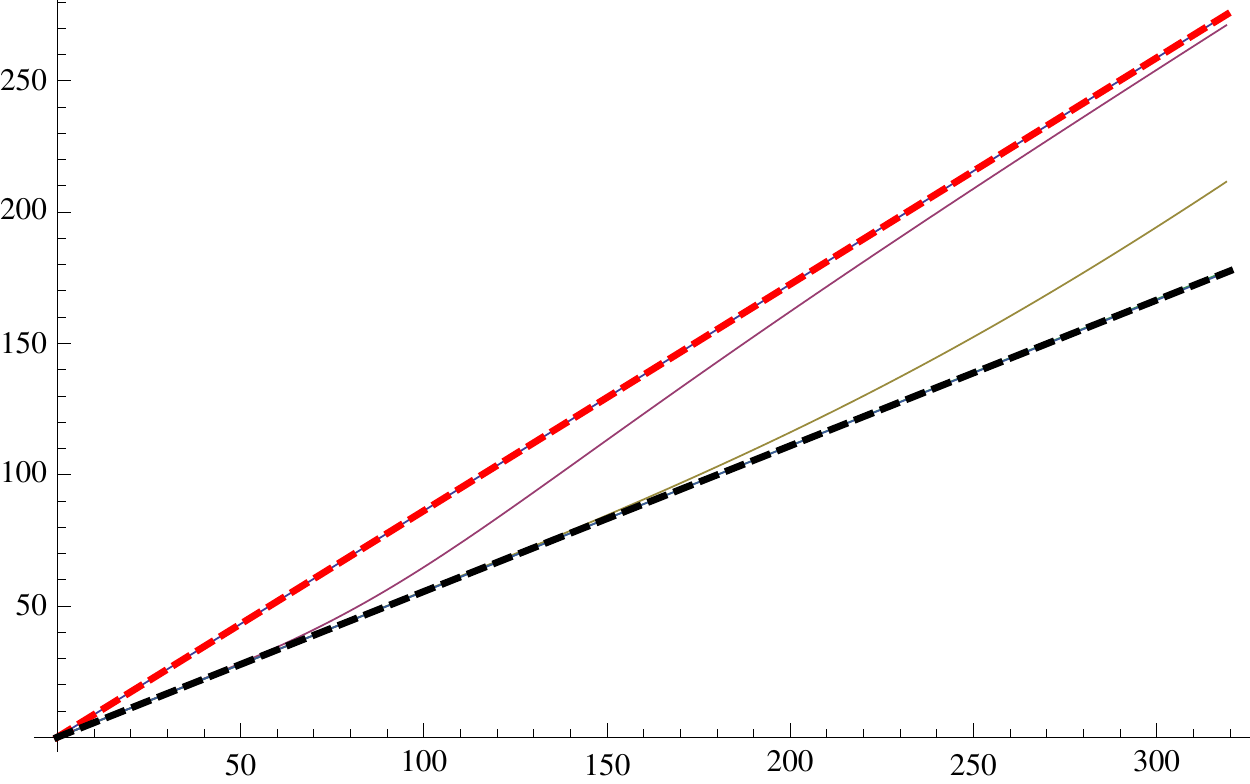}
  \begin{picture}(0,0)(0,0)
\put(-220,110){ $ L  $}
\put(-25,-8){ $ \sigma_0 $}
\end{picture}	
\end{subfigure}%
\begin{subfigure}{.5\textwidth}
  \centering
  \includegraphics[width=.8\linewidth]{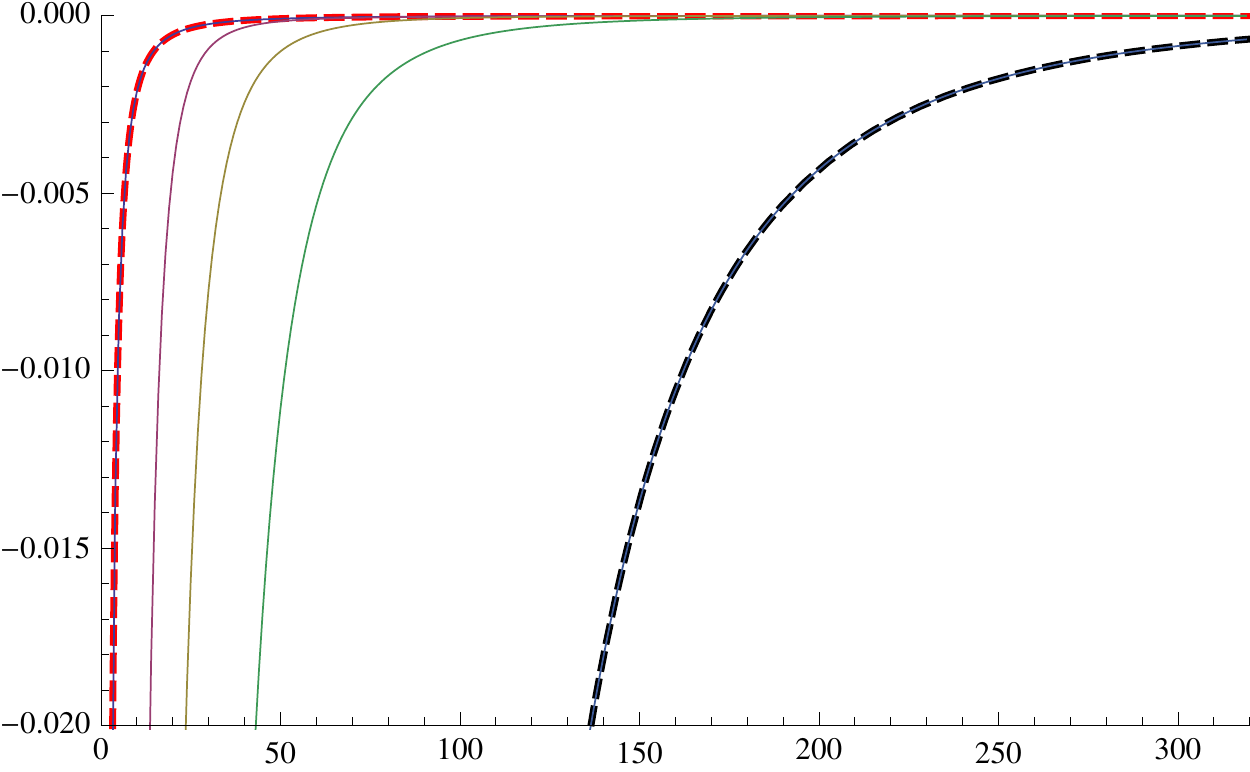}  
  \begin{picture}(0,0)(0,0)
  \put(-250,110){ $\frac{\mathcal{F}_{\text{surf}}\pi}{4\mathbb{L}\sqrt{\lambda_M}N_{(2)}^{3/2}} $}
  \put(-30,-5){ $\sigma_0$}
  \end{picture}	
\end{subfigure}
\caption{On the left we display the behaviour of the distance $L$ as function of $\sigma_0$ at $T=0$. The dashed solid red line corresponds to the AdS result given by the first term in \eqref{ltm} while the dashed solid black line corresponds to the result \eqref{rext1}. The blue line represents $(\ell/2^{1/4})=1$, the magenta line $(\ell/2^{1/4})=100$, the yellow line $(\ell/2^{1/4})=300$, the green line $(\ell/2^{1/4})=1000$ and the cyan line $(\ell/2^{1/4})=10000$. On the right is the behaviour of the free energy as a function of $\sigma_0$ at $T=0$. The dashed solid red line is the first term obtained in \eqref{fsurf} while the dash solid black line is the result obtained in \eqref{rext1}.} \label{ddc}
\end{figure}
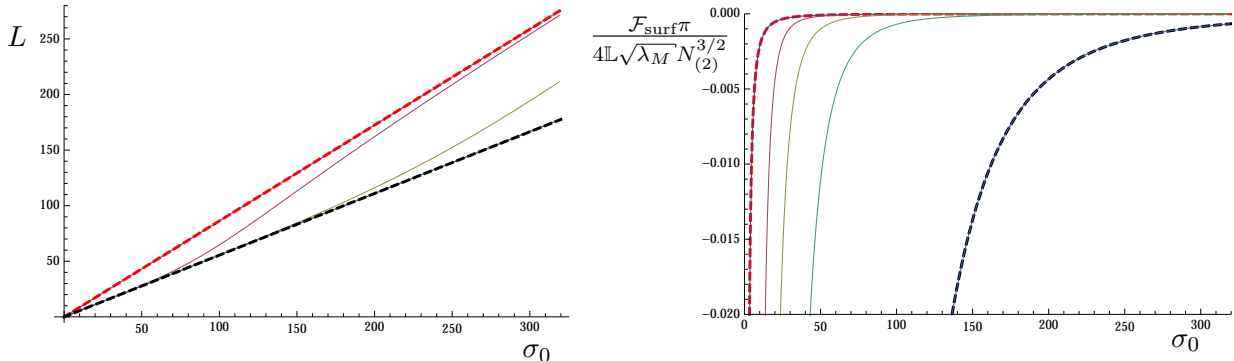
As one can see, the behaviour exhibited in the above plots resembles the case of black strings in Sec.~\ref{interpolstring}. The red dashed lines correspond to the leading order results for the length and free energy obtained in \eqref{ltm} and \eqref{fsurf} respectively. The black dashed lines correspond to the results which will be obtained in the next section and to leading order read
\beq \label{rext1}
L|_{\ell\to\infty}=2\sqrt{\pi}\sigma_0\frac{\Gamma(\frac{3}{5})}{\Gamma(\frac{1}{10})}~~,~~\frac{\mathcal{F}_{\text{surf}}\pi}{4\mathbb{L}\sqrt{\lambda_M}N_{(2)}^{3/2}}|_{\ell\to\infty}=\frac{\sqrt{\pi}\ell^2}{10\sigma_0^4}\frac{\Gamma(-\frac{2}{5})}{\Gamma(\frac{1}{10})}~~.
\eeq
In particular, values of the length $L$ and the free energy are bound between the AdS values and a purely Schr\"{o}dinger configuration which will be analysed in the next section. Furthermore, we observe that $\mathcal{F}_{\text{surf}}$ is proportional to $L^{-2}$ for small $ \ell$ and to $L^{-4}$ for large $ \ell$, since the relation between $L$ and $\sigma_0$ is roughly linear.

We can now consider the finite temperature versions of these geometries. However, all the qualitative features are the same as those presented in Sec.~\ref{interpolstring}. We will therefore only consider in detail the heated up version of the configuration in the limit $\ell\to\infty$ in the next section. Nevertheless, we note that at finite temperature we can define a parameter $\hat\ell$ such that
\beq
y(\hat\sigma)=1+\frac{\hat\ell^4}{\hat\sigma^4}~~,~~\hat\ell^4=\frac{2^{3}\pi^4 T^4\ell^4}{3^{4}}~~.
\eeq
In terms of this parameter and of the parameter $\bar\kappa$ defined \eqref{barkappa} we can find the critical distance for these configurations, however, its form is too cumbersome to be presented here. The resulting behaviour of the free energy is similar to what is presented in Fig.~\ref{Fiv}, namely, the solution terminates for a certain value of $\hat\sigma_c$.

%%%%%%%%%%%%%%%%%%%%%%%%%%%%%%%%%%%%%%%%%%%%%%%%%%%%%%%%%%%%%%

\subsubsection*{Membranes with spatially extended and spacelike separated boundary strings}
In this section we construct the membrane geometry which in the limit $\ell\to\infty$ connects with the interpolating membrane solution found in the previous section. In order to do so, we choose the embedding map 
\beq
t=\tau~~,~~z=\sigma~~,~~x_1=x(\sigma)~~,~~x_2=\sigma_2~~,~~\xi=x_i=0~,~\forall i=3,4~~,~~d\Omega_{(4)}=0~~,
\eeq
which is the analogous configuration in AdS$_7\times S^{4}$ to the one constructed in Sec.~\ref{blackspatial}. The world volume inherits the metric
\beq
\gamma_{ab}d\sigma^{a}d\sigma^{b}=\frac{R^2}{\sigma^2}\left(-\frac{\ell^4}{\sigma^4}d\tau^2+(1+x'(\sigma)^2)d\sigma^2+d\sigma_2^2\right)~~,
\eeq
which when introduced in \eqref{freeenergy} and after careful variation leads to the solution
\beq \label{x11}
x'(\sigma)=\left(\frac{K(\sigma)^2}{K(\sigma_0)^2}-1\right)^{\frac{1}{2}}~~,~~K(\sigma)=\frac{R^9\ell^{14}}{\sigma^{23}}\frac{1+6\sinh^2\alpha}{\cosh^6\alpha}~~,
\eeq
where we have imposed the same boundary conditions as in the previous section. The equations of motion, as usual, also allow for Polyakov surfaces with $x'(\sigma)=0$. 

Since these configurations are valid only for small $\bar\kappa$ we proceed and analyse their properties in a small $\bar\kappa$ expansion. We consider first the distance between the two strings on the boundary. With the help of the expansion
\beq  \label{expsn}
\cosh^2\alpha=\frac{1}{\tilde\sigma^9\sqrt{\bar\kappa}}-\frac{1}{4}-\frac{5}{32}\tilde\sigma^9\sqrt{\bar\kappa}+\mathcal{O}(\bar\kappa)~~,
\eeq
we find, to leading order in the bulk depth $\tilde\sigma_0$, the dimensionless distance
\beq \label{lf1}
\textbf{L}\textbf{T}^{\frac{1}{3}}\left(\frac{\pi}{12}\right)^{\frac{1}{3}}=\frac{\sqrt{\pi}\Gamma(\frac{3}{5})}{\Gamma(\frac{1}{10})}\tilde\sigma_0+\frac{\pi}{30} \left(\frac{2 \Gamma \left(\frac{3}{5}\right)}{\sqrt{\pi }
   \Gamma \left(\frac{1}{10}\right)}-1\right)\tilde\sigma_0^{10}\sqrt{\bar\kappa}+\mathcal{O}(\bar\kappa)~~.
\eeq
The leading order term, when written in terms of $\sigma_0$ leads to the result presented in \eqref{rext1}. We now take a look at the free energy. After implementing a similar regularisation scheme as in the previous sections we arrive at the regularised free energy
\beq \label{f11}
\boldsymbol{\mathcal{F}}_{(1)}[x(\tilde\sigma)]=c_n \tilde{\mathbb{L}} N_{(2)}^{\frac{3}{2}}\sqrt{\lambda_M}\textbf{T}^{\frac{4}{3}}\left(-\frac{1}{4\tilde\sigma_0^4}+\int_{0}^{\tilde\sigma_0}\frac{d\tilde\sigma}{\tilde\sigma^5}\left((1-X)\sqrt{1+x'(\tilde\sigma)^2}-1\right)\right)~~,
\eeq
where we have defined the constant $c_n=(2^{10}\pi/3^{4})^{1/3}$ and introduced the dimensionless quantities defined in \eqref{dim2}. The Polyakov surface, corresponding to the solution $x'(\sigma)=0$, can be argued, following the same reasoning as the previous section, to have zero free energy. Expanding now in small $\bar\kappa$ and small distance $L$, and inverting \eqref{lf1}, we find
\beq \label{fc1}
\boldsymbol{\mathcal{F}}_{(1)}=c_n \frac{\tilde{\mathbb{L}}}{\textbf{L}^{4}} N_{(2)}^{\frac{3}{2}}\sqrt{\lambda_M}\left(\frac{3^{4/3}\pi^{\frac{7}{6}}\Gamma \left(-\frac{2}{5}\right) \Gamma \left(\frac{3}{5}\right)^4}{2^{\frac{1}{3}}12500 \Gamma\left(\frac{1}{10}\right) \Gamma \left(\frac{11}{10}\right)^4}+C_1\sqrt{\bar\kappa}\textbf{L}^{9}\textbf{T}^{3}\right)+\mathcal{O}(\bar\kappa)+\mathcal{O}(\textbf{L}^{10})~~,
\eeq
where $C_1$ is a real numerical constant which we give in App.~\ref{app}. The leading order term in this equation, when written in terms of $\sigma_0$ leads to the result presented in \eqref{rext1}. From \eqref{fc1} we can see that the combination $\boldsymbol{\mathcal{F}}_{(1)}\textbf{L}^4/(\tilde{\mathbb{L}})$ is scale invariant as well as $\textbf{L}\textbf{T}^{1/3}$, hence exhibiting the scale invariance of the dual gauge theory. We can also depict the entire solution space. It exhibits the same qualitative features as those encountered in Sec.~\ref{blackspatial} for black strings in Schr\"{o}dinger with spacelike separated boundary endpoints. In particular there is no phase transition and this configuration is always the preferred one with respect to the Polyakov surface. We provide details in App.~\ref{app}.

%%%%%%%%%%%%%%%%%%%%%%%%%%%%%%%%%%%%%%%%%%%%%%%%%%%%%%%%%%%%%%

\subsubsection*{Membranes with null extended and spacelike separated boundary strings}
In order to construct these configurations we choose the embedding map
\beq
t=\tau~~,~~z=\sigma~~,~~x_1=x(\sigma)~~,~~\xi=\sigma_2~~,~~x_i=0~,~\forall i=2,3,4~~,~~d\Omega_{(4)}=0~~,
\eeq
leading to the stationary induced metric
\beq \label{dsnn}
\gamma_{ab}d\sigma^{a}d\sigma^{b}=\frac{R^2}{\sigma^2}\left(-\frac{\ell^4}{\sigma^4}d\tau^2+2d\tau d\sigma_2+(1+x'(\sigma)^2)d\sigma^2\right)~~.
\eeq
The boundary strings now lie in the null $\xi$-direction. Since these configurations are translational invariant along $\xi$, this direction does not play a role and can be integrated out leading to an overall factor of $\mathbb{L}$ in the free energy.  Using the induced metric \eqref{dsnn} and solving the equation of motion obtained by varying \eqref{freeenergy} we obtain the solution
\beq
x'(\sigma)=\left(\frac{Y(\sigma)^2}{Y(\sigma_0)^2}-1\right)^{\frac{1}{2}}~~,~~Y(\sigma)=\frac{R^9\ell^{12}}{\sigma^{21}}\frac{1+6\sinh^2\alpha}{\cosh^6\alpha}~~.
\eeq
We now study this solution in the small $\bar\kappa$ regime using the expansion \eqref{expsn}. To leading order in $\bar\kappa$ we find the dimensionless length
\beq \label{lf2}
\textbf{L}\textbf{T}^{\frac{1}{3}}\left(\frac{\pi}{12}\right)^{\frac{1}{3}}=\frac{\sqrt{\pi}\Gamma(\frac{2}{3})}{\Gamma(\frac{1}{6})}\tilde\sigma_0+\frac{\Gamma \left(\frac{2}{3}\right) \Gamma \left(\frac{5}{6}\right)+\sqrt{3} \Gamma \left(-\frac{2}{3}\right) \Gamma
   \left(\frac{13}{6}\right)}{18 \sqrt{\pi}}\sqrt{\bar\kappa}\tilde\sigma_0^{10}+\mathcal{O}(\bar\kappa)~~.
\eeq
The regularised free energy in turn reads
\beq \label{fcs}
\boldsymbol{\mathcal{F}}_{(2)}[x(\tilde\sigma)]=c_s \tilde{\mathbb{L}} N_{(2)}^{\frac{3}{2}}\sqrt{\lambda_M}\textbf{T}^{\frac{2}{3}}\left(-\frac{1}{2\tilde\sigma_0^2}+\int_{0}^{\tilde\sigma_0}\frac{d\tilde\sigma}{\tilde\sigma^3}\left((1-X)\sqrt{1+x'(\tilde\sigma)^2}-1\right)\right)~~,
\eeq
where we have defined the constant $c_s=(2^{8}/(3^2\pi))^{1/3}$. Note that the background subtraction here is different than the one used to regularise \eqref{f11}. Using the expansion \eqref{expsn} and inverting \eqref{lf2} we obtain
\beq  \label{fc2}
\boldsymbol{\mathcal{F}}_{(2)}=c_s \frac{\tilde{\mathbb{L}}}{\textbf{L}^2} N_{(2)}^{\frac{3}{2}}\sqrt{\lambda_M}\left(\frac{2^{\frac{1}{3}}\Gamma(-\frac{1}{3})^3}{3^{\frac{1}{3}}9\Gamma(\frac{1}{6})^3}+C_2\sqrt{\bar\kappa}\textbf{L}^{9}\textbf{T}^{3}\right)+\mathcal{O}(\bar\kappa)+\mathcal{O}(L^{10})~~,
\eeq
where $C_2$ is a real numerical constant that we give in App.~\ref{app}. The general form of $\boldsymbol{\mathcal{F}}_{(2)}$ is in accordance with scale invariance of the dual theory. The full solution space of these configurations is also very similar to the black strings constructed in Sec.~\ref{blackspatial} and hence we leave these details to App.~\ref{app}. However, we note that there is no phase transition with respect to the Polyakov surface and that this is always the preferred configuration. Also note that since we have made a different regularisation procedure in \eqref{f11} and \eqref{fcs} what is meant here by Polyakov surface is also different for both configurations.

%%%%%%%%%%%%%%%%%%%%%%%%%%%%%%%%%%%%%%%%%%%%%%%%%%%%%%%%%%%%%%
\subsubsection*{Membranes with spatially extended and lightlike separated boundary strings}
The last configuration we construct in this section is obtained by choosing the embedding map
\beq
t=\tau~~,~~z=\sigma~~,~~\xi=\xi(\sigma)~~,~~x_1=\sigma_2~~,~~x_i=0~,~\forall i=2,3,4~~,~~d\Omega_{(4)}=0~~,
\eeq
which leads to the induced metric
\beq
\gamma_{ab}d\sigma^{a}d\sigma^{b}=\frac{R^2}{\sigma^2}\left(-\frac{\ell^4}{\sigma^4}d\tau^2+2\xi'(\sigma)d\tau d\sigma +d\sigma^2+d\sigma_2^2\right)~~.
\eeq
Using this in the free energy \eqref{freeenergy} and solving the resultant equation of motion we find the solution
\beq
\xi'(\sigma)=\left(\frac{\sigma^8 K(\sigma)^2}{\ell^4 \tilde\sigma_0^4K(\sigma_0)^2}-\frac{\sigma^4}{\ell^4}\right)^{-\frac{1}{2}}~~,
\eeq
where the function $K(\sigma)$ has been defined in \eqref{x11}. Using now the expansion \eqref{expsn} we find the length
\beq
\textbf{L}\textbf{T}^{-\frac{1}{3}}\left(\frac{3}{2\pi}\right)^{\frac{1}{3}}=-\frac{\sqrt{\pi}\Gamma \left(\frac{1}{3}\right)}{\Gamma \left(-\frac{1}{6}\right)\tilde\sigma_0} -\frac{ \Gamma \left(\frac{1}{3}\right) \Gamma \left(\frac{7}{6}\right) \Gamma
   \left(\frac{4}{3}\right)+2 \pi  \Gamma \left(\frac{11}{6}\right)}{18\sqrt{\pi } \Gamma \left(\frac{4}{3}\right)}\tilde\sigma_0^8 \sqrt{\bar\kappa}+\mathcal{O}(\bar\kappa)~~.
\eeq
As we can see, the dimensionless distance $\textbf{L}\textbf{T}^{-\frac{1}{3}}$ is proportional to $\tilde\sigma_0^{-1}$ to leading order, meaning that similarly to the black strings with lightlike separated boundary endpoints  analysed in Sec.~\ref{nullstring}, the distance $\textbf{L}\textbf{T}^{-\frac{1}{3}}$ decreases for increasing bulk depth $\tilde\sigma_0$. In fact, numerically we can obtain how the distance $\textbf{L}\textbf{T}^{-\frac{1}{3}}$ depends on $\tilde\sigma_0$ for small values of $\bar\kappa$, which is depicted in the figure below.
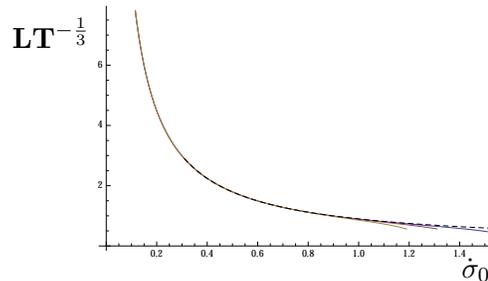
\begin{figure}[H]
\centering
  \includegraphics[width=0.3\linewidth]{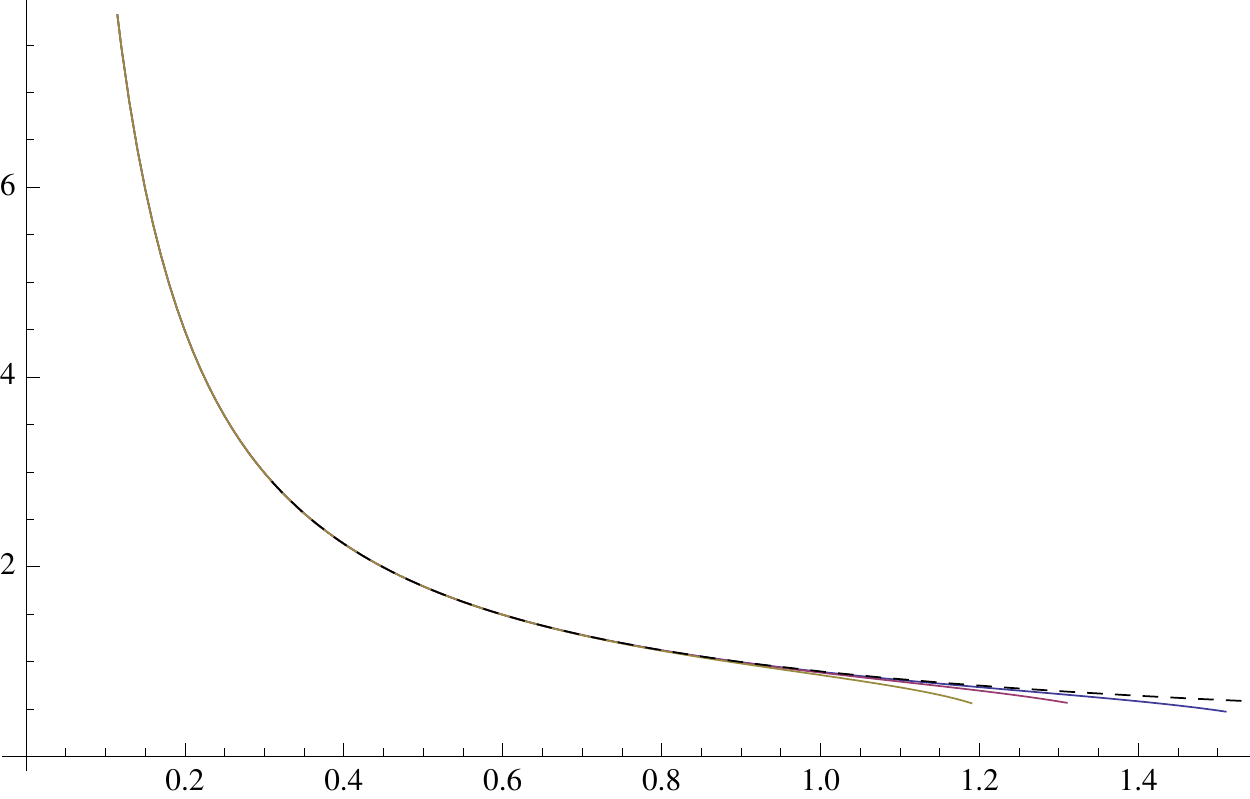}
  \begin{picture}(0,0)(0,0)
\put(-190,80){ $ \textbf{L}\textbf{T}^{-\frac{1}{3}}  $}
\put(-20,-7){ $ \dot \sigma_0 $}
\end{picture}	
\caption{The dimensionless distance $\textbf{L}\textbf{T}^{-\frac{1}{3}}$ as a function of $\tilde\sigma_0$ obtained numerically for $\bar\kappa=0.01$ (yellow curve), $\bar\kappa=0.001$ (red curve) and $\bar\kappa=0.0001$ (blue curve). The black dashed line represents $\bar\kappa=0$.}
\end{figure}
As seen from the plot above, the difference between this configuration and those which have been previously analysed, is that the dimensionless distance $\textbf{L}\textbf{T}^{-\frac{1}{3}}$ becomes unbounded as $\tilde\sigma_0\to0$. This means that in order to probe the effect of having a small bulk depth $\tilde\sigma_0$ one must study the behaviour of the free energy as $\textbf{L}\textbf{T}^{-\frac{1}{3}}\to\infty$. We now look at the regularised free energy for these configurations, which can be written as
\beq \label{ff3}
\boldsymbol{\mathcal{F}}_{(3)}[\xi(\tilde\sigma)]=c_n \tilde{\mathbb{L}} N_{(2)}^{\frac{3}{2}}\sqrt{\lambda_M}\textbf{T}^{\frac{4}{3}}\left(-\frac{1}{4\tilde\sigma_0^4}+\int_{0}^{\tilde\sigma_0}\frac{d\tilde\sigma}{\tilde\sigma^5}\left((1-X)\sqrt{1+\tilde\sigma^4\xi'(\tilde\sigma)^2}-1\right)\right)~~,
\eeq
where $c_n$ has been defined below Eq.~\eqref{f11}. Expanding this for small $\bar\kappa$ and large $\textbf{L}$ we find
\beq  \label{fc3}
\boldsymbol{\mathcal{F}}_{(3)}=c_n \tilde{\mathbb{L}} N_{(2)}^{\frac{3}{2}}\sqrt{\lambda_M}\textbf{L}^4\left(\frac{19683\times 3^{\frac{5}{6}}\Gamma(-\frac{1}{6})^2}{20000\pi^{\frac{4}{3}}\Gamma(-\frac{5}{3})^3\Gamma(\frac{1}{3})}+C_{3}\sqrt{\bar\kappa}\textbf{L}^{-9}\textbf{T}^{3}\right)+\mathcal{O}(\bar\kappa)+\mathcal{O}(\textbf{L}^{-10})~~,
\eeq
where $C_{3}$ is a real constant defined in App.~\ref{app}. The behaviour exhibited here also differs from the previous configurations. In particular, in the case $\bar\kappa=0$ the free energy is independent of $T$ and increasing for increasing distance $\textbf{L}$. Going to higher orders in $\bar\kappa$, for which the details are given in App.~\ref{app}, we can depict the solution space for these solutions. The result is given in the figures below.
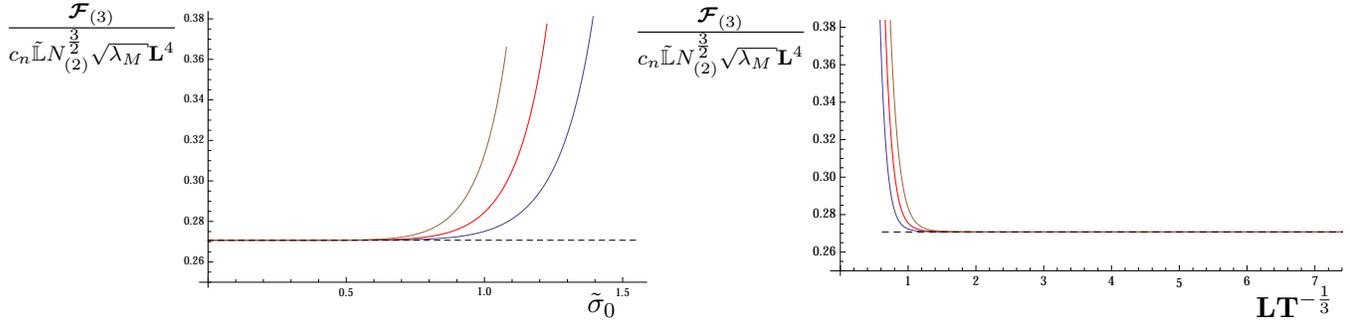
\begin{figure}[H]
\centering
\begin{subfigure}{.5\textwidth}
  \centering
  \includegraphics[width=0.7\linewidth]{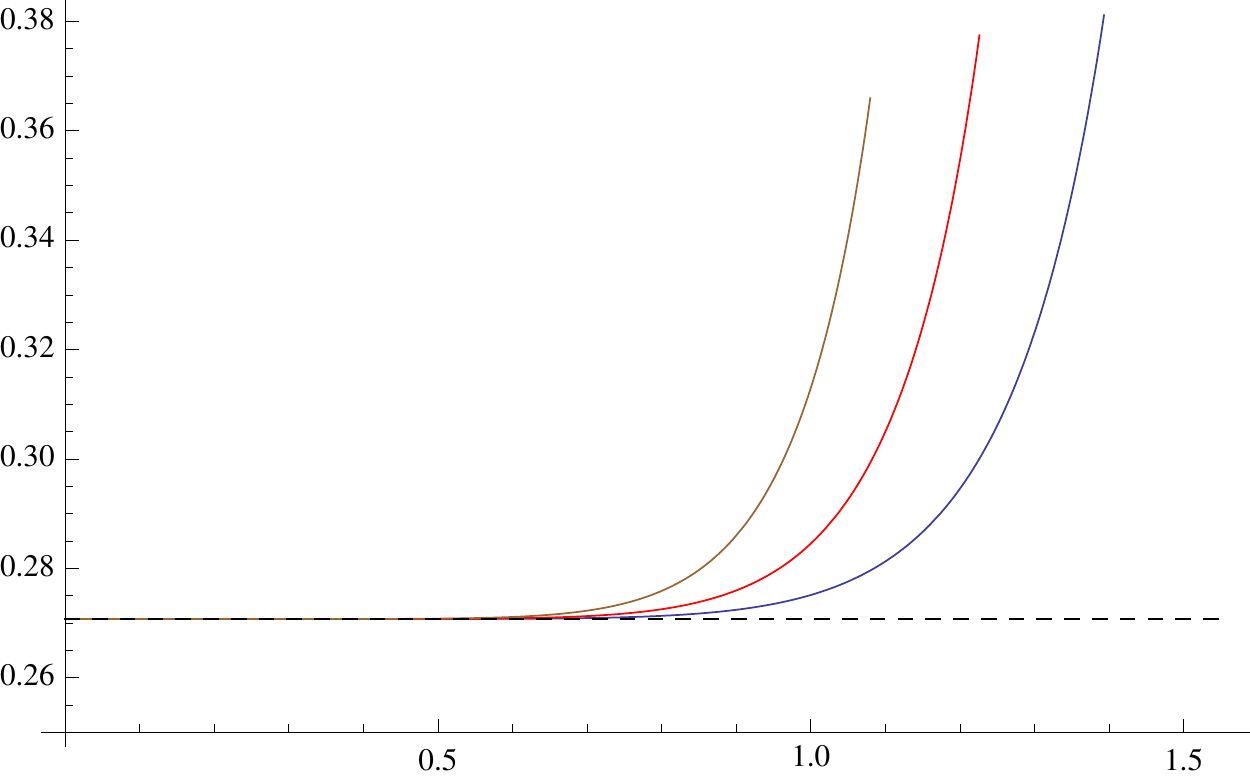}
  \begin{picture}(0,0)(0,0)
\put(-250,100){ $\frac{\boldsymbol{\mathcal{F}}_{(3)}}{c_n\tilde{\mathbb{L}}N^{\frac{3}{2}}_{(2)}\sqrt{\lambda_M}\textbf{L}^4}$}
\put(-30,-5){ $ \tilde\sigma_0 $}
\end{picture}	
\end{subfigure}%
\begin{subfigure}{.5\textwidth}
  \centering
  \includegraphics[width=0.8\linewidth]{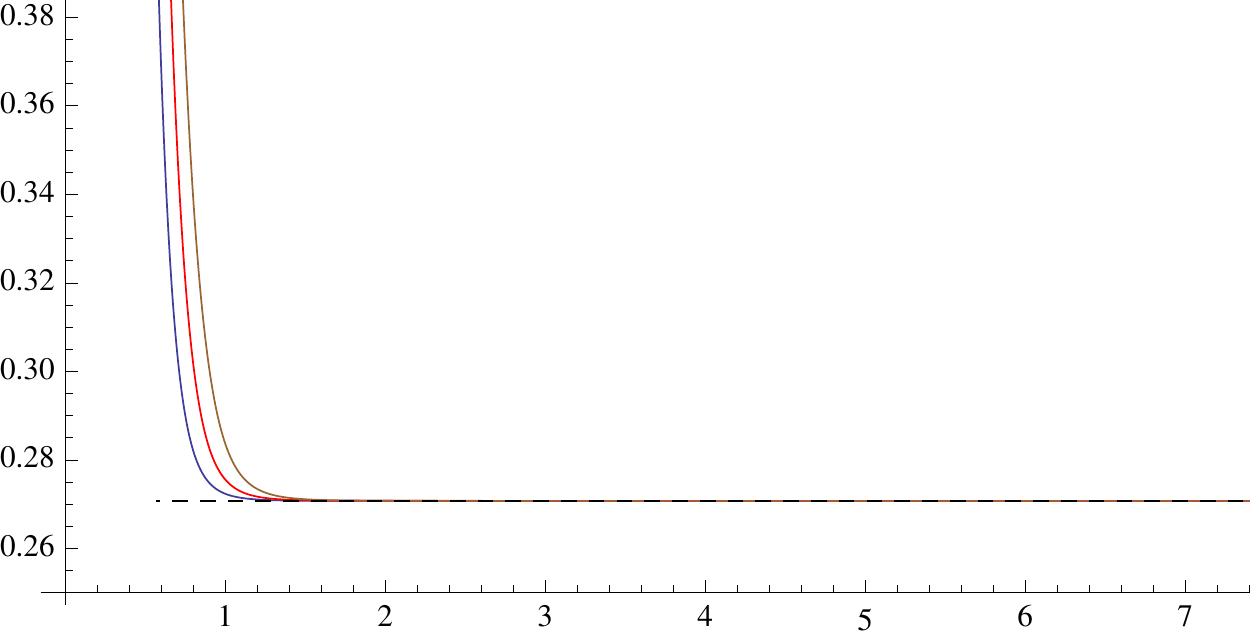}  
  \begin{picture}(0,0)(0,0)
  \put(-275,95){ $ \frac{\boldsymbol{\mathcal{F}}_{(3)}}{c_n\tilde{\mathbb{L}}N^{\frac{3}{2}}_{(2)}\sqrt{\lambda_M}\textbf{L}^4}$ }
  \put(-40,-10){ $\textbf{L}\textbf{T}^{-\frac{1}{3}}$}
  \end{picture}	
\end{subfigure}
\caption{On the left we have plotted the free energy $\boldsymbol{\mathcal{F}}_{(3)}$ a function of $\tilde\sigma_0$ and on the right as a function of the distance $\textbf{L}\textbf{T}^{-\frac{1}{3}}$. The colour coding is the same as in the previous figure.}
\end{figure}
Noting that the Polyakov surface has zero free energy, then the figure above represents the energy difference between these two configurations. Since this difference is always positive, one concludes that this configuration is not the preferred one with respect to the Polyakov surface. Since now we have several competing configurations at a given temperature $T$ it is interesting to compare which of the configurations is the preferred one. The free energies \eqref{ff1} and \eqref{ff3} have been regularised via the same background subtraction procedure so it is sensible to compare the two. Below we show the behaviour of the two free energies for $\bar\kappa=0.0001$ on the left while on the right we show the behaviour of $\Delta\boldsymbol{\mathcal{F}}_{(13)}=\boldsymbol{\mathcal{F}}_{(1)}-\boldsymbol{\mathcal{F}}_{(3)}$.
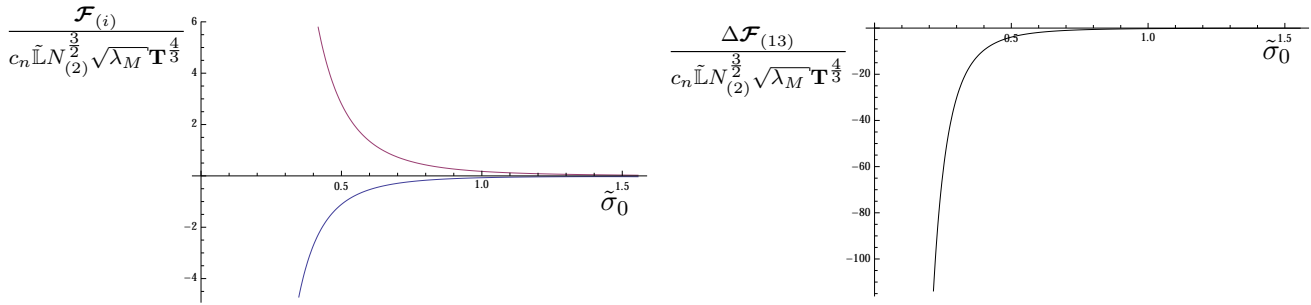
\begin{figure}[H]
\centering
\begin{subfigure}{.5\textwidth}
  \centering
  \includegraphics[width=0.7\linewidth]{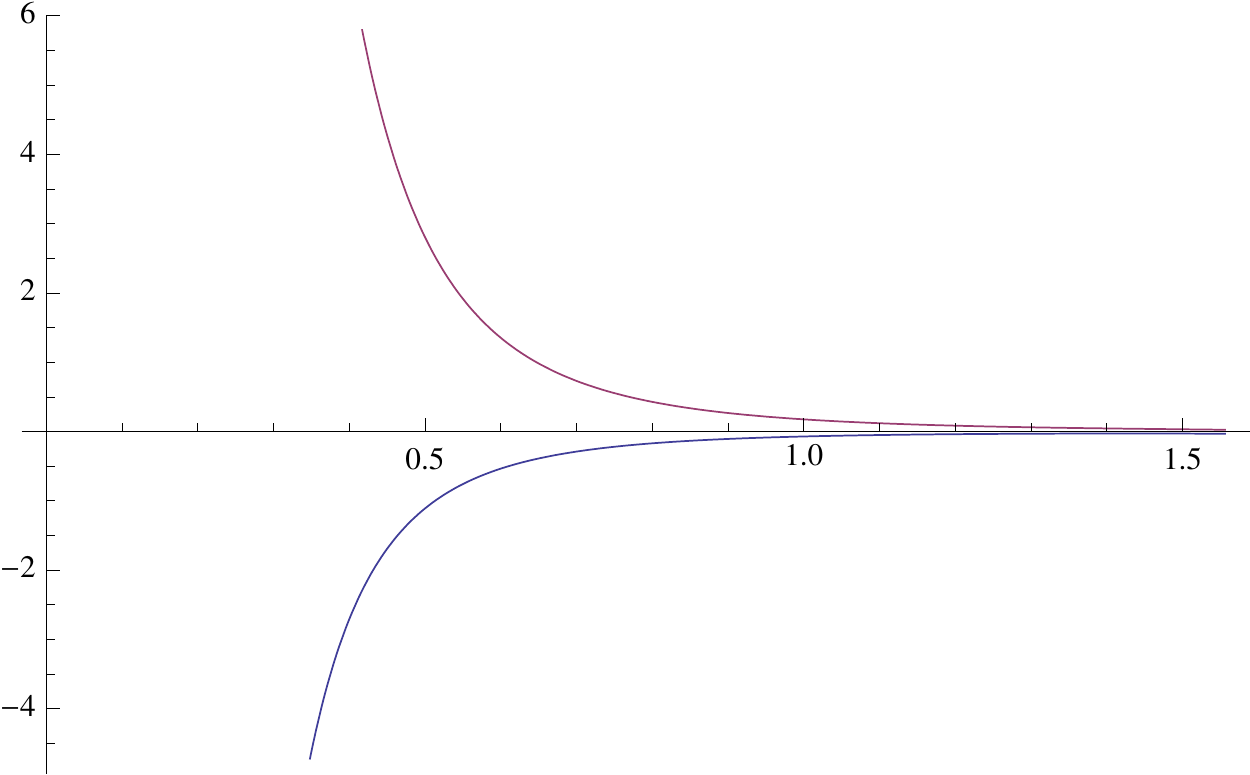}
  \begin{picture}(0,0)(0,0)
\put(-250,100){ $\frac{\boldsymbol{\mathcal{F}}_{(i)} }{c_n\tilde{\mathbb{L}}N^{\frac{3}{2}}_{(2)}\sqrt{\lambda_M}\textbf{T}^{\frac{4}{3}}}$}
\put(-25,35){ $ \tilde\sigma_0 $}
\end{picture}	
\end{subfigure}%
\begin{subfigure}{.5\textwidth}
  \centering
  \includegraphics[width=0.7\linewidth]{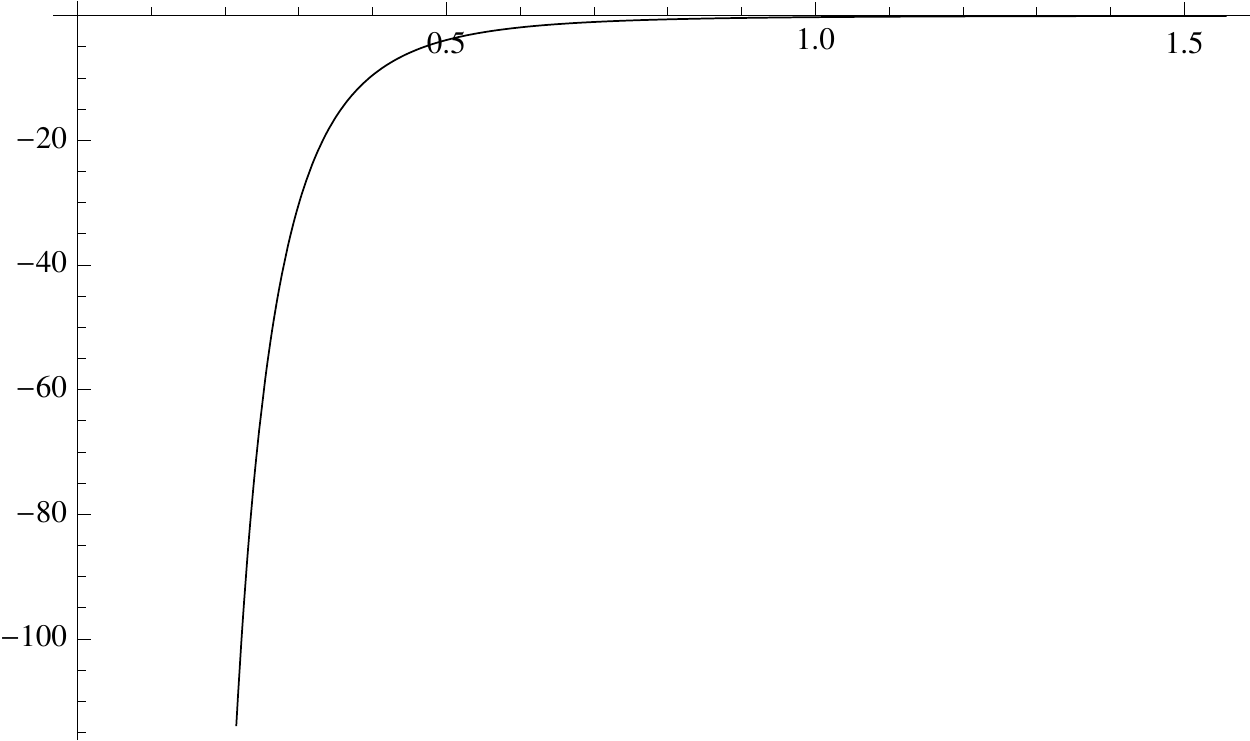}  
  \begin{picture}(0,0)(0,0)
  \put(-250,90){ $\frac{\Delta\boldsymbol{\mathcal{F}}_{(13)} }{c_n\tilde{\mathbb{L}}N^{\frac{3}{2}}_{(2)}\sqrt{\lambda_M}\textbf{T}^{\frac{4}{3}}}$ }
  \put(-25,90){ $\tilde\sigma_0$}
  \end{picture}	
\end{subfigure}
\caption{On the left as we display the free energy $\boldsymbol{\mathcal{F}}_{(1)}$ given by \eqref{f11} (blue curve) and  $\boldsymbol{\mathcal{F}}_{(3)}$ given by \eqref{ff3} (red curve) as a function of $\tilde\sigma_0$ for $\bar\kappa=0.0001$. On the right we have the free energy difference $\Delta\boldsymbol{\mathcal{F}}_{(13)}$ as a function of $\tilde\sigma_0$ for $\bar\kappa=0.0001$.}
\end{figure}
We see from the figures above that the free energy \eqref{f11} is always negative while \eqref{ff3} is always positive. The free energy difference satisfies $\Delta\boldsymbol{\mathcal{F}}_{(13)}<0$ and therefore the configuration with free energy $\boldsymbol{\mathcal{F}}_{(1)}$ given by \eqref{f11} is always the preferred configuration among the two for a given $\tilde{\mathbb{L}}, \textbf{T}, N_{(2)},\lambda_M, \tilde\sigma_0$ and $\bar\kappa$.

%%%%%%%%%%%%%%%%%%%%%%%%%%%%%%%%%%%%%%%%%%%%%%%%%%%%%%%%%%%%%%
\section{Discussion} \label{discussion}
In this paper we have constructed several new configurations of stretched black strings and black membranes in the probe approximation both in AdS$_5\times S^{5}$, Sch$_5\times S^{5}$ and in M-theory. According to the method we used, namely the blackfold approach, these solutions are valid supergravity solutions for which their metrics can be constructed perturbatively using a matched asymptotic expansion which takes into account back reaction effects. This paper has provided the first application of this method to the case of Schr\"{o}dinger backgrounds.

In the case of black strings in AdS$_5\times S^{5}$, we have reviewed previously constructed configurations in a black hole background \cite{Grignani:2012iw} and generalised it to thermal AdS. These configurations represent Wilson loop operators at finite temperature in the dual gauge theory at strong 't Hooft coupling. In this context the free energy of the black string yields the quark-antiquark potential which scales with the number of fundamental strings $k$ and is inversely proportional to the length $L$ of the pair. Following the black probe method leads to dominant finite temperature corrections in the quark-antiquark potential and to new physics, which can be observed already in thermal AdS, without the presence of a black hole.

In Sec.~\ref{stringschrodinger} we have constructed black strings in Sch$_5\times S^{5}$, whose near horizon geometry is composed of black, asymptotically flat, fundamental strings. In this setting we constructed the analogous configuration of the one found in AdS$_5\times S^{5}$ and compared the two cases. We found that there exists a string configuration that interpolates between the AdS Wilson loop and a new configuration in Sch$_5\times S^{5}$. Due to the presence of the null direction in Sch$_5\times S^{5}$ we also saw that it was possible, not only to have a configuration analogous to the one found in AdS$_5\times S^{5}$, but also to have a black string configuration in which its boundary endpoints are lightlike separated. This configuration turns out not to be the preferred one, having a higher value of its free energy. The differences between null AdS spaces and Schr\"{o}dinger spacetimes are subtle, namely when probing the spacetimes with particle or field probes \cite{Blau:2010fh}, but here we have seen that the black probe method allows to distinguish between these two spacetimes by placing probes along the null direction of Sch$_5\times S^{5}$ and hence leading to a configuration with no AdS counterpart.

According to the works of \cite{Maldacena:2008wh, Adams:2008wt, Herzog:2008wg}, backgrounds obtained via the null Melvin twist of supergravity branes should have well defined quantum field theory duals. In particular, in the case of Sch$_5\times S^{5}$ studied here, the dual quantum field theory is supposed to be obtained via the DLCQ$_\beta$ of $\mathcal{N}=4$ SYM and to lead to a non-relativistic CFT. In this line of thought it is relevant to ask if the black string geometries that we constructed here may be considered to be dual to Wilson loop operators defined in the gauge theory. In this context we note that the interpolating black string geometry in Sch$_5\times S^{5}$ in Sec.~\ref{interpolstring} analogous to the one in AdS$_5\times S^{5}$, has a free energy which scales with the total number of fundamental strings $k$ but now is inversely proportional to a certain power of $L$ of the quark-antiquark pair which can vary between $L^{-1}$ to $L^{-2}$ depending on the magnitude of the deformation parameter $\ell$ and for large enough $\ell$ it becomes proportional to $\ell$ itself. We have seen that for the case of the configurations of Sec.~\ref{blackspatial}, which connect at large $\ell$ to the interpolating strings, the combination $\mathcal{F}_{\text{loop}}L^2/\ell$ appearing in the free energy \eqref{ff1} is scale invariant and only depends on the scale invariant combination $L (T/\ell)^{1/2}$. This indeed suggests that the dual gauge theory is conformally invariant as previously suspected \cite{Maldacena:2008wh, Adams:2008wt, Herzog:2008wg} and furthermore that the deformation parameter $\ell$ has a physical role to play in the gauge theory. It is worth noting that the blackness of the probe, characterised by corrections in powers of $\kappa$, enters at the same order in the quark-antiquark potential as in its AdS counterpart and it is as sensitive to temperature corrections, which enter at order $T^3$, but more sensitive to small distances between the quark and the antiquark, scaling with $L^{-2}$ at zero temperature. One may also wonder about the dual interpretation of the black strings in Sch$_5\times S^{5}$ with boundary endpoints lightlike separated. However, the issue of whether or not  the null direction should be compactified for the purposes of holography is not yet settled, which leads to lack of intuition about the role of these configurations. Related to this, the exact details of the dual field theory action obtained by the DLCQ$_\beta$ of $\mathcal{N}=4$ SYM are still lacking but they would consequently allow for the study of expectation values of Wilson loop operators at zero temperature and to check if the corresponding result can be matched with the predictions made here. This endeavour would lead to a better understanding of holography in these spacetimes.

In Sec.~\ref{Mtheory} we have studied the case of stretched black membranes in AdS$_7\times S^{4}$ and found that, similarly to the case of AdS$_5\times S^{5}$, the black probe method leads to new physics and dominant contributions to the free energy which were not seen in the work of \cite{Chen:2008ds}. In particular, we see that there is a contribution of order $(TL)^3$ in the free energy of the black membrane \eqref{fm1} which is dominant over the contribution due to the black hole background entering at order $(TL)^6$. Since in this background the black membrane is less sensitive to the corrections due to the black hole, corrections due to the blackness of the probe become even more relevant and, for example, change the qualitatively behaviour of the onset of the Debye screening effect. In Sch$_7\times S^{4}$ we constructed analogous configurations and also three other black membrane configurations, depending on how we place its endpoints on the boundary geometry, and which have no AdS counterpart. We have in particular seen that the interpolating black membrane with an AdS counterpart, the free energy scales with the number of probe N2-branes $N_{(2)}^{\frac{3}{2}}$ and with a power of $L$ that varies from $L^{-2}$ to $L^{-4}$ as the deformation parameter $\ell$ is increased continuously to higher and higher values. Again, assuming the existence of a dual gauge theory and that these configurations represent Wilson surfaces, we see that the free energy \eqref{fc1} suggests that the dual non-realivistic gauge theory is conformally invariant. One may also wonder if it is possible to check this dependence in the expectation values of Wilson surface operators in Sch$_7\times S^{4}$ as it was recently done in \cite{Mori:2014tca} for the same operators in AdS$_7\times S^{4}$.

We also note that all the black strings and black membranes constructed in Schr\"{o}dinger spacetimes in this paper were done in thermal Schr\"{o}dinger, without the presence of a black hole. Schr\"{o}dinger black branes can be obtained via the null Melvin twist of supergravity black branes but generate non-trivial dilatonic fields \cite{Maldacena:2008wh, Adams:2008wt, Herzog:2008wg}. Within the blackfold approach, coupling black probes to dilatonic background fields has not been yet properly understood, though currently under study \cite{Armasnew}. Therefore, we have not considered such background geometries in this work but we admit that it would be worthwhile studying them as one would most likely observe non-trivial phase transitions and Debye screening effects. It is worth mentioning that all the configurations that we constructed both in thermal AdS and thermal Schr\"{o}dinger do not exhibit Debye screening effects, suggesting that it would be interesting to understand whether a general argument ruling out the existence of these effects exists. 

Finally, we note that applying the null Melvin twist on supergravity branes leads to black branes whose near horizon geometry exhibits the Schr\"{o}dinger symmetry group \cite{Maldacena:2008wh, Adams:2008wt, Herzog:2008wg, Mazzucato:2008tr}. It would be interesting to take these branes and use them to probe the spacetime and in the process construct new black hole geometries with non-relativistic near-horizon geometry. We leave this possibility for future studies.

%%%%%%%%%%%%%%%%%%%%%%%%%%%%%%%%%%%%%%%%%%%%%%%%%%%%%%%%%%%%%%
\section*{Acknowledgements}
We would like to thank Bin Chen, Troels Harmark, Niels Obers and Donovan Young for useful discussions. We specially thank Jelle Hartong for many useful discussions and for the careful reading of an early draft of this paper. JA is very thankful to Irene Amado for basic lessons in numerical integration and to Adolfo Guarino for important tips. JA would like to thank the organisers of the workshop \textbf{Black Holes and Quantum Information} at the Weizmann Institute of Science (2014), the organisers of the workshop \textbf{New Frontiers in Dynamical Gravity} at the University of Cambridge (2014) and the organisers of the workshop \textbf{Holography for Black Holes and Cosmology} at ULB (2014), where part of this work was carried out. This research has been supported by the Swiss National Science Foundation and the `Innovations- und Kooperationsprojekt C-13' of the Schweizerische Universit\"{a}tskonferenz SUK/CUS.

\appendix

\section{Higher order expansions for black membranes in Schr\"{o}dinger} \label{app}
In this appendix we collect some details on the solutions presented in Sec.~\ref{smembranes}. In particular we give higher order expansions in $\bar\kappa$ for the dimensionless distances and free energies characterising each solution and which allowed us to analyse the solution space.

\subsection*{Membranes with boundary strings spatially extended and spacelike separated}
We have solved for the dimensionless distance numerically and seen that it is well approximated for small $\bar\kappa$ and up to values $\sigma\sim\sigma_c$ by the expansion.
\beq \label{a1}
\begin{split}
\textbf{L}\textbf{T}^{\frac{1}{3}}\left(\frac{\pi}{12}\right)^{1/3}=&\frac{\sqrt{\pi}\Gamma(\frac{3}{5})}{\Gamma(\frac{1}{10})}\tilde\sigma_0+\frac{\pi}{30} \left(\frac{2 \Gamma \left(\frac{3}{5}\right)}{\sqrt{\pi }
   \Gamma \left(\frac{1}{10}\right)}-1\right)\sqrt{\bar\kappa}\tilde\sigma_0^{10}+\frac{\pi\left(\frac{\frac{276 \Gamma \left(\frac{8}{5}\right)}{\Gamma \left(\frac{1}{10}\right)}+\frac{11 \Gamma
   \left(\frac{17}{5}\right)}{\Gamma \left(\frac{19}{10}\right)}}{\sqrt{\pi }}-96\right)}{4320}\bar\kappa\tilde\sigma_0^{19}\\
   &+\frac{\sqrt{\pi } \left(-346500 \sqrt{\pi }+\frac{950675 \Gamma \left(\frac{8}{5}\right)}{\Gamma
   \left(\frac{1}{10}\right)}+\frac{193116 \Gamma \left(\frac{7}{5}\right)}{\Gamma \left(\frac{19}{10}\right)}-\frac{4650 \Gamma
   \left(\frac{43}{10}\right)}{\Gamma \left(\frac{14}{5}\right)}\right)}{2\times8910000 }\bar\kappa^{3/2}\tilde\sigma_0^{28}+\mathcal{O}(\bar\kappa^2)~~.
   \end{split}
\eeq
Using the above expansion and a similar one for the free energy \eqref{f11},
\beq\label{bb1}
\begin{split}
\frac{\boldsymbol{\mathcal{F}}_{(1)}}{c_n\tilde{\mathbb{L}}N^{\frac{3}{2}}_{(2)}\sqrt{\lambda_M}\textbf{T}^{\frac{4}{3}}}=& \frac{\sqrt{\pi } \Gamma \left(-\frac{2}{5}\right)}{10  \Gamma \left(\frac{1}{10}\right)\tilde\sigma_0^4}-\frac{1}{135} \sqrt{\pi }  \left(9 \sqrt{\pi }+\frac{10 \Gamma \left(\frac{3}{5}\right)}{\Gamma
   \left(-\frac{9}{10}\right)}\right)\sqrt{\bar\kappa }\tilde\sigma_0^5 \\
   &-\frac{\sqrt{\pi } \left(720 \sqrt{\pi }-\frac{418 \Gamma \left(\frac{7}{5}\right)}{\Gamma
   \left(\frac{9}{10}\right)}-\frac{147 \Gamma \left(\frac{8}{5}\right)}{\Gamma \left(\frac{11}{10}\right)}\right)}{32400}\bar\kappa\tilde\sigma_0^{14}\\
   &-\frac{\sqrt{\pi }  \left(61200 \sqrt{\pi }+\frac{8463 \Gamma \left(\frac{13}{10}\right)}{\Gamma
   \left(\frac{4}{5}\right)}-\frac{46816 \Gamma \left(\frac{7}{5}\right)}{\Gamma \left(\frac{9}{10}\right)}-\frac{13422 \Gamma
   \left(\frac{8}{5}\right)}{\Gamma \left(\frac{11}{10}\right)}\right)}{3888000}\bar\kappa^{3/2}\tilde\sigma_0^{23}+\mathcal{O}(\bar\kappa^2)~~,
\end{split}
\eeq
where $c_n$ was defined below \eqref{f11}, we are able to analyse the solution space. Inverting \eqref{a1} and introducing in \eqref{bb1} leads to the value of $C_1$ introduced in \eqref{fc1},
\beq
C_1=\frac{25 \left(9 \sqrt{\pi } \Gamma \left(-\frac{9}{10}\right)+10 \Gamma \left(\frac{3}{5}\right)\right) \Gamma
   \left(\frac{11}{10}\right)^4}{9\times 3^{\frac{2}{3}} (2\pi)^{\frac{1}{3}}\Gamma \left(\frac{3}{5}\right)^5}  ~~.
    \eeq

\subsection*{Membranes with boundary strings null extended and spacelike separated}
For these configurations we have also studied numerically the dimensionless distance and seen that it is well approximated by the following expansion,
   \beq \label{a2}
\begin{split}
\textbf{L}\textbf{T}^{\frac{1}{3}}&\left(\frac{\pi}{12}\right)^{1/3}=\frac{\sqrt{\pi }\Gamma \left(\frac{2}{3}\right)}{ \Gamma \left(\frac{1}{6}\right)}\tilde\sigma_0+\frac{\Gamma \left(\frac{2}{3}\right) \Gamma \left(\frac{5}{6}\right)+\sqrt{3} \Gamma \left(-\frac{2}{3}\right) \Gamma
   \left(\frac{13}{6}\right)}{18 \sqrt{\pi}}\sqrt{\bar\kappa}\tilde\sigma_0^{10}\\
  &+ \sqrt{\pi }\frac{\frac{82779 \Gamma \left(\frac{2}{3}\right)}{\Gamma \left(\frac{1}{6}\right)}-\frac{25480 \Gamma \left(\frac{7}{6}\right)}{\Gamma
   \left(\frac{5}{3}\right)}}{2\times88452}\bar\kappa\tilde\sigma_0^{19}+\sqrt{\pi }\frac{\frac{24316558464 \Gamma \left(\frac{2}{3}\right)}{\Gamma \left(\frac{1}{6}\right)}-\frac{6522475505 \Gamma
   \left(\frac{7}{6}\right)}{\Gamma \left(\frac{5}{3}\right)}}{2\times13450364928 }\bar\kappa^{3/2}\tilde\sigma_0^{28}+\mathcal{O}(\bar\kappa^2)~~.
   \end{split}
   \eeq
Expanding the free energy \eqref{fcs} in powers of $\bar\kappa$,
\beq \label{b2}
\begin{split}
&\frac{\boldsymbol{\mathcal{F}}_{(2)}}{c_s\tilde{\mathbb{L}}N^{\frac{3}{2}}_{(2)}\sqrt{\lambda_M}\textbf{T}^{\frac{2}{3}}}=\frac{\sqrt{\pi } \Gamma \left(-\frac{1}{3}\right)}{6 \Gamma \left(\frac{1}{6}\right)\tilde\sigma_0^2 }+\frac{\sqrt{\pi }  \left(\frac{2 \Gamma \left(\frac{2}{3}\right)^2}{\Gamma \left(\frac{1}{6}\right)}-5 \Gamma
   \left(\frac{7}{6}\right)\right)}{18 \Gamma \left(\frac{2}{3}\right)}\sqrt{\bar\kappa}\tilde\sigma_0^7\\
   &+\frac{\sqrt{\pi }  \left(\frac{92823 \Gamma \left(\frac{2}{3}\right)}{\Gamma \left(\frac{1}{6}\right)}-\frac{25480
   \Gamma \left(\frac{7}{6}\right)}{\Gamma \left(\frac{5}{3}\right)}\right)}{176904}\bar\kappa\tilde\sigma_0^{16}
 +\frac{\sqrt{\pi }  \left(\frac{29744167200 \Gamma \left(\frac{2}{3}\right)}{\Gamma
   \left(\frac{1}{6}\right)}-\frac{7686566251 \Gamma \left(\frac{7}{6}\right)}{\Gamma \left(\frac{5}{3}\right)}\right)}{33625912320}\bar\kappa^{3/2}\tilde\sigma_0^{25}+\mathcal{O}(\bar\kappa^2)~~,
\end{split}
\eeq
where $c_s$ was defined below \eqref{fcs}, allows us to probe the solution space. Inverting \eqref{a2} and using it in \eqref{b2} allows us to derive the constant $C_2$ introduced in \eqref{fc2},
\beq
C_2=\frac{3^{\frac{1}{6}} \Gamma \left(\frac{1}{6}\right)^7 \left(45\times 2^{\frac{1}{3}} \sqrt{3 \pi } \Gamma \left(\frac{7}{6}\right)-7 \Gamma
   \left(-\frac{2}{3}\right)^2\right)}{64 \pi ^{7/6} \Gamma \left(-\frac{1}{3}\right)^7 \Gamma \left(\frac{2}{3}\right)}~~.
   \eeq
   
\subsection*{Membranes with boundary strings spatially extended and lightlike separated}
At the end of Sec.~\ref{smembranes} we have studied numerically the dimensionless length $\mathbf{L}$. We have checked that its behaviour is well approximated for small $\bar\kappa$ by the expansion
\beq \label{a3}
\begin{split}
\textbf{L}\textbf{T}^{-\frac{1}{3}}\left(\frac{3}{2\pi}\right)^{\frac{1}{3}}&=-\frac{\sqrt{\pi} \Gamma \left(\frac{1}{3}\right)}{ \Gamma \left(-\frac{1}{6}\right)\tilde\sigma_0} -\frac{ \left(\Gamma \left(\frac{1}{3}\right) \Gamma \left(\frac{7}{6}\right) \Gamma
   \left(\frac{4}{3}\right)+2 \pi  \Gamma \left(\frac{11}{6}\right)\right)}{18\sqrt{\pi } \Gamma \left(\frac{4}{3}\right)}\sqrt{\bar\kappa}\tilde\sigma_0^8 \\
   &+\frac{\sqrt{\pi}  \left(4400 \Gamma \left(-\frac{1}{6}\right)^2-13713 \Gamma
   \left(\frac{1}{3}\right)^2\right)}{2\times53460 \Gamma \left(-\frac{1}{6}\right) \Gamma \left(\frac{1}{3}\right)} \bar\kappa\tilde\sigma_0^{17} \\
   &+\left(\frac{56722777231 \Gamma \left(\frac{7}{6}\right) \Gamma \left(\frac{4}{3}\right)}{2\sqrt{\pi }\times58193562240}-\frac{151202389099
   \sqrt{\frac{1 }{3}} \Gamma \left(\frac{5}{6}\right) \Gamma \left(\frac{5}{3}\right)}{2\sqrt{\pi}\times92748644352}\right)\bar\kappa^{3/2}\tilde\sigma_0^{26}+\mathcal{O}(\bar\kappa^2) ~~.
   \end{split}
\eeq
A similar expansion in the free energy \eqref{ff3},
\beq\label{b3}
\begin{split}
&\frac{\boldsymbol{\mathcal{F}}_{(3)}}{c_n\tilde{\mathbb{L}}N^{\frac{3}{2}}_{(2)}\sqrt{\lambda_M}\textbf{T}^{\frac{4}{3}}}=\frac{\sqrt{\pi } \Gamma \left(-\frac{2}{3}\right)}{6  \Gamma \left(-\frac{1}{6}\right)\tilde\sigma_0^4}-\frac{ \left(8 \pi  \Gamma \left(\frac{5}{6}\right)+\Gamma \left(\frac{1}{3}\right)^2 \Gamma
   \left(\frac{7}{6}\right)\right)}{18 \sqrt{\pi } \Gamma \left(\frac{1}{3}\right)}\sqrt{\bar\kappa } \tilde\sigma_0^5\\
   &-\frac{\sqrt{\pi } \left(\frac{19869 \Gamma \left(\frac{1}{3}\right)}{\Gamma
   \left(-\frac{1}{6}\right)}+\frac{8800 \Gamma \left(\frac{5}{6}\right)}{\Gamma \left(\frac{4}{3}\right)}\right)}{106920}\bar\kappa  \tilde\sigma_0^{14} -\frac{\sqrt{\pi }  \left(\frac{1645699776 \Gamma \left(\frac{1}{3}\right)}{\Gamma
   \left(-\frac{1}{6}\right)}+\frac{629014375 \Gamma \left(\frac{5}{6}\right)}{\Gamma \left(\frac{4}{3}\right)}\right)}{5173217280}\bar\kappa^{3/2}\tilde\sigma_0^{23}+\mathcal{O}(\bar\kappa^2)~~,
   \end{split}
\eeq
has allowed us to depict the solution space in Sec.~\ref{smembranes}. Inverting \eqref{a3} and plugging it into \eqref{b3} yields the constant $C_3$ introduced in \eqref{fc3},
\beq
C_3=\frac{1309000000 \left(\frac{2}{3}\right)^{2/3} \pi ^{11/3} \Gamma \left(-\frac{20}{3}\right) \Gamma
   \left(-\frac{5}{3}\right)^3 \Gamma \left(-\frac{5}{6}\right)}{387420489 \Gamma \left(-\frac{1}{6}\right)^3}~~.
   \eeq

\addcontentsline{toc}{section}{References}
\footnotesize
\providecommand{\href}[2]{#2}\begingroup\raggedright\endgroup

\end{document}